\documentclass[
 reprint,
 showkeys,
 amsmath,amssymb,
 aps,
 floatfix,
]{revtex4-2}

\usepackage{graphicx}% Include figure files
\usepackage{dcolumn}% Align table columns on decimal point
\usepackage{bm}% bold math
\usepackage{multirow} % Required for multirows
\newcommand{\subfigimg}[3][,]{%
  \setbox1=\hbox{\includegraphics[#1]{#3}}% Store image in box
  \leavevmode\rlap{\usebox1}% Print image
  \rlap{\hspace*{5pt}\raisebox{\dimexpr\ht1-1\baselineskip}{#2}}% Print label
  \phantom{\usebox1}% Insert appropriate spacing
}

\begin{document}

\title{Exchanging replicas with unequal cost, infinitely and permanently}% Force line breaks with \\

\author{Sander Roet}
\author{Daniel T. Zhang}%
\author{Titus S. van Erp}
 \email{titus.van.erp@ntnu.no}
\affiliation{%
Department of Chemistry, Norwegian University of Science and Technology (NTNU),\\ N-7491 Trondheim, Norway
}

\date{\today}

\begin{abstract}
We developed a replica exchange method that is effectively parallelizable even if the computational cost of the Monte Carlo moves in the parallel replicas are considerably different, for instance, because the replicas run on different type of processor units or 
because of the algorithmic complexity. 
To prove detailed-balance, we make a paradigm shift 
from the common conceptual viewpoint in which the set of parallel replicas represents a high-dimensional superstate, to 
an ensemble based criterion in which the other ensembles represent 
an environment that might or might not participate in the Monte Carlo move.
In addition, 
based on a recent algorithm for computing permanents,
we effectively increase the exchange rate to infinite
without
the steep factorial scaling as function of the number of replicas. 
We illustrate the effectiveness of the replica exchange methodology by combining it with 
a quantitative path sampling method, 
replica exchange transition interface sampling (RETIS), 
in which the costs for a Monte Carlo move can vary enormously as paths in a RETIS algorithm do not have the same length
 and the average path lengths tend to vary considerably for the different path ensembles that run in parallel.
  This combination, coined $\infty$RETIS, was tested on three model 
  systems. 
\end{abstract}


\maketitle

The Markov chain Monte Carlo (MC) method
is one of the most important 
numerical techniques for computing averages in high-dimensional 
spaces, like the configuration space of a many particle system. 
The approach has applications in a wide variety of fields ranging from
computational physics, theoretical chemistry, economics, and genetics.
The MC algorithm effectively generates
a selective random walk through 
state space
in which the artificial steps are designed 
such to ensure that the frequency of visiting any particular state is proportional to
the equilibrium probability of that state. 
The Metropolis~\cite{Metroplois} or the more general Metropolis-Hastings~\cite{Hastings} algorithms
are the most common approaches for 
designing such random steps (MC moves)
 based on the detailed-balance principle.
That is, the MC moves should 
be constructed such that the
number of transition from an old state $s^{(o)}$ to a 
new state $s^{(n)}$
is exactly balanced by the number of transitions from the new to the old state: 
 $\rho(s^{(o)})\pi(s^{(o)} \rightarrow  s^{(n)})=
 \rho(s^{(n)})\pi(s^{(n)} \rightarrow  s^{(o)})$
where, $\rho(\cdot)$  is the 
state space
equilibrium probability density and
$\pi(\cdot)$ are the probabilities to make a transition between the two states given the  set of possible MC moves. 
Further, the transition is split into  
a generation and an acceptance/rejection step
such that $\pi(s\rightarrow  s')=
P_{\rm gen}(s\rightarrow  s') P_{\rm acc}(s\rightarrow  s')$.
In the case that the sampled state space is the configuration space
of
a molecular system at constant temperature, 
$P_{\rm gen}$ might relate
to moving a randomly picked particle 
in a random direction over a 
small random distance, and $\rho(s)$ is proportional to the Boltzmann weight $e^{-\beta E(s)}$ with $\beta=1/k_BT$ the inverse temperature and $E(s)$ the state's energy.
The Metropolis-Hastings algorithm
takes a specific solution for
the acceptance probability
\begin{align}
P_{\rm acc}(
s^{(o)}\rightarrow s^{(n)})
={\rm min}\left[ 1, \frac{\rho(s^{(n)}) 
P_{\rm gen}(s^{(n)}\rightarrow s^{(o)})
}{
\rho(s^{(o)}) P_{\rm gen}(s^{(o)}\rightarrow s^{(n)})
}\right]
\label{eq:MetroHast}
\end{align}

The generation probabilities 
will cancel in the above expression if they are symmetric, 
$P_{\rm gen}(s\rightarrow  s')=P_{\rm gen}(s'\rightarrow  s)$ as in the less generic Metropolis scheme.
At each MC step, the new state is either accepted or rejected
based on the probability above. In case of a rejection, the old state is maintained and resampled. This scheme obeys  detailed-balance and if, in addition, the set of MC moves are ergodic, 
equilibrium sampling is guaranteed. When ergodic sampling, even if mathematically obeyed, is 
slowed down by a rough (free) energy landscape, Replica exchange MC becomes useful.

Replica exchange MC (or replica exchange molecular dynamics)
is based on the idea to simulate several copies of the system with different ensemble definitions~\cite{RE2, marinari92, Sugita1999REMD}, most commonly 
 ensembles with increasing temperature (parallel tempering). 
By performing “swaps” between adjacent replicas,
the low-temperature replicas gain access to the
broader space region that are explored by the high-temperature
replicas.
The detailed-balance and corresponding acceptance-rejection step
can be derived by viewing the
set of states in the different 
ensembles (replicas) as a single high-dimensional superstate 
$S=(s_1,s_2, \ldots, s_N)$ 
representing 
the system in a set of $N$ independent "parallel universes".
The Metropolis scheme  applied to the superstate 
yields
\begin{align}
P_{\rm acc}(
S^{(o)}\rightarrow S^{(n)}
)={\rm min}\left[ 1, \frac{\rho(S^{(n)})}{
\rho(S^{(o)})
}\right]
\label{eq:PaccSuperState}
\end{align}
in which the probability of the superstate equals
\begin{align}
\rho(S^{})=\rho(s_1,s_2, \ldots, s_N)=\prod_{i=1}^N
\rho_i(s_i) \label{eq:PS}
\end{align}
where $\rho_i(\cdot)$ is the specific 
probability density of ensemble $i$.
For example,  the  move that attempts 
to swap the first two states, implying 
$S^{(o)}=(s_1^{}, s_2^{}, \ldots,  s_N)$ and 
$S^{(n)}=(s_2^{}, s_1^{}, \ldots,  s_N)$,
will be accepted with a probability
\begin{align}
P_{\rm acc}
=
{\rm min}\left[ 1, \frac{\rho_1(s_2^{}) \rho_2(s_1^{})}{
\rho_1(s_1^{}) \rho_2(s_2^{})
}\right] \label{eq:PaccRE}
\end{align}

In a replica exchange simulation, swapping moves and standard MC or MD steps 
are applied alternately. Parallel computing will typically distribute
the same number of processing units per ensemble to carry out
the computational intensive standard moves. The swapping 
move is cheap, but it requires that the ensembles involved in the swap
have completed their 
previous
move.
If the standard moves in each ensemble require different computing times, then several processing units have to wait for the slow ones to finish.
If the disbalance per move is relatively constant, the replicas could effectively be made to progress in cohort by trying to differentiate the number of processing units per ensemble or the relative frequency of doing replica exchange versus standard moves per ensemble. However, in several MC methods this disbalance is not constant, such as with configurational bias MC~\cite{Siepmann92,Vlugt99,FrenkelBook} or path sampling~\cite{TPS98}.
The number of elementary steps to grow 
a polymer in configurational bias MC obviously depends on the polymer's length that is being grown, but also early rejections lead to a broad distribution of the time it takes to complete a single MC move even in uniform polymer systems.
Analogously, the time required to complete a MC move in 
path sampling simulations 
will depend on the length of the path being created.
Other examples of complex Monte Carlo methods
with a fluctuating CPU cost per move are cluster Monte Carlo algorithms~\cite{SwendsenWang} and event-chain Monte Carlo~\cite{Peters2012, Michel2014}.

We will show that the standard acceptance Eqs.~\ref{eq:MetroHast}
and \ref{eq:PaccRE}
can be applied in a parallel scheme in which ensembles are 
updated irregularly in time and the average frequency
of MC moves is different for the ensembles. In addition, we show that we can apply an infinite swapping~\cite{plattner2011} 
scheme
between the available ensembles.
For this,
we develop a new protocol based on 
the evaluation of
permanents  
that circumvents the steep 
 factorial scaling. This last development is also useful 
for standard replica exchange.

\section*{Methods}
{\bf Finite swapping.}
In the following,
we will assume that we have two types of MC moves. One move 
that is CPU intensive and can be carried out within a single ensemble, and replica exchange moves between ensembles which are relatively cheap to execute. 
The CPU intensive move will be carried out by a single worker (one processor unit, one node or a group of nodes) and these
workers perform their task in parallel on the different ensembles.
One essential part of our algorithm is that we have less workers than ensembles such that whenever the worker is finished and produced a new state for one ensemble, this state can directly be swapped with the states of any
of the available ensembles (the ones not occupied by a worker).
After that, the worker will randomly switch to another unoccupied ensemble for
performing a CPU intensive move.

In its most basic form, the algorithm consists of the following steps:
\begin{enumerate}
    \item Define $N$ ensembles and let $\rho_i(\cdot)$ be the 
    probability distribution of ensemble $i$. 
    We also define  $P_{\rm RE}$ which is the probability for 
    a replica exchange move.
     \item Assign $K<N$ 
     'workers' (processing units) to $K$ of the $N$ ensembles for performing
     a CPU intensive MC move. Each  ensemble is at all times
     occupied by either 1 or 0 workers.
     The following steps are identical for all the workers.
     \item \label{finished} If the worker is finished with its MC move in ensemble $i$, the new state is accepted or rejected according to 
     Eq.~\ref{eq:MetroHast} (with $\rho_i$ for $\rho$).
     Ensemble $i$ is updated with the new state (or by resampling the old state in case of rejection) and is then considered to be free.
     \item \label{startloop} 
     Take a uniform random number $\nu$ between 0 and 1.
     %\item 
     If $\nu> P_{\rm RE}$ go to step \ref{new_work}.
     %Otherwise continue with the next step.
         \item 
     Among the available ensembles, pick a random pair $(i,j)$.
     \item \label{Tryswap} Try to swap the states of ensembles $i$ and $j$ using 
     Eq.~\ref{eq:PaccRE} (with labels $i,j$ instead of $1,2$).
     Update ensembles $i,j$ with the swapped state or the old state 
     in case of a rejection. Return to step~\ref{startloop}.
     \item \label{new_work} Select one of the free ensembles at random and assign 
     the worker to that ensemble for performing a new 
     standard move.  Go to step~\ref{finished}.
\end{enumerate}

In this algorithm ensembles are not updated in cohort like in standard replica exchange, but 
updates occur at irregular intervals. In addition, the different ensemble conditions can result in systematic differences in the number of states 
that are being created over time. 
To prove that the above scheme actually samples the correct distributions requires a fundamentally new conceptual view as
the superstate picture is no longer applicable. 
Despite that
the algorithm uses the same type of Eqs. \ref{eq:MetroHast} and \ref{eq:PaccRE}, as one would use in standard replica exchange, it does not rely on Eqs.~\ref{eq:PaccSuperState} and \ref{eq:PS} that are no longer valid.
In the Supplementary Information (SI) we provide a proof 
from the individual ensemble's perspective in which 
the other ensembles provide an "environment" ${\mathcal E}$
that might, or might not, participate in the move of the ensemble considered. 
By doing so, we no longer require
that the number of transitions from old to new,
$S^{(o)} \rightarrow S^{(n)}$,
is  the same as from new to old, $S^{(n)} \rightarrow S^{(o)}$.
Instead, by writing $S=(s_1,{\mathcal E})$, from ensemble 1's perspective,   
we have that 
the number of 
$(s_1^{(o)}, {\mathcal E}^{(o)}) \rightarrow 
(s_1^{(n)},
\textrm{$^a \hskip -2pt {\mathcal E}$}^{(n)}
)$ transitions should be equal to 
the number of 
$(s_1^{(n)}, {\mathcal E}^{(o)}) \rightarrow 
(s_1^{(o)}, 
\textrm{$^a \hskip -2pt {\mathcal E}$}^{(n)}
)$ transitions when the standard move is applied
where
$
\textrm{$^a \hskip -2pt {\mathcal E}$}^{(n)}
$ refers to {\it any} new environment.
The SI shows a similar detailed-balance
condition for 
the replica exchange moves.  
At step \ref{Tryswap} we sample only ensemble $i$ and $j$ or, alternatively, all free ensembles 
get a sample update. This
would mean resampling the existing state of those not involved in a swap  ("null move").
This makes the approach more similar to the superstate sampling albeit using only free ensembles, as described in the SI. 
The null move does not reduce the statistical
uncertainty, but we mention it here as
it makes it easier to explain the infinite swapping approach.
But for the detailed-balance conditions to be
valid it is imperative that occupied ensembles
are not sampled.

An essential aspect of the efficiency of our algorithm is 
that the number of workers $K$ is less than the number of ensembles $N$. The case $K=N$ is valid but would reduce the number of replica exchange moves to zero as only one ensemble is free at the maximum. Reducing the $K/N$ ratio will generally imply a higher acceptance
in the replica exchange moves as we can expect a higher number 
of free ensembles whose distributions have significant 
overlap. What gives the optimum number of workers is therefore a non-trivial question that we will further explore in the 
Results and Discussion section. However, for case $K<N$ we
can maximize the effect of the replica exchange moves by taking the
$P_{\rm RE}$ parameter as high as possible. In fact, we can 
simulate the effect of the limit $P_{\rm RE}\rightarrow 1$
without having to do an 
infinite number of replica exchange moves explicitly.
This lead to an infinite swapping~\cite{plattner2011} version
of our algorithm.

{\bf Infinite swapping.}
If in the 
previously described algorithm
we 
take $P_{\rm RE}=1-\delta$, 
 we will loop 
through the steps \ref{startloop}-\ref{Tryswap} 
for many iterations ($n_{\rm it}=\sum_{n=0}^\infty n (1-\delta)^n \delta=1/\delta$ in the limit $\delta \rightarrow 0$)
before getting to step \ref{new_work}. When $\delta$ vanishes and 
$n_{\rm it}$ becomes infinitely large, we expect that all possible swaps will be executed an 
infinite number of times. Since the swaps obey detailed balance
between unoccupied ensembles, these will essentially sample the distribution
of Eq.~\ref{eq:PS} (for the subset $S^*$ of unoccupied ensembles). Hence, when the loop is exited, each possible permutation $\sigma \in S^*$ has been sampled 
$n_{\rm it}\times\rho(\sigma)/\sum_{\sigma}  \rho(\sigma)$ times. By lumping all the times that the
same permutation was sampled and 
normalizing by division
with $n_{\rm it}$, we simply sample all the possible
permutations in one go using fractional weights that sum up to 1. This is then the only sampling step, as the  single update in step~\ref{finished} can be skipped due to its negligible 
$1/n_{\rm it}$ weight.

The idea of doing an "infinite number" of swapping moves has been proposed 
before~\cite{plattner2011,infswap2013,Lu_2019}, but here we give a different flavor to this approach by
a convenient reformulation of the problem into permanents that allows us to beat the 
steep factorial scaling reported in earlier works~\cite{plattner2011}. 
The permanents formulation goes as follows. Supposed that after step~\ref{finished}, there 
are 4 free ensembles (we name them $e_1, e_2, e_3, e_4$) containing 4 states ($s_1$, $s_2$, $s_3$, $s_4$). Which state is in which ensemble after this step is irrelevant. We can now define a weight-matrix $W$:\\
$$
W =
\bordermatrix{ & e_1 & e_2 & e_3 & e_3 \cr
       s_1 & W_{11} & W_{12} & W_{13} & W_{14} \cr
       s_2 & W_{21} & W_{22} & W_{23} & W_{24} \cr
       s_3 & W_{31} & W_{32} & W_{33} & W_{34} \cr
       s_4 & W_{41} & W_{42} & W_{43} & W_{44} }
$$\\
where $W_{ij}\propto \rho_{j}(s_i)$. Essential to our approach is the computation of the permanent of the $W$ matrix, ${\rm perm}(W)$, and that of the 
$W{\{ij\}}$-matrices in which the row $i$ and
column 
$j$ are removed.

The permanent of a matrix is similar to the determinant, but without alternating signs. We can, henceforth, write ${\rm perm}(W)=\sum_{j=1}^4 W_{1j}{\rm perm}(W{\{1j\}})$.
As the permanent of the $1\times 1$ matrix
is obviously equal to the single matrix value, the permanent of arbitrary dimension could in principle be solved recursively using this relation.
Based on the permanents of $W$, we will construct a probability matrix $P$:\\
$$
P =
\bordermatrix{ & e_1 & e_2 & e_3 & e_4 \cr
       s_1 & P_{11} & P_{12} & P_{13} & P_{14} \cr
       s_2 & P_{21} & P_{22} & P_{23} & P_{24} \cr
       s_3 & P_{31} & P_{32} & P_{33} & P_{34} \cr
       s_4 & P_{41} & P_{42} & P_{43} & P_{44} }
$$\\
where $P_{ij}$
 is the chance to find state $s_i$ in ensemble $e_j$. As  
 for each permutation each state is in one ensemble and each ensemble contains 
 one state, the $P$-matrix is bistochastic: both the columns and the rows sum up to 1.
 If we consider $S^*_{ij}$ the set of permutations in which state $s_i$ is in $e_j$, we can write 
 $P_{ij}=\sum_{\sigma \in S^*_{ij}} 
 \rho(\sigma)/\sum_{\sigma' \in S^*} 
 \rho(\sigma')$.
 We can, however, also use the permanent 
 representation in which
 \begin{align}
     P_{ij}=
     \frac{
     W_{ij} {\rm perm}(W\{ij\})
     }{
        {\rm perm}(W)
     } \label{eq:Pmat}
 \end{align}
 So far we have not won anything as computing the 
 the permanent via the recursive relation mentioned above has still the factorial 
 scaling.
 The 
 Gaussian elimination approach, 
 that allows an order ${\mathcal O}(n^3)$
 computation for   determinants of $n\times n$ matrices, 
 won't work for permanents as only some but not all
 row- and column-operations have the same 
 effect to a permanent as to a determinant. 
  One can for instance swap rows and columns without changing the permanent.
 Multiplying a row by a nonzero scalar multiplies the permanent by the same scalar. Hence,
 this will
 not affect the P-matrix based on 
 Eq.~\ref{eq:Pmat}. 
 Unlike the determinant, adding or subtracting to a row a 
 scalar multiple of another row, an essential part of the Gaussian elimination method,
 does change the permanent. 
 This makes the
 permanent computation of a 
 large 
 matrix
 excessively
 more expensive than the computation of a determinant. 
 Yet, recent algorithms 
 based on the 
 Balasubramanian–Bax–Franklin–Glynn (BBFG) formula~\cite{balasubramanian_1984, bax_1998, Glynn_2010} 
scale as ${\mathcal O}(2^{n})$.
This means that the computation of the
full $P$-matrix scales as ${\mathcal O}(2^n \times n^2)$,
which seems still steep but is nevertheless a dramatic improvement compared
to factorial scaling.

For our target time of 1 second, for instance, we could only run 
the algorithm up to $N=7$ in the factorial approach, 
while we reach $N=12$ in the BBFG method using a
 mid-to-high-end laptop (DELL XPS 15 with an Intel Core i7-8750H).
If matrix size of $N=20$ is the target, 
the BBFG method can perform a full
$P$-matrix determination in
$\sim 711$ seconds, while it would take
$\sim 15.3\times10^6$ years in the factorial approach.
The BBFG method
is the fastest completely general solution for the problem 
of computing a $P$-matrix from any $W$-matrix.
For several algorithms, the $W$-matrix has special characteristics 
that can be exploited to further increase efficiency.
For instance, if by shuffling the rows and columns the $W$-matrix can be
made into a block form, where squared blocks at the diagonal have only zero's at their right and upper side, the permanent is equal to the product 
of the block's permanents. For instance, if $W_{14}=W_{24}=W_{34}=0$ we have two blocks, $3\times 3$ and  $1\times 1$. If  $W_{13}=W_{14}=W_{23}=
W_{24}=0$, we can identify 2 blocks of $2\times 2$ etc. Identification of blocks
can hugely decrease the computation of a large permanent.
Another 
speed-up can be made if  
 all rows in the $W$-matrix are a sequence of ones 
 followed by all zeros, or can be made into that form after previously mentioned column and row operations. 
This makes an
order ${\mathcal O}(n^2)$ approach possible. 
We will further 
discuss this in Sec. 
Application: $\infty$RETIS. 

 The infinite swapping approach 
 changes the aforementioned algorithm  from step 3:
 \begin{enumerate}
  \setcounter{enumi}{2}
     \item \label{finished2} If the worker is finished with its MC move in a specific ensemble, the new state is accepted or rejected (but not yet sampled) according to 
     Eq.~\ref{eq:MetroHast}. The ensemble is free.
     \item Determine the $W$-matrix based on all unoccupied 
     ensembles, calculate the $P$-matrix based on Eq.~\ref{eq:Pmat}, and update all the unoccupied ensembles 
     by sampling all free states with the fractional
     probabilities corresponding 
     to the columns in the  $P$-matrix.
     \item \label{new_work2} 
     Pick randomly one of the free ensembles 
     $e_j$. 
     \item  
     Pick one of the available states $(s_1, s_2, \ldots)$ 
     based on a weighted random selection in which state $s_i$
     has a probability of $P_{ij}$ to be selected.
     \item The worker is 
     assigned  to do  a new 
     standard move in ensemble $e_j$ based on previous state $s_i$.  
     Go to step~\ref{finished2}.
\end{enumerate}

\section*{Application: $\infty$RETIS}

Replica Exchange Transition interface sampling (RETIS)~\cite{RETIS, Raffa} is a 
quantitative
path sampling algorithm
in which the sampled states are short molecular trajectories (paths) with certain start- and end-conditions, and a minimal progress condition.
New paths are being generated by 
a Monte Carlo move in path space, such as the shooting move~\cite{shoot} in which
a randomly selected phase point of the previous path is randomly modified and 
then integrated backward and forward in time by means of molecular dynamics (MD).  
The required minimal progress 
increases with the rank of the ensemble such that the final ensemble
contains a reasonable fraction of transition trajectories. 
The start- and end-conditions, as well as the minimal progress, 
are administered by the crossings of 
interfaces $(\lambda_0, \lambda_1, \ldots, \lambda_M)$ with $\lambda_{k+1}>\lambda_{k}$, that  
can be viewed as non-intersecting hypersurfaces 
in phase space having a fixed value of the reaction coordinate.
A MC move that generates a trial path not fulfilling the path ensemble's criteria is always rejected.
RETIS defines
different path ensembles  based on the 
the direction of the paths and the interface that has to be crossed, but all paths start by crossing $\lambda_0$ (near the reactant state/state $A$) and they end by either
crossing $\lambda_0$ again or reaching the last interface $\lambda_M$ (near the product state/state $B$). There is one special 
path ensemble, called $[0^-]$, that explores
the left side of
$\lambda_0$, the reactant well, while all other path
ensembles, called $[k^+]$ with $k=0,1, \ldots M-1$, start by moving to the right from $\lambda_0$ reaching at least $\lambda_k$.

A central concept in RETIS is the so-called overall crossing probability, the chance that a path that crosses $\lambda_0$ in the positive direction
reaches $\lambda_M$ without recrossing $\lambda_0$. 
It provides the rate of the process when
multiplied with the flux 
through $\lambda_0$ (obtained from the path lengths in $[0^-]$ and $[0^+]$~\cite{Raffa}) and is usually an extremely small number. 
The chance that any of the sampled paths in the $[0^+]$ path ensemble
crosses $\lambda_M$ is generally negligible, but a decent fraction of
those ($\sim 0.1-0.5$) will cross $\lambda_1$ and some even  
$\lambda_2$. Likewise, paths in the $[k^+]$, $k>0$, path ensembles
have a much higher chance to cross $\lambda_{k+1}$ than a
$[0^+]$-path as they already cross $\lambda_k$. This leads to the
 calculation of $M$ local conditional crossing probabilities,
 the chance to cross $\lambda_{k+1}$
 given $\lambda_{k}$ was crossed for $k=0,1 \ldots M-1$,
 whose product gives an
exact expression for the overall crossing probability with an exponentially reduced CPU cost compared to %brute force 
MD.

The efficiency is further hugely improved 
by executing replica exchange moves between the path ensembles.
 These swaps are essentially 
cost-free since there is no need to simulate additional ensembles that 
are not already required.
An accepted swapping move in RETIS provides new paths in two ensembles  without the expense of having to do 
MD steps. The enhancement in efficiency is generally even larger 
than one would expect based on those arguments alone
as path ensembles
 higher-up the barrier 
provide a
 similar effect  as the high-temperature
ensembles in parallel tempering.
In addition, point-exchange moves between the 
 $[0^-]$ and  $[0^+]$ are performed by exchanging the end- and 
start-points of these path that are then continued by MD at the opposite
site of the $\lambda_0$ interface.

While TIS~\cite{TIS} (without replica exchange)
can run 
all path ensembles %independently in an 
embarrassingly parallel, % fashion.
the RETIS algorithm increases the CPU-time efficiency, but 
is difficult to parallelize and
open source path-sampling codes, like OpenPathSampling~\cite{OPS2} and PyRETIS~\cite{PyRETIS2}, 
implement RETIS as a fully sequential algorithm.
The path length distributions are generally broad 
with an increasing average path length as function of the ensemble's rank.
 This becomes 
 increasingly problematic
  the more ensembles you have as they all have to wait for the slowest ensemble.
This means that while RETIS will give you the best statistics per CPU-hour, it might not give you the best statistics in wall-time.
With the continuous increase in computing power, trading some CPU-time efficiency for wall-time efficiency, getting the answer faster while spending more CPU-cycles, might be preferential.
Our
parallel scheme can 
effectively deal with the unequal CPU cost of the 
replicas,  
which allows us
to
increase the wall-time efficiency 
with no or minimal reduction
in 
CPU-time efficiency.

{\bf The $W$-matrix in RETIS.}
If there are $M+1$ interfaces, $\lambda_0, \lambda_1, \ldots, \lambda_M$,
there are also $N=M+1$ ensembles, $[0^-], [0^+], [1^+], \ldots, [(M-1)^+]$. 
For $K$ workers, the size of the 
$W$-matrix is, hence, either $(N-K+1) \times (N-K+1)$ or  $(N-K) \times (N-K)$ as swappings are executed when 1 of the $K$ workers is free,
while the remaining $K-1$ workers occupy path ensembles that are  locked and do not participate in the swap. 
The smallest matrix occurs when one worker 
is occupying both $[0^-]$ and $[0^+]$ during 
the  point exchange move, as described in the simulation methods. 

Paths  
can be represented by 
a sequence of time slices, the phase points visited by the MD trajectory. 
For a path of length $L+1$,
$X=(x_0,x_1, \ldots, x_L)$, the plain path probability density 
$\rho(X)$ 
is given by the probability of the initial phase point times 
the dynamical transition probabilities to go from one phase point to the next:
$\rho(X)=\rho(x_0) \phi(x_0 \rightarrow x_1) \phi(x_1 \rightarrow x_2)
\ldots \phi(x_{L-1} \rightarrow x_L)$.
Here, the transition probabilities depend on the type of dynamics (deterministic, Langevin, Nos\'e-Hoover dynamics, etc). 
The weight of a path within a specific path ensemble $\rho_{j}(X)$ can be
expressed as the plain path density times the indicator function ${\bf 1}_{e_j}$
and possibly an additional weight function $w_j(X)$: 
$\rho_{j}(X)=\rho(X) \times {\bf 1}_{e_j}(X) \times w_j(X)$. 
The indicator function equals 1 if the path $X$ belongs to ensemble $e_j$. Otherwise it is 0. The additional weight function $w_j(X)$ is part of the 
high-acceptance protocol that is used in combination with the
more recent
path generation MC moves such as stone skipping~\cite{riccardi2017fast} and 
wire-fencing~\cite{wf}.
Using these "high-acceptance weights", nearly 
all the CPU intensive moves can be
accepted as they are tuned
to cancel the
$P_{\rm gen}$-terms in Metropolis-Hastings scheme, Eq.~\ref{eq:MetroHast},
and the effect of the non-physical weights is undone  
in the analysis by weighting each sampled path with the inverse of $w_j(X)$.
 
While the path probability $\rho(s_i=X)$ is difficult to compute, determining 
${\bf 1}_{j}(s_i)$ and $w_j(s_i)$ is trivial. 
It is therefore a fortunate coincidence that we can replace 
$W_{ij}= \rho_{j}(s_i)$
with
\begin{align}
    W_{ij}=  {\bf 1}_{e_j}(s_i)  w_j(s_i)
\end{align}
because the $P$-matrix does not change if we divide or multiply a row by the same number, as mentioned in Sec. Methods.
Except for $[0^-]$,
all path ensembles have the same start and end condition
and only differ with respect to the interface crossing condition.
A path that crosses interface $\lambda_k$ automatically crosses
all lower interfaces $\lambda_{l<k}$. Reversely, if the path does not 
cross $\lambda_k$, it won't cross any of the higher interfaces
$\lambda_{l>k}$. 
This implies that if  
the columns of $W_{ij}$ are ordered such that 
the 1st column ($e_1$) is the first available
ensemble from the sequence $([0^-], [0^+], [1^+], \ldots, [(M-1)^+])$, the 2nd column ($e_2$) is the second available ensemble etc, most rows will end 
with a series of zeros.

Reordering the rows 
with respect to the number of trailing zeros,
  almost always ensures that the $W$-matrix 
can be brought into a block-form  such that the permanent can be computed
faster
based on smaller matrices.
In particular, 
if $[0^-]$ is part of the free ensembles,
it will always form a $1 \times 1$ block as 
there is always one and no more than one available path that fits in this ensemble.

If high-acceptance is not applied,  we have $w_j(X)=1$ and 
each row in the 
$W$-matrix (after separating the $[0^-]$
ensemble if it is part of the free ensembles) 
is a sequence of ones 
 followed by all zeros. 
The $W$-matrix can hence be represented by an array $(n_1, n_2, n_3, 
\ldots n_{n})$ 
where each 
integer $n_i$ indicates the number of ones in row $i$. As we show in the SI, the  permanent of such a $W$-matrix is simply the product of $(n_i+1-i)$: ${\rm perm}(W)=\prod_i (n_i+1-i)$. Further, the $P$-matrix can be constructed from
following order ${\mathcal O}(n^2)$ method.

The first step is to order 
the rows of the $W$-matrix such that
$n_1 \leq n_2 
\leq \ldots \leq n_n$.
We then fill in the $P$-matrix
from top to bottom for each row using
\begin{align}
P_{ij} =
\begin{cases}
  0,  
  %&
  \text{ if } W_{ij}=0 \\
  \frac{1}{n_i+1-i}, 
  %&  
  \text{ if }  W_{ij}=1 \text{ and } [W_{(i-1)j}=0
  \text{ or } i=1] \\
  \left( \frac{n_{i-1}+1-i}{n_i-i}\right) P_{(i-1)j}, 
  %&  
  \text{ otherwise } 
  \end{cases}
\end{align}
The approach is extremely fast and allows the computation of 
$P$-matrices from a 
large
$W$-matrix, up to several thousands, 
within a second of CPU-time.
The above method applies
whenever 
the rows of the 
$W$-matrix can be 
transformed into 
 sequence of ones 
 followed by all zeros. 
 Besides 
 RETIS without high-acceptance, 
 this would apply to other MC methods like subset-sampling~\cite{subset} or
 umbrella sampling~\cite{US} with semi-infinite
 rectangular windows.

\section*{Results and Discussion} \label{sec:resdis}
To test our algorithms we ran 3 types of simulations. 
First a memoryless single variable stochastic (MSVS)
process was simulated
in order to mimic
a RETIS simulation in which the average path length increases linearly with the rank of the ensemble. 
A "path" is created by drawing 2 random numbers
where the first determines how much progress
a path makes and the second determines the path length.
These two outcomes are variable and depend 
on the rank of the ensemble such that the fictitious
path in ensemble $[k^+]$ has a 0.1 probability to 
cross $\lambda_{k+1}$ and has an average path length of approximately $k/10$ seconds (see Section Materials and Methods).
The worker is paused for a number of seconds equal to the path length before
it can participate in replica moves
to mimic the time it would take to do all the necessary MD steps.
While this artificial simulation
allows us to investigate the 
potential strength of the method 
to tackle extremely rare events, it cannot 
reveal the
effect of correlations between accepted paths 
when fast exploration of 
the reaction coordinate's
orthogonal directions  are crucial.
To analyze this effect,
we also ran a 2D membrane permeation system with 
two
slightly asymmetric channels~\cite{permeability}. 
Lastly, 
to study
our algorithm 
with a more generic 
$W$-matrix that 
needs to be solved via 
BBFG formula,
we also ran a set of 
underdamped Langevin simulations 
of a particle in a double well 
potential~\cite{TitusRev} using the
recent  wire fencing algorithm with the 
high acceptance protocol~\cite{wf}.
All simulation results were performed
using 5 independent runs of 12 hours.
Errors were based  on the standard deviations from these 5 simulations, except for the MSVS process, where a more reliable statistical error was desired for the comparison with analytical results. Here, block errors were determined on each of the five simulations based on the running average of the overall crossing probability. 
The block errors were finally combined
to obtain the statistical error in the average of the five simulations.

\subsection*{Memoryless single variable stochastic (MSVS) process}
Table~\ref{tab:table1} reports the overall crossing probabilities
and their statistical errors
for a system with 50 interfaces and
1, 5, 10, 15, 20, 25, 30, 35, 40, 45, and 50 workers.
%every number of workers. 
%%%%%%%%%%%%%%%%%%%%%%%%%%%%%%%%%%%
%\documentclass{article}
%\usepackage{siunitx} % Required for alignment
%\usepackage{multirow} % Required for multirows
%\sisetup{
%round-mode          = places, % Rounds numbers
%round-precision     = 2, % to 2 places
%}
%\begin{document}
\begin{table*}[htb!]
\footnotesize
\begin{center}
\caption{Results of the 3 model systems showing crossing probabilities ($P_{\rm cross}$), permeabilities (perm.), and rates for different number of workers (\#w). All results are 
shown in dimensionless units.
Errors are based on single standard deviations. Values shown in the lower part are a: exact result, b: Ref.~\cite{permeability}, c: approximated value based on Kramers' therory (see SI), d: Ref.~\cite{TitusRev}, and e: Ref.~\cite{wf}.
 }
\label{tab:table1}
\begin{tabular}{
%alignment column:1
l 
%alignment column:2
l 
%alignment column:3
l 
%alignment column:4
l 
%alignment column:5
l 
%alignment column:6
l 
%alignment column:7
l 
%alignment column:8
l 
}
%%%%%%%%%%%%%%%%%%%%%%%%%%%%%%%%%%%%%%%%%%%%%%%%%%
%horizontal line:0
%subline:(0,1)
\cline{1-1}
%subline:(0,2)
\cline{2-2}
%subline:(0,3)
\cline{3-3}
%subline:(0,4)
\cline{4-4}
%subline:(0,5)
\cline{5-5}
%subline:(0,6)
\cline{6-6}
%subline:(0,7)
\cline{7-7}
%subline:(0,8)
\cline{8-8}
%%%%%%%%%%%%%%%%%%%%%%%%%%%%%%%%%%%%%%%%%%%%%%%%%%
%%%%%%%%%%%%%%%%%%%%%%%%%%%%%%%%%%%%%%%%%%%%%%%%%%
%  row:1
%%%%%%%%%%%%%%%%%%%%%%%%%%%%%%%%%%%%%%%%%%%%%%%%%%
%%%%%%%%%%%%%%%%%%%%%%%%%%%%%%%%%%%%%%%%%%%%%%%%%%
%subvertical line:(1,0)
     %%%%%%%%%%%%%%%%%%%%%%%%%%%%%%%%%%%%%%%%%%%%%
     %  column:1
     %%%%%%%%%%%%%%%%%%%%%%%%%%%%%%%%%%%%%%%%%%%%%
\multicolumn{2}{c|}{\multirow{1}{*}{MSVS}}&
     %%%%%%%%%%%%%%%%%%%%%%%%%%%%%%%%%%%%%%%%%%%%%
     %  column:2
     %%%%%%%%%%%%%%%%%%%%%%%%%%%%%%%%%%%%%%%%%%%%%
% \multicolumn{2}{c|}{\multirow{1}{*}{(1, 2)}}&
     %%%%%%%%%%%%%%%%%%%%%%%%%%%%%%%%%%%%%%%%%%%%%
     %  column:3
     %%%%%%%%%%%%%%%%%%%%%%%%%%%%%%%%%%%%%%%%%%%%%
\multicolumn{3}{c|}{\multirow{1}{*}{two-channel system}}&
     %%%%%%%%%%%%%%%%%%%%%%%%%%%%%%%%%%%%%%%%%%%%%
     %  column:4
     %%%%%%%%%%%%%%%%%%%%%%%%%%%%%%%%%%%%%%%%%%%%%
% \multicolumn{3}{c|}{\multirow{1}{*}{(1, 4)}}&
     %%%%%%%%%%%%%%%%%%%%%%%%%%%%%%%%%%%%%%%%%%%%%
     %  column:5
     %%%%%%%%%%%%%%%%%%%%%%%%%%%%%%%%%%%%%%%%%%%%%
% \multicolumn{3}{c|}{\multirow{1}{*}{(1, 5)}}&
     %%%%%%%%%%%%%%%%%%%%%%%%%%%%%%%%%%%%%%%%%%%%%
     %  column:6
     %%%%%%%%%%%%%%%%%%%%%%%%%%%%%%%%%%%%%%%%%%%%%
\multicolumn{3}{c}{\multirow{1}{*}{double well with wire fencing}}\\
     %%%%%%%%%%%%%%%%%%%%%%%%%%%%%%%%%%%%%%%%%%%%%
     %  column:7
     %%%%%%%%%%%%%%%%%%%%%%%%%%%%%%%%%%%%%%%%%%%%%
% \multicolumn{3}{c}{\multirow{1}{*}{(1, 7)}}\\
     %%%%%%%%%%%%%%%%%%%%%%%%%%%%%%%%%%%%%%%%%%%%%
     %  column:8
     %%%%%%%%%%%%%%%%%%%%%%%%%%%%%%%%%%%%%%%%%%%%%
% \multicolumn{3}{c}{\multirow{1}{*}{(1, 8)}}\\
%%%%%%%%%%%%%%%%%%%%%%%%%%%%%%%%%%%%%%%%%%%%%%%%%%
%horizontal line:1
%subline:(1,1)
%\cline{1-1}
%subline:(1,2)
%\cline{2-2}
%subline:(1,3)
%\cline{3-3}
%subline:(1,4)
%\cline{4-4}
%subline:(1,5)
%\cline{5-5}
%subline:(1,6)
%\cline{6-6}
%subline:(1,7)
%\cline{7-7}
%subline:(1,8)
%\cline{8-8}
%%%%%%%%%%%%%%%%%%%%%%%%%%%%%%%%%%%%%%%%%%%%%%%%%%
%%%%%%%%%%%%%%%%%%%%%%%%%%%%%%%%%%%%%%%%%%%%%%%%%%
%  row:2
%%%%%%%%%%%%%%%%%%%%%%%%%%%%%%%%%%%%%%%%%%%%%%%%%%
%%%%%%%%%%%%%%%%%%%%%%%%%%%%%%%%%%%%%%%%%%%%%%%%%%
%subvertical line:(2,0)
     %%%%%%%%%%%%%%%%%%%%%%%%%%%%%%%%%%%%%%%%%%%%%
     %  column:1
     %%%%%%%%%%%%%%%%%%%%%%%%%%%%%%%%%%%%%%%%%%%%%
\multicolumn{1}{r}{\multirow{1}{*}{\#w}}&
     %%%%%%%%%%%%%%%%%%%%%%%%%%%%%%%%%%%%%%%%%%%%%
     %  column:2
     %%%%%%%%%%%%%%%%%%%%%%%%%%%%%%%%%%%%%%%%%%%%%
\multicolumn{1}{l|}{\multirow{1}{*}{$P_{\rm cross}/10^{-50}$}}&
     %%%%%%%%%%%%%%%%%%%%%%%%%%%%%%%%%%%%%%%%%%%%%
     %  column:3
     %%%%%%%%%%%%%%%%%%%%%%%%%%%%%%%%%%%%%%%%%%%%%
\multicolumn{1}{r}{\multirow{1}{*}{\#w}}&
     %%%%%%%%%%%%%%%%%%%%%%%%%%%%%%%%%%%%%%%%%%%%%
     %  column:4
     %%%%%%%%%%%%%%%%%%%%%%%%%%%%%%%%%%%%%%%%%%%%%
\multicolumn{1}{l}{\multirow{1}{*}{$P_{\rm cross}/10^{-5}$}}&
     %%%%%%%%%%%%%%%%%%%%%%%%%%%%%%%%%%%%%%%%%%%%%
     %  column:5
     %%%%%%%%%%%%%%%%%%%%%%%%%%%%%%%%%%%%%%%%%%%%%
\multicolumn{1}{l|}{\multirow{1}{*}{perm./$10^{-6}$}}&
     %%%%%%%%%%%%%%%%%%%%%%%%%%%%%%%%%%%%%%%%%%%%%
     %  column:6
     %%%%%%%%%%%%%%%%%%%%%%%%%%%%%%%%%%%%%%%%%%%%%
\multicolumn{1}{r}{\multirow{1}{*}{\#w}}&
     %%%%%%%%%%%%%%%%%%%%%%%%%%%%%%%%%%%%%%%%%%%%%
     %  column:7
     %%%%%%%%%%%%%%%%%%%%%%%%%%%%%%%%%%%%%%%%%%%%%
\multicolumn{1}{l}{\multirow{1}{*}{$P_{\rm cross}/10^{-7}$}}&
     %%%%%%%%%%%%%%%%%%%%%%%%%%%%%%%%%%%%%%%%%%%%%
     %  column:8
     %%%%%%%%%%%%%%%%%%%%%%%%%%%%%%%%%%%%%%%%%%%%%
\multicolumn{1}{l}{\multirow{1}{*}{rate/$10^{-7}$}}\\
%%%%%%%%%%%%%%%%%%%%%%%%%%%%%%%%%%%%%%%%%%%%%%%%%%
%horizontal line:2
%subline:(2,1)
\cline{1-1}
%subline:(2,2)
\cline{2-2}
%subline:(2,3)
\cline{3-3}
%subline:(2,4)
\cline{4-4}
%subline:(2,5)
\cline{5-5}
%subline:(2,6)
\cline{6-6}
%subline:(2,7)
\cline{7-7}
%subline:(2,8)
\cline{8-8}
%%%%%%%%%%%%%%%%%%%%%%%%%%%%%%%%%%%%%%%%%%%%%%%%%%
%%%%%%%%%%%%%%%%%%%%%%%%%%%%%%%%%%%%%%%%%%%%%%%%%%
%  row:3
%%%%%%%%%%%%%%%%%%%%%%%%%%%%%%%%%%%%%%%%%%%%%%%%%%
%%%%%%%%%%%%%%%%%%%%%%%%%%%%%%%%%%%%%%%%%%%%%%%%%%
%subvertical line:(3,0)
     %%%%%%%%%%%%%%%%%%%%%%%%%%%%%%%%%%%%%%%%%%%%%
     %  column:1
     %%%%%%%%%%%%%%%%%%%%%%%%%%%%%%%%%%%%%%%%%%%%%
\multicolumn{1}{r}{\multirow{1}{*}{1}}&
     %%%%%%%%%%%%%%%%%%%%%%%%%%%%%%%%%%%%%%%%%%%%%
     %  column:2
     %%%%%%%%%%%%%%%%%%%%%%%%%%%%%%%%%%%%%%%%%%%%%
\multicolumn{1}{l|}{\multirow{1}{*}{$0.61 \pm 0.33$}}&
     %%%%%%%%%%%%%%%%%%%%%%%%%%%%%%%%%%%%%%%%%%%%%
     %  column:3
     %%%%%%%%%%%%%%%%%%%%%%%%%%%%%%%%%%%%%%%%%%%%%
\multicolumn{1}{r}{\multirow{1}{*}{1}}&
     %%%%%%%%%%%%%%%%%%%%%%%%%%%%%%%%%%%%%%%%%%%%%
     %  column:4
     %%%%%%%%%%%%%%%%%%%%%%%%%%%%%%%%%%%%%%%%%%%%%
\multicolumn{1}{l}{\multirow{1}{*}{$1.52 \pm 0.17$}}&
     %%%%%%%%%%%%%%%%%%%%%%%%%%%%%%%%%%%%%%%%%%%%%
     %  column:5
     %%%%%%%%%%%%%%%%%%%%%%%%%%%%%%%%%%%%%%%%%%%%%
\multicolumn{1}{l|}{\multirow{1}{*}{$1.28 \pm 0.14$}}&
     %%%%%%%%%%%%%%%%%%%%%%%%%%%%%%%%%%%%%%%%%%%%%
     %  column:6
     %%%%%%%%%%%%%%%%%%%%%%%%%%%%%%%%%%%%%%%%%%%%%
\multicolumn{1}{r}{\multirow{1}{*}{1}}&
     %%%%%%%%%%%%%%%%%%%%%%%%%%%%%%%%%%%%%%%%%%%%%
     %  column:7
     %%%%%%%%%%%%%%%%%%%%%%%%%%%%%%%%%%%%%%%%%%%%%
\multicolumn{1}{l}{\multirow{1}{*}{$5.91 \pm 0.18$}}&
     %%%%%%%%%%%%%%%%%%%%%%%%%%%%%%%%%%%%%%%%%%%%%
     %  column:8
     %%%%%%%%%%%%%%%%%%%%%%%%%%%%%%%%%%%%%%%%%%%%%
\multicolumn{1}{l}{\multirow{1}{*}{$2.59 \pm 0.07$}}\\
%%%%%%%%%%%%%%%%%%%%%%%%%%%%%%%%%%%%%%%%%%%%%%%%%%
%horizontal line:3
%subline:(3,1)
%\cline{1-1}
%subline:(3,2)
%\cline{2-2}
%subline:(3,3)
%\cline{3-3}
%subline:(3,4)
%\cline{4-4}
%subline:(3,5)
%\cline{5-5}
%subline:(3,6)
%\cline{6-6}
%subline:(3,7)
%\cline{7-7}
%subline:(3,8)
%\cline{8-8}
%%%%%%%%%%%%%%%%%%%%%%%%%%%%%%%%%%%%%%%%%%%%%%%%%%
%%%%%%%%%%%%%%%%%%%%%%%%%%%%%%%%%%%%%%%%%%%%%%%%%%
%  row:4
%%%%%%%%%%%%%%%%%%%%%%%%%%%%%%%%%%%%%%%%%%%%%%%%%%
%%%%%%%%%%%%%%%%%%%%%%%%%%%%%%%%%%%%%%%%%%%%%%%%%%
%subvertical line:(4,0)
     %%%%%%%%%%%%%%%%%%%%%%%%%%%%%%%%%%%%%%%%%%%%%
     %  column:1
     %%%%%%%%%%%%%%%%%%%%%%%%%%%%%%%%%%%%%%%%%%%%%
\multicolumn{1}{r}{\multirow{1}{*}{5}}&
     %%%%%%%%%%%%%%%%%%%%%%%%%%%%%%%%%%%%%%%%%%%%%
     %  column:2
     %%%%%%%%%%%%%%%%%%%%%%%%%%%%%%%%%%%%%%%%%%%%%
\multicolumn{1}{l|}{\multirow{1}{*}{$1.47 \pm 1.04$}}&
     %%%%%%%%%%%%%%%%%%%%%%%%%%%%%%%%%%%%%%%%%%%%%
     %  column:3
     %%%%%%%%%%%%%%%%%%%%%%%%%%%%%%%%%%%%%%%%%%%%%
\multicolumn{1}{r}{\multirow{1}{*}{2}}&
     %%%%%%%%%%%%%%%%%%%%%%%%%%%%%%%%%%%%%%%%%%%%%
     %  column:4
     %%%%%%%%%%%%%%%%%%%%%%%%%%%%%%%%%%%%%%%%%%%%%
\multicolumn{1}{l}{\multirow{1}{*}{$1.63 \pm 0.24$}}&
     %%%%%%%%%%%%%%%%%%%%%%%%%%%%%%%%%%%%%%%%%%%%%
     %  column:5
     %%%%%%%%%%%%%%%%%%%%%%%%%%%%%%%%%%%%%%%%%%%%%
\multicolumn{1}{l|}{\multirow{1}{*}{$1.37 \pm 0.20$}}&
     %%%%%%%%%%%%%%%%%%%%%%%%%%%%%%%%%%%%%%%%%%%%%
     %  column:6
     %%%%%%%%%%%%%%%%%%%%%%%%%%%%%%%%%%%%%%%%%%%%%
\multicolumn{1}{r}{\multirow{1}{*}{2}}&
     %%%%%%%%%%%%%%%%%%%%%%%%%%%%%%%%%%%%%%%%%%%%%
     %  column:7
     %%%%%%%%%%%%%%%%%%%%%%%%%%%%%%%%%%%%%%%%%%%%%
\multicolumn{1}{l}{\multirow{1}{*}{$5.70 \pm 0.13$}}&
     %%%%%%%%%%%%%%%%%%%%%%%%%%%%%%%%%%%%%%%%%%%%%
     %  column:8
     %%%%%%%%%%%%%%%%%%%%%%%%%%%%%%%%%%%%%%%%%%%%%
\multicolumn{1}{l}{\multirow{1}{*}{$2.51 \pm 0.06$}}\\
%%%%%%%%%%%%%%%%%%%%%%%%%%%%%%%%%%%%%%%%%%%%%%%%%%
%horizontal line:4
%subline:(4,1)
%\cline{1-1}
%subline:(4,2)
%\cline{2-2}
%subline:(4,3)
%\cline{3-3}
%subline:(4,4)
%\cline{4-4}
%subline:(4,5)
%\cline{5-5}
%subline:(4,6)
%\cline{6-6}
%subline:(4,7)
%\cline{7-7}
%subline:(4,8)
%\cline{8-8}
%%%%%%%%%%%%%%%%%%%%%%%%%%%%%%%%%%%%%%%%%%%%%%%%%%
%%%%%%%%%%%%%%%%%%%%%%%%%%%%%%%%%%%%%%%%%%%%%%%%%%
%  row:5
%%%%%%%%%%%%%%%%%%%%%%%%%%%%%%%%%%%%%%%%%%%%%%%%%%
%%%%%%%%%%%%%%%%%%%%%%%%%%%%%%%%%%%%%%%%%%%%%%%%%%
%subvertical line:(5,0)
     %%%%%%%%%%%%%%%%%%%%%%%%%%%%%%%%%%%%%%%%%%%%%
     %  column:1
     %%%%%%%%%%%%%%%%%%%%%%%%%%%%%%%%%%%%%%%%%%%%%
\multicolumn{1}{r}{\multirow{1}{*}{10}}&
     %%%%%%%%%%%%%%%%%%%%%%%%%%%%%%%%%%%%%%%%%%%%%
     %  column:2
     %%%%%%%%%%%%%%%%%%%%%%%%%%%%%%%%%%%%%%%%%%%%%
\multicolumn{1}{l|}{\multirow{1}{*}{$0.86 \pm 0.51$}}&
     %%%%%%%%%%%%%%%%%%%%%%%%%%%%%%%%%%%%%%%%%%%%%
     %  column:3
     %%%%%%%%%%%%%%%%%%%%%%%%%%%%%%%%%%%%%%%%%%%%%
\multicolumn{1}{r}{\multirow{1}{*}{3}}&
     %%%%%%%%%%%%%%%%%%%%%%%%%%%%%%%%%%%%%%%%%%%%%
     %  column:4
     %%%%%%%%%%%%%%%%%%%%%%%%%%%%%%%%%%%%%%%%%%%%%
\multicolumn{1}{l}{\multirow{1}{*}{$1.52 \pm 0.07$}}&
     %%%%%%%%%%%%%%%%%%%%%%%%%%%%%%%%%%%%%%%%%%%%%
     %  column:5
     %%%%%%%%%%%%%%%%%%%%%%%%%%%%%%%%%%%%%%%%%%%%%
\multicolumn{1}{l|}{\multirow{1}{*}{$1.28 \pm 0.06$}}&
     %%%%%%%%%%%%%%%%%%%%%%%%%%%%%%%%%%%%%%%%%%%%%
     %  column:6
     %%%%%%%%%%%%%%%%%%%%%%%%%%%%%%%%%%%%%%%%%%%%%
\multicolumn{1}{r}{\multirow{1}{*}{3}}&
     %%%%%%%%%%%%%%%%%%%%%%%%%%%%%%%%%%%%%%%%%%%%%
     %  column:7
     %%%%%%%%%%%%%%%%%%%%%%%%%%%%%%%%%%%%%%%%%%%%%
\multicolumn{1}{l}{\multirow{1}{*}{$5.57 \pm 0.19$}}&
     %%%%%%%%%%%%%%%%%%%%%%%%%%%%%%%%%%%%%%%%%%%%%
     %  column:8
     %%%%%%%%%%%%%%%%%%%%%%%%%%%%%%%%%%%%%%%%%%%%%
\multicolumn{1}{l}{\multirow{1}{*}{$2.45 \pm 0.08$}}\\
%%%%%%%%%%%%%%%%%%%%%%%%%%%%%%%%%%%%%%%%%%%%%%%%%%
%horizontal line:5
%subline:(5,1)
%\cline{1-1}
%subline:(5,2)
%\cline{2-2}
%subline:(5,3)
%\cline{3-3}
%subline:(5,4)
%\cline{4-4}
%subline:(5,5)
%\cline{5-5}
%subline:(5,6)
%\cline{6-6}
%subline:(5,7)
%\cline{7-7}
%subline:(5,8)
%\cline{8-8}
%%%%%%%%%%%%%%%%%%%%%%%%%%%%%%%%%%%%%%%%%%%%%%%%%%
%%%%%%%%%%%%%%%%%%%%%%%%%%%%%%%%%%%%%%%%%%%%%%%%%%
%  row:6
%%%%%%%%%%%%%%%%%%%%%%%%%%%%%%%%%%%%%%%%%%%%%%%%%%
%%%%%%%%%%%%%%%%%%%%%%%%%%%%%%%%%%%%%%%%%%%%%%%%%%
%subvertical line:(6,0)
     %%%%%%%%%%%%%%%%%%%%%%%%%%%%%%%%%%%%%%%%%%%%%
     %  column:1
     %%%%%%%%%%%%%%%%%%%%%%%%%%%%%%%%%%%%%%%%%%%%%
\multicolumn{1}{r}{\multirow{1}{*}{15}}&
     %%%%%%%%%%%%%%%%%%%%%%%%%%%%%%%%%%%%%%%%%%%%%
     %  column:2
     %%%%%%%%%%%%%%%%%%%%%%%%%%%%%%%%%%%%%%%%%%%%%
\multicolumn{1}{l|}{\multirow{1}{*}{$0.68 \pm 0.08$}}&
     %%%%%%%%%%%%%%%%%%%%%%%%%%%%%%%%%%%%%%%%%%%%%
     %  column:3
     %%%%%%%%%%%%%%%%%%%%%%%%%%%%%%%%%%%%%%%%%%%%%
\multicolumn{1}{r}{\multirow{1}{*}{4}}&
     %%%%%%%%%%%%%%%%%%%%%%%%%%%%%%%%%%%%%%%%%%%%%
     %  column:4
     %%%%%%%%%%%%%%%%%%%%%%%%%%%%%%%%%%%%%%%%%%%%%
\multicolumn{1}{l}{\multirow{1}{*}{$1.42 \pm 0.10$}}&
     %%%%%%%%%%%%%%%%%%%%%%%%%%%%%%%%%%%%%%%%%%%%%
     %  column:5
     %%%%%%%%%%%%%%%%%%%%%%%%%%%%%%%%%%%%%%%%%%%%%
\multicolumn{1}{l|}{\multirow{1}{*}{$1.19 \pm 0.08$}}&
     %%%%%%%%%%%%%%%%%%%%%%%%%%%%%%%%%%%%%%%%%%%%%
     %  column:6
     %%%%%%%%%%%%%%%%%%%%%%%%%%%%%%%%%%%%%%%%%%%%%
\multicolumn{1}{r}{\multirow{1}{*}{4}}&
     %%%%%%%%%%%%%%%%%%%%%%%%%%%%%%%%%%%%%%%%%%%%%
     %  column:7
     %%%%%%%%%%%%%%%%%%%%%%%%%%%%%%%%%%%%%%%%%%%%%
\multicolumn{1}{l}{\multirow{1}{*}{$5.20 \pm 0.30$}}&
     %%%%%%%%%%%%%%%%%%%%%%%%%%%%%%%%%%%%%%%%%%%%%
     %  column:8
     %%%%%%%%%%%%%%%%%%%%%%%%%%%%%%%%%%%%%%%%%%%%%
\multicolumn{1}{l}{\multirow{1}{*}{$2.34 \pm 0.12$}}\\
%%%%%%%%%%%%%%%%%%%%%%%%%%%%%%%%%%%%%%%%%%%%%%%%%%
%horizontal line:6
%subline:(6,1)
%\cline{1-1}
%subline:(6,2)
%\cline{2-2}
%subline:(6,3)
%\cline{3-3}
%subline:(6,4)
%\cline{4-4}
%subline:(6,5)
%\cline{5-5}
%subline:(6,6)
%\cline{6-6}
%subline:(6,7)
%\cline{7-7}
%subline:(6,8)
%\cline{8-8}
%%%%%%%%%%%%%%%%%%%%%%%%%%%%%%%%%%%%%%%%%%%%%%%%%%
%%%%%%%%%%%%%%%%%%%%%%%%%%%%%%%%%%%%%%%%%%%%%%%%%%
%  row:7
%%%%%%%%%%%%%%%%%%%%%%%%%%%%%%%%%%%%%%%%%%%%%%%%%%
%%%%%%%%%%%%%%%%%%%%%%%%%%%%%%%%%%%%%%%%%%%%%%%%%%
%subvertical line:(7,0)
     %%%%%%%%%%%%%%%%%%%%%%%%%%%%%%%%%%%%%%%%%%%%%
     %  column:1
     %%%%%%%%%%%%%%%%%%%%%%%%%%%%%%%%%%%%%%%%%%%%%
\multicolumn{1}{r}{\multirow{1}{*}{20}}&
     %%%%%%%%%%%%%%%%%%%%%%%%%%%%%%%%%%%%%%%%%%%%%
     %  column:2
     %%%%%%%%%%%%%%%%%%%%%%%%%%%%%%%%%%%%%%%%%%%%%
\multicolumn{1}{l|}{\multirow{1}{*}{$1.02 \pm 0.13$}}&
     %%%%%%%%%%%%%%%%%%%%%%%%%%%%%%%%%%%%%%%%%%%%%
     %  column:3
     %%%%%%%%%%%%%%%%%%%%%%%%%%%%%%%%%%%%%%%%%%%%%
\multicolumn{1}{r}{\multirow{1}{*}{5}}&
     %%%%%%%%%%%%%%%%%%%%%%%%%%%%%%%%%%%%%%%%%%%%%
     %  column:4
     %%%%%%%%%%%%%%%%%%%%%%%%%%%%%%%%%%%%%%%%%%%%%
\multicolumn{1}{l}{\multirow{1}{*}{$1.40 \pm 0.12$}}&
     %%%%%%%%%%%%%%%%%%%%%%%%%%%%%%%%%%%%%%%%%%%%%
     %  column:5
     %%%%%%%%%%%%%%%%%%%%%%%%%%%%%%%%%%%%%%%%%%%%%
\multicolumn{1}{l|}{\multirow{1}{*}{$1.18 \pm 0.10$}}&
     %%%%%%%%%%%%%%%%%%%%%%%%%%%%%%%%%%%%%%%%%%%%%
     %  column:6
     %%%%%%%%%%%%%%%%%%%%%%%%%%%%%%%%%%%%%%%%%%%%%
\multicolumn{1}{r}{\multirow{1}{*}{5}}&
     %%%%%%%%%%%%%%%%%%%%%%%%%%%%%%%%%%%%%%%%%%%%%
     %  column:7
     %%%%%%%%%%%%%%%%%%%%%%%%%%%%%%%%%%%%%%%%%%%%%
\multicolumn{1}{l}{\multirow{1}{*}{$5.05 \pm 0.41$}}&
     %%%%%%%%%%%%%%%%%%%%%%%%%%%%%%%%%%%%%%%%%%%%%
     %  column:8
     %%%%%%%%%%%%%%%%%%%%%%%%%%%%%%%%%%%%%%%%%%%%%
\multicolumn{1}{l}{\multirow{1}{*}{$2.23 \pm 0.18$}}\\
%%%%%%%%%%%%%%%%%%%%%%%%%%%%%%%%%%%%%%%%%%%%%%%%%%
%horizontal line:7
%subline:(7,1)
%\cline{1-1}
%subline:(7,2)
%\cline{2-2}
%subline:(7,3)
%\cline{3-3}
%subline:(7,4)
%\cline{4-4}
%subline:(7,5)
%\cline{5-5}
%subline:(7,6)
%\cline{6-6}
%subline:(7,7)
%\cline{7-7}
%subline:(7,8)
%\cline{8-8}
%%%%%%%%%%%%%%%%%%%%%%%%%%%%%%%%%%%%%%%%%%%%%%%%%%
%%%%%%%%%%%%%%%%%%%%%%%%%%%%%%%%%%%%%%%%%%%%%%%%%%
%  row:8
%%%%%%%%%%%%%%%%%%%%%%%%%%%%%%%%%%%%%%%%%%%%%%%%%%
%%%%%%%%%%%%%%%%%%%%%%%%%%%%%%%%%%%%%%%%%%%%%%%%%%
%subvertical line:(8,0)
     %%%%%%%%%%%%%%%%%%%%%%%%%%%%%%%%%%%%%%%%%%%%%
     %  column:1
     %%%%%%%%%%%%%%%%%%%%%%%%%%%%%%%%%%%%%%%%%%%%%
\multicolumn{1}{r}{\multirow{1}{*}{25}}&
     %%%%%%%%%%%%%%%%%%%%%%%%%%%%%%%%%%%%%%%%%%%%%
     %  column:2
     %%%%%%%%%%%%%%%%%%%%%%%%%%%%%%%%%%%%%%%%%%%%%
\multicolumn{1}{l|}{\multirow{1}{*}{$1.02 \pm 0.17$}}&
     %%%%%%%%%%%%%%%%%%%%%%%%%%%%%%%%%%%%%%%%%%%%%
     %  column:3
     %%%%%%%%%%%%%%%%%%%%%%%%%%%%%%%%%%%%%%%%%%%%%
\multicolumn{1}{r}{\multirow{1}{*}{6}}&
     %%%%%%%%%%%%%%%%%%%%%%%%%%%%%%%%%%%%%%%%%%%%%
     %  column:4
     %%%%%%%%%%%%%%%%%%%%%%%%%%%%%%%%%%%%%%%%%%%%%
\multicolumn{1}{l}{\multirow{1}{*}{$1.54 \pm 0.06$}}&
     %%%%%%%%%%%%%%%%%%%%%%%%%%%%%%%%%%%%%%%%%%%%%
     %  column:5
     %%%%%%%%%%%%%%%%%%%%%%%%%%%%%%%%%%%%%%%%%%%%%
\multicolumn{1}{l|}{\multirow{1}{*}{$1.30 \pm 0.05$}}&
     %%%%%%%%%%%%%%%%%%%%%%%%%%%%%%%%%%%%%%%%%%%%%
     %  column:6
     %%%%%%%%%%%%%%%%%%%%%%%%%%%%%%%%%%%%%%%%%%%%%
\multicolumn{1}{r}{\multirow{1}{*}{6}}&
     %%%%%%%%%%%%%%%%%%%%%%%%%%%%%%%%%%%%%%%%%%%%%
     %  column:7
     %%%%%%%%%%%%%%%%%%%%%%%%%%%%%%%%%%%%%%%%%%%%%
\multicolumn{1}{l}{\multirow{1}{*}{$5.49 \pm 0.29$}}&
     %%%%%%%%%%%%%%%%%%%%%%%%%%%%%%%%%%%%%%%%%%%%%
     %  column:8
     %%%%%%%%%%%%%%%%%%%%%%%%%%%%%%%%%%%%%%%%%%%%%
\multicolumn{1}{l}{\multirow{1}{*}{$2.42 \pm 0.13$}}\\
%%%%%%%%%%%%%%%%%%%%%%%%%%%%%%%%%%%%%%%%%%%%%%%%%%
%horizontal line:8
%subline:(8,1)
%\cline{1-1}
%subline:(8,2)
%\cline{2-2}
%subline:(8,3)
%\cline{3-3}
%subline:(8,4)
%\cline{4-4}
%subline:(8,5)
%\cline{5-5}
%subline:(8,6)
%\cline{6-6}
%subline:(8,7)
%\cline{7-7}
%subline:(8,8)
%\cline{8-8}
%%%%%%%%%%%%%%%%%%%%%%%%%%%%%%%%%%%%%%%%%%%%%%%%%%
%%%%%%%%%%%%%%%%%%%%%%%%%%%%%%%%%%%%%%%%%%%%%%%%%%
%  row:9
%%%%%%%%%%%%%%%%%%%%%%%%%%%%%%%%%%%%%%%%%%%%%%%%%%
%%%%%%%%%%%%%%%%%%%%%%%%%%%%%%%%%%%%%%%%%%%%%%%%%%
%subvertical line:(9,0)
     %%%%%%%%%%%%%%%%%%%%%%%%%%%%%%%%%%%%%%%%%%%%%
     %  column:1
     %%%%%%%%%%%%%%%%%%%%%%%%%%%%%%%%%%%%%%%%%%%%%
\multicolumn{1}{r}{\multirow{1}{*}{30}}&
     %%%%%%%%%%%%%%%%%%%%%%%%%%%%%%%%%%%%%%%%%%%%%
     %  column:2
     %%%%%%%%%%%%%%%%%%%%%%%%%%%%%%%%%%%%%%%%%%%%%
\multicolumn{1}{l|}{\multirow{1}{*}{$1.26 \pm 0.24$}}&
     %%%%%%%%%%%%%%%%%%%%%%%%%%%%%%%%%%%%%%%%%%%%%
     %  column:3
     %%%%%%%%%%%%%%%%%%%%%%%%%%%%%%%%%%%%%%%%%%%%%
\multicolumn{1}{r}{\multirow{1}{*}{7}}&
     %%%%%%%%%%%%%%%%%%%%%%%%%%%%%%%%%%%%%%%%%%%%%
     %  column:4
     %%%%%%%%%%%%%%%%%%%%%%%%%%%%%%%%%%%%%%%%%%%%%
\multicolumn{1}{l}{\multirow{1}{*}{$1.48 \pm 0.08$}}&
     %%%%%%%%%%%%%%%%%%%%%%%%%%%%%%%%%%%%%%%%%%%%%
     %  column:5
     %%%%%%%%%%%%%%%%%%%%%%%%%%%%%%%%%%%%%%%%%%%%%
\multicolumn{1}{l|}{\multirow{1}{*}{$1.24 \pm 0.07$}}&
     %%%%%%%%%%%%%%%%%%%%%%%%%%%%%%%%%%%%%%%%%%%%%
     %  column:6
     %%%%%%%%%%%%%%%%%%%%%%%%%%%%%%%%%%%%%%%%%%%%%
\multicolumn{1}{r}{\multirow{1}{*}{7}}&
     %%%%%%%%%%%%%%%%%%%%%%%%%%%%%%%%%%%%%%%%%%%%%
     %  column:7
     %%%%%%%%%%%%%%%%%%%%%%%%%%%%%%%%%%%%%%%%%%%%%
\multicolumn{1}{l}{\multirow{1}{*}{$4.99 \pm 0.39$}}&
     %%%%%%%%%%%%%%%%%%%%%%%%%%%%%%%%%%%%%%%%%%%%%
     %  column:8
     %%%%%%%%%%%%%%%%%%%%%%%%%%%%%%%%%%%%%%%%%%%%%
\multicolumn{1}{l}{\multirow{1}{*}{$2.21 \pm 0.17$}}\\
%%%%%%%%%%%%%%%%%%%%%%%%%%%%%%%%%%%%%%%%%%%%%%%%%%
%horizontal line:9
%subline:(9,1)
%\cline{1-1}
%subline:(9,2)
%\cline{2-2}
%subline:(9,3)
%\cline{3-3}
%subline:(9,4)
%\cline{4-4}
%subline:(9,5)
%\cline{5-5}
%subline:(9,6)
%\cline{6-6}
%subline:(9,7)
%\cline{7-7}
%subline:(9,8)
%\cline{8-8}
%%%%%%%%%%%%%%%%%%%%%%%%%%%%%%%%%%%%%%%%%%%%%%%%%%
%%%%%%%%%%%%%%%%%%%%%%%%%%%%%%%%%%%%%%%%%%%%%%%%%%
%  row:10
%%%%%%%%%%%%%%%%%%%%%%%%%%%%%%%%%%%%%%%%%%%%%%%%%%
%%%%%%%%%%%%%%%%%%%%%%%%%%%%%%%%%%%%%%%%%%%%%%%%%%
%subvertical line:(10,0)
     %%%%%%%%%%%%%%%%%%%%%%%%%%%%%%%%%%%%%%%%%%%%%
     %  column:1
     %%%%%%%%%%%%%%%%%%%%%%%%%%%%%%%%%%%%%%%%%%%%%
\multicolumn{1}{r}{\multirow{1}{*}{35}}&
     %%%%%%%%%%%%%%%%%%%%%%%%%%%%%%%%%%%%%%%%%%%%%
     %  column:2
     %%%%%%%%%%%%%%%%%%%%%%%%%%%%%%%%%%%%%%%%%%%%%
\multicolumn{1}{l|}{\multirow{1}{*}{$1.05 \pm 0.15$}}&
     %%%%%%%%%%%%%%%%%%%%%%%%%%%%%%%%%%%%%%%%%%%%%
     %  column:3
     %%%%%%%%%%%%%%%%%%%%%%%%%%%%%%%%%%%%%%%%%%%%%
\multicolumn{1}{r}{\multirow{1}{*}{8}}&
     %%%%%%%%%%%%%%%%%%%%%%%%%%%%%%%%%%%%%%%%%%%%%
     %  column:4
     %%%%%%%%%%%%%%%%%%%%%%%%%%%%%%%%%%%%%%%%%%%%%
\multicolumn{1}{l}{\multirow{1}{*}{$1.46 \pm 0.08$}}&
     %%%%%%%%%%%%%%%%%%%%%%%%%%%%%%%%%%%%%%%%%%%%%
     %  column:5
     %%%%%%%%%%%%%%%%%%%%%%%%%%%%%%%%%%%%%%%%%%%%%
\multicolumn{1}{l|}{\multirow{1}{*}{$1.23 \pm 0.06$}}&
     %%%%%%%%%%%%%%%%%%%%%%%%%%%%%%%%%%%%%%%%%%%%%
     %  column:6
     %%%%%%%%%%%%%%%%%%%%%%%%%%%%%%%%%%%%%%%%%%%%%
\multicolumn{1}{r}{\multirow{1}{*}{8}}&
     %%%%%%%%%%%%%%%%%%%%%%%%%%%%%%%%%%%%%%%%%%%%%
     %  column:7
     %%%%%%%%%%%%%%%%%%%%%%%%%%%%%%%%%%%%%%%%%%%%%
\multicolumn{1}{l}{\multirow{1}{*}{$4.88 \pm 0.43$}}&
     %%%%%%%%%%%%%%%%%%%%%%%%%%%%%%%%%%%%%%%%%%%%%
     %  column:8
     %%%%%%%%%%%%%%%%%%%%%%%%%%%%%%%%%%%%%%%%%%%%%
\multicolumn{1}{l}{\multirow{1}{*}{$2.15 \pm 0.19$}}\\
%%%%%%%%%%%%%%%%%%%%%%%%%%%%%%%%%%%%%%%%%%%%%%%%%%
%horizontal line:10
%subline:(10,1)
%\cline{1-1}
%subline:(10,2)
%\cline{2-2}
%subline:(10,3)
%\cline{3-3}
%subline:(10,4)
%\cline{4-4}
%subline:(10,5)
%\cline{5-5}
%subline:(10,6)
%\cline{6-6}
%subline:(10,7)
%\cline{7-7}
%subline:(10,8)
%\cline{8-8}
%%%%%%%%%%%%%%%%%%%%%%%%%%%%%%%%%%%%%%%%%%%%%%%%%%
%%%%%%%%%%%%%%%%%%%%%%%%%%%%%%%%%%%%%%%%%%%%%%%%%%
%  row:11
%%%%%%%%%%%%%%%%%%%%%%%%%%%%%%%%%%%%%%%%%%%%%%%%%%
%%%%%%%%%%%%%%%%%%%%%%%%%%%%%%%%%%%%%%%%%%%%%%%%%%
%subvertical line:(11,0)
     %%%%%%%%%%%%%%%%%%%%%%%%%%%%%%%%%%%%%%%%%%%%%
     %  column:1
     %%%%%%%%%%%%%%%%%%%%%%%%%%%%%%%%%%%%%%%%%%%%%
\multicolumn{1}{r}{\multirow{1}{*}{40}}&
     %%%%%%%%%%%%%%%%%%%%%%%%%%%%%%%%%%%%%%%%%%%%%
     %  column:2
     %%%%%%%%%%%%%%%%%%%%%%%%%%%%%%%%%%%%%%%%%%%%%
\multicolumn{1}{l|}{\multirow{1}{*}{$1.05 \pm 0.14$}}&
     %%%%%%%%%%%%%%%%%%%%%%%%%%%%%%%%%%%%%%%%%%%%%
     %  column:3
     %%%%%%%%%%%%%%%%%%%%%%%%%%%%%%%%%%%%%%%%%%%%%
\multicolumn{1}{r}{\multirow{1}{*}{9}}&
     %%%%%%%%%%%%%%%%%%%%%%%%%%%%%%%%%%%%%%%%%%%%%
     %  column:4
     %%%%%%%%%%%%%%%%%%%%%%%%%%%%%%%%%%%%%%%%%%%%%
\multicolumn{1}{l}{\multirow{1}{*}{$1.42 \pm 0.10$}}&
     %%%%%%%%%%%%%%%%%%%%%%%%%%%%%%%%%%%%%%%%%%%%%
     %  column:5
     %%%%%%%%%%%%%%%%%%%%%%%%%%%%%%%%%%%%%%%%%%%%%
\multicolumn{1}{l|}{\multirow{1}{*}{$1.20 \pm 0.08$}}&
     %%%%%%%%%%%%%%%%%%%%%%%%%%%%%%%%%%%%%%%%%%%%%
     %  column:6
     %%%%%%%%%%%%%%%%%%%%%%%%%%%%%%%%%%%%%%%%%%%%%
\multicolumn{1}{r}{\multirow{1}{*}{}}&
     %%%%%%%%%%%%%%%%%%%%%%%%%%%%%%%%%%%%%%%%%%%%%
     %  column:7
     %%%%%%%%%%%%%%%%%%%%%%%%%%%%%%%%%%%%%%%%%%%%%
\multicolumn{1}{l}{\multirow{1}{*}{}}&
     %%%%%%%%%%%%%%%%%%%%%%%%%%%%%%%%%%%%%%%%%%%%%
     %  column:8
     %%%%%%%%%%%%%%%%%%%%%%%%%%%%%%%%%%%%%%%%%%%%%
\multicolumn{1}{l}{\multirow{1}{*}{}}\\
%%%%%%%%%%%%%%%%%%%%%%%%%%%%%%%%%%%%%%%%%%%%%%%%%%
%horizontal line:11
%subline:(11,1)
%\cline{1-1}
%subline:(11,2)
%\cline{2-2}
%subline:(11,3)
%\cline{3-3}
%subline:(11,4)
%\cline{4-4}
%subline:(11,5)
%\cline{5-5}
%subline:(11,6)
%\cline{6-6}
%subline:(11,7)
%\cline{7-7}
%subline:(11,8)
%\cline{8-8}
%%%%%%%%%%%%%%%%%%%%%%%%%%%%%%%%%%%%%%%%%%%%%%%%%%
%%%%%%%%%%%%%%%%%%%%%%%%%%%%%%%%%%%%%%%%%%%%%%%%%%
%  row:12
%%%%%%%%%%%%%%%%%%%%%%%%%%%%%%%%%%%%%%%%%%%%%%%%%%
%%%%%%%%%%%%%%%%%%%%%%%%%%%%%%%%%%%%%%%%%%%%%%%%%%
%subvertical line:(12,0)
     %%%%%%%%%%%%%%%%%%%%%%%%%%%%%%%%%%%%%%%%%%%%%
     %  column:1
     %%%%%%%%%%%%%%%%%%%%%%%%%%%%%%%%%%%%%%%%%%%%%
\multicolumn{1}{r}{\multirow{1}{*}{45}}&
     %%%%%%%%%%%%%%%%%%%%%%%%%%%%%%%%%%%%%%%%%%%%%
     %  column:2
     %%%%%%%%%%%%%%%%%%%%%%%%%%%%%%%%%%%%%%%%%%%%%
\multicolumn{1}{l|}{\multirow{1}{*}{$0.93 \pm 0.09$}}&
     %%%%%%%%%%%%%%%%%%%%%%%%%%%%%%%%%%%%%%%%%%%%%
     %  column:3
     %%%%%%%%%%%%%%%%%%%%%%%%%%%%%%%%%%%%%%%%%%%%%
\multicolumn{1}{r}{\multirow{1}{*}{10}}&
     %%%%%%%%%%%%%%%%%%%%%%%%%%%%%%%%%%%%%%%%%%%%%
     %  column:4
     %%%%%%%%%%%%%%%%%%%%%%%%%%%%%%%%%%%%%%%%%%%%%
\multicolumn{1}{l}{\multirow{1}{*}{$1.44 \pm 0.08$}}&
     %%%%%%%%%%%%%%%%%%%%%%%%%%%%%%%%%%%%%%%%%%%%%
     %  column:5
     %%%%%%%%%%%%%%%%%%%%%%%%%%%%%%%%%%%%%%%%%%%%%
\multicolumn{1}{l|}{\multirow{1}{*}{$1.21 \pm 0.07$}}&
     %%%%%%%%%%%%%%%%%%%%%%%%%%%%%%%%%%%%%%%%%%%%%
     %  column:6
     %%%%%%%%%%%%%%%%%%%%%%%%%%%%%%%%%%%%%%%%%%%%%
\multicolumn{1}{r}{\multirow{1}{*}{}}&
     %%%%%%%%%%%%%%%%%%%%%%%%%%%%%%%%%%%%%%%%%%%%%
     %  column:7
     %%%%%%%%%%%%%%%%%%%%%%%%%%%%%%%%%%%%%%%%%%%%%
\multicolumn{1}{l}{\multirow{1}{*}{}}&
     %%%%%%%%%%%%%%%%%%%%%%%%%%%%%%%%%%%%%%%%%%%%%
     %  column:8
     %%%%%%%%%%%%%%%%%%%%%%%%%%%%%%%%%%%%%%%%%%%%%
\multicolumn{1}{l}{\multirow{1}{*}{}}\\
%%%%%%%%%%%%%%%%%%%%%%%%%%%%%%%%%%%%%%%%%%%%%%%%%%
%horizontal line:12
%subline:(12,1)
%\cline{1-1}
%subline:(12,2)
%\cline{2-2}
%subline:(12,3)
%\cline{3-3}
%subline:(12,4)
%\cline{4-4}
%subline:(12,5)
%\cline{5-5}
%subline:(12,6)
%\cline{6-6}
%subline:(12,7)
%\cline{7-7}
%subline:(12,8)
%\cline{8-8}
%%%%%%%%%%%%%%%%%%%%%%%%%%%%%%%%%%%%%%%%%%%%%%%%%%
%%%%%%%%%%%%%%%%%%%%%%%%%%%%%%%%%%%%%%%%%%%%%%%%%%
%  row:13
%%%%%%%%%%%%%%%%%%%%%%%%%%%%%%%%%%%%%%%%%%%%%%%%%%
%%%%%%%%%%%%%%%%%%%%%%%%%%%%%%%%%%%%%%%%%%%%%%%%%%
%subvertical line:(13,0)
     %%%%%%%%%%%%%%%%%%%%%%%%%%%%%%%%%%%%%%%%%%%%%
     %  column:1
     %%%%%%%%%%%%%%%%%%%%%%%%%%%%%%%%%%%%%%%%%%%%%
\multicolumn{1}{r}{\multirow{1}{*}{50}}&
     %%%%%%%%%%%%%%%%%%%%%%%%%%%%%%%%%%%%%%%%%%%%%
     %  column:2
     %%%%%%%%%%%%%%%%%%%%%%%%%%%%%%%%%%%%%%%%%%%%%
\multicolumn{1}{l|}{\multirow{1}{*}{$1.00 \pm 0.07$}}&
     %%%%%%%%%%%%%%%%%%%%%%%%%%%%%%%%%%%%%%%%%%%%%
     %  column:3
     %%%%%%%%%%%%%%%%%%%%%%%%%%%%%%%%%%%%%%%%%%%%%
\multicolumn{1}{r}{\multirow{1}{*}{11}}&
     %%%%%%%%%%%%%%%%%%%%%%%%%%%%%%%%%%%%%%%%%%%%%
     %  column:4
     %%%%%%%%%%%%%%%%%%%%%%%%%%%%%%%%%%%%%%%%%%%%%
\multicolumn{1}{l}{\multirow{1}{*}{$1.41 \pm 0.09$}}&
     %%%%%%%%%%%%%%%%%%%%%%%%%%%%%%%%%%%%%%%%%%%%%
     %  column:5
     %%%%%%%%%%%%%%%%%%%%%%%%%%%%%%%%%%%%%%%%%%%%%
\multicolumn{1}{l|}{\multirow{1}{*}{$1.19 \pm 0.08$}}&
     %%%%%%%%%%%%%%%%%%%%%%%%%%%%%%%%%%%%%%%%%%%%%
     %  column:6
     %%%%%%%%%%%%%%%%%%%%%%%%%%%%%%%%%%%%%%%%%%%%%
\multicolumn{1}{r}{\multirow{1}{*}{}}&
     %%%%%%%%%%%%%%%%%%%%%%%%%%%%%%%%%%%%%%%%%%%%%
     %  column:7
     %%%%%%%%%%%%%%%%%%%%%%%%%%%%%%%%%%%%%%%%%%%%%
\multicolumn{1}{l}{\multirow{1}{*}{}}&
     %%%%%%%%%%%%%%%%%%%%%%%%%%%%%%%%%%%%%%%%%%%%%
     %  column:8
     %%%%%%%%%%%%%%%%%%%%%%%%%%%%%%%%%%%%%%%%%%%%%
\multicolumn{1}{l}{\multirow{1}{*}{}}\\
%%%%%%%%%%%%%%%%%%%%%%%%%%%%%%%%%%%%%%%%%%%%%%%%%%
%horizontal line:13
%subline:(13,1)
%\cline{1-1}
%subline:(13,2)
%\cline{2-2}
%subline:(13,3)
%\cline{3-3}
%subline:(13,4)
%\cline{4-4}
%subline:(13,5)
%\cline{5-5}
%subline:(13,6)
%\cline{6-6}
%subline:(13,7)
%\cline{7-7}
%subline:(13,8)
%\cline{8-8}
%%%%%%%%%%%%%%%%%%%%%%%%%%%%%%%%%%%%%%%%%%%%%%%%%%
%%%%%%%%%%%%%%%%%%%%%%%%%%%%%%%%%%%%%%%%%%%%%%%%%%
%  row:14
%%%%%%%%%%%%%%%%%%%%%%%%%%%%%%%%%%%%%%%%%%%%%%%%%%
%%%%%%%%%%%%%%%%%%%%%%%%%%%%%%%%%%%%%%%%%%%%%%%%%%
%subvertical line:(14,0)
     %%%%%%%%%%%%%%%%%%%%%%%%%%%%%%%%%%%%%%%%%%%%%
     %  column:1
     %%%%%%%%%%%%%%%%%%%%%%%%%%%%%%%%%%%%%%%%%%%%%
\multicolumn{1}{r}{\multirow{1}{*}{}}&
     %%%%%%%%%%%%%%%%%%%%%%%%%%%%%%%%%%%%%%%%%%%%%
     %  column:2
     %%%%%%%%%%%%%%%%%%%%%%%%%%%%%%%%%%%%%%%%%%%%%
\multicolumn{1}{l|}{\multirow{1}{*}{}}&
     %%%%%%%%%%%%%%%%%%%%%%%%%%%%%%%%%%%%%%%%%%%%%
     %  column:3
     %%%%%%%%%%%%%%%%%%%%%%%%%%%%%%%%%%%%%%%%%%%%%
\multicolumn{1}{r}{\multirow{1}{*}{12}}&
     %%%%%%%%%%%%%%%%%%%%%%%%%%%%%%%%%%%%%%%%%%%%%
     %  column:4
     %%%%%%%%%%%%%%%%%%%%%%%%%%%%%%%%%%%%%%%%%%%%%
\multicolumn{1}{l}{\multirow{1}{*}{$1.30 \pm 0.15$}}&
     %%%%%%%%%%%%%%%%%%%%%%%%%%%%%%%%%%%%%%%%%%%%%
     %  column:5
     %%%%%%%%%%%%%%%%%%%%%%%%%%%%%%%%%%%%%%%%%%%%%
\multicolumn{1}{l|}{\multirow{1}{*}{$1.09 \pm 0.12$}}&
     %%%%%%%%%%%%%%%%%%%%%%%%%%%%%%%%%%%%%%%%%%%%%
     %  column:6
     %%%%%%%%%%%%%%%%%%%%%%%%%%%%%%%%%%%%%%%%%%%%%
\multicolumn{1}{r}{\multirow{1}{*}{}}&
     %%%%%%%%%%%%%%%%%%%%%%%%%%%%%%%%%%%%%%%%%%%%%
     %  column:7
     %%%%%%%%%%%%%%%%%%%%%%%%%%%%%%%%%%%%%%%%%%%%%
\multicolumn{1}{l}{\multirow{1}{*}{}}&
     %%%%%%%%%%%%%%%%%%%%%%%%%%%%%%%%%%%%%%%%%%%%%
     %  column:8
     %%%%%%%%%%%%%%%%%%%%%%%%%%%%%%%%%%%%%%%%%%%%%
\multicolumn{1}{l}{\multirow{1}{*}{}}\\
%%%%%%%%%%%%%%%%%%%%%%%%%%%%%%%%%%%%%%%%%%%%%%%%%%
%horizontal line:14
%subline:(14,1)
\cline{1-1}
%subline:(14,2)
\cline{2-2}
%subline:(14,3)
\cline{3-3}
%subline:(14,4)
\cline{4-4}
%subline:(14,5)
\cline{5-5}
%subline:(14,6)
\cline{6-6}
%subline:(14,7)
\cline{7-7}
%subline:(14,8)
\cline{8-8}
%%%%%%%%%%%%%%%%%%%%%%%%%%%%%%%%%%%%%%%%%%%%%%%%%%
%%%%%%%%%%%%%%%%%%%%%%%%%%%%%%%%%%%%%%%%%%%%%%%%%%
%  row:15
%%%%%%%%%%%%%%%%%%%%%%%%%%%%%%%%%%%%%%%%%%%%%%%%%%
%%%%%%%%%%%%%%%%%%%%%%%%%%%%%%%%%%%%%%%%%%%%%%%%%%
%subvertical line:(15,0)
     %%%%%%%%%%%%%%%%%%%%%%%%%%%%%%%%%%%%%%%%%%%%%
     %  column:1
     %%%%%%%%%%%%%%%%%%%%%%%%%%%%%%%%%%%%%%%%%%%%%
\multicolumn{8}{c}{\multirow{1}{*}{literature/theoretical result}}\\
     %%%%%%%%%%%%%%%%%%%%%%%%%%%%%%%%%%%%%%%%%%%%%
     %  column:2
     %%%%%%%%%%%%%%%%%%%%%%%%%%%%%%%%%%%%%%%%%%%%%
% \multicolumn{8}{r}{\multirow{1}{*}{}}\\
     %%%%%%%%%%%%%%%%%%%%%%%%%%%%%%%%%%%%%%%%%%%%%
     %  column:3
     %%%%%%%%%%%%%%%%%%%%%%%%%%%%%%%%%%%%%%%%%%%%%
% \multicolumn{8}{r}{\multirow{1}{*}{}}\\
     %%%%%%%%%%%%%%%%%%%%%%%%%%%%%%%%%%%%%%%%%%%%%
     %  column:4
     %%%%%%%%%%%%%%%%%%%%%%%%%%%%%%%%%%%%%%%%%%%%%
% \multicolumn{8}{r}{\multirow{1}{*}{(15, 4)}}\\
     %%%%%%%%%%%%%%%%%%%%%%%%%%%%%%%%%%%%%%%%%%%%%
     %  column:5
     %%%%%%%%%%%%%%%%%%%%%%%%%%%%%%%%%%%%%%%%%%%%%
% \multicolumn{8}{r}{\multirow{1}{*}{(15, 5)}}\\
     %%%%%%%%%%%%%%%%%%%%%%%%%%%%%%%%%%%%%%%%%%%%%
     %  column:6
     %%%%%%%%%%%%%%%%%%%%%%%%%%%%%%%%%%%%%%%%%%%%%
% \multicolumn{8}{r}{\multirow{1}{*}{}}\\
     %%%%%%%%%%%%%%%%%%%%%%%%%%%%%%%%%%%%%%%%%%%%%
     %  column:7
     %%%%%%%%%%%%%%%%%%%%%%%%%%%%%%%%%%%%%%%%%%%%%
% \multicolumn{8}{r}{\multirow{1}{*}{}}\\
     %%%%%%%%%%%%%%%%%%%%%%%%%%%%%%%%%%%%%%%%%%%%%
     %  column:8
     %%%%%%%%%%%%%%%%%%%%%%%%%%%%%%%%%%%%%%%%%%%%%
% \multicolumn{8}{r}{\multirow{1}{*}{(15, 8)}}\\
%%%%%%%%%%%%%%%%%%%%%%%%%%%%%%%%%%%%%%%%%%%%%%%%%%
%horizontal line:15
%subline:(15,1)
\cline{1-1}
%subline:(15,2)
\cline{2-2}
%subline:(15,3)
\cline{3-3}
%subline:(15,4)
\cline{4-4}
%subline:(15,5)
\cline{5-5}
%subline:(15,6)
\cline{6-6}
%subline:(15,7)
\cline{7-7}
%subline:(15,8)
\cline{8-8}
%%%%%%%%%%%%%%%%%%%%%%%%%%%%%%%%%%%%%%%%%%%%%%%%%%
%%%%%%%%%%%%%%%%%%%%%%%%%%%%%%%%%%%%%%%%%%%%%%%%%%
%  row:16
%%%%%%%%%%%%%%%%%%%%%%%%%%%%%%%%%%%%%%%%%%%%%%%%%%
%%%%%%%%%%%%%%%%%%%%%%%%%%%%%%%%%%%%%%%%%%%%%%%%%%
%subvertical line:(16,0)
     %%%%%%%%%%%%%%%%%%%%%%%%%%%%%%%%%%%%%%%%%%%%%
     %  column:1
     %%%%%%%%%%%%%%%%%%%%%%%%%%%%%%%%%%%%%%%%%%%%%
\multicolumn{1}{r}{\multirow{1}{*}{}}&
     %%%%%%%%%%%%%%%%%%%%%%%%%%%%%%%%%%%%%%%%%%%%%
     %  column:2
     %%%%%%%%%%%%%%%%%%%%%%%%%%%%%%%%%%%%%%%%%%%%%
\multicolumn{1}{l|}{\multirow{1}{*}{}}&
     %%%%%%%%%%%%%%%%%%%%%%%%%%%%%%%%%%%%%%%%%%%%%
     %  column:3
     %%%%%%%%%%%%%%%%%%%%%%%%%%%%%%%%%%%%%%%%%%%%%
\multicolumn{1}{r}{\multirow{1}{*}{}}&
     %%%%%%%%%%%%%%%%%%%%%%%%%%%%%%%%%%%%%%%%%%%%%
     %  column:4
     %%%%%%%%%%%%%%%%%%%%%%%%%%%%%%%%%%%%%%%%%%%%%
\multicolumn{1}{l}{\multirow{1}{*}{$1.23\pm 0.16^{\rm b}$}}&
     %%%%%%%%%%%%%%%%%%%%%%%%%%%%%%%%%%%%%%%%%%%%%
     %  column:5
     %%%%%%%%%%%%%%%%%%%%%%%%%%%%%%%%%%%%%%%%%%%%%
\multicolumn{1}{l|}{\multirow{1}{*}{$1.06\pm 0.14^{\rm b}$}}&
     %%%%%%%%%%%%%%%%%%%%%%%%%%%%%%%%%%%%%%%%%%%%%
     %  column:6
     %%%%%%%%%%%%%%%%%%%%%%%%%%%%%%%%%%%%%%%%%%%%%
\multicolumn{1}{r}{\multirow{1}{*}{}}&
     %%%%%%%%%%%%%%%%%%%%%%%%%%%%%%%%%%%%%%%%%%%%%
     %  column:7
     %%%%%%%%%%%%%%%%%%%%%%%%%%%%%%%%%%%%%%%%%%%%%
\multicolumn{1}{l}{\multirow{1}{*}{$$}}&
     %%%%%%%%%%%%%%%%%%%%%%%%%%%%%%%%%%%%%%%%%%%%%
     %  column:8
     %%%%%%%%%%%%%%%%%%%%%%%%%%%%%%%%%%%%%%%%%%%%%
\multicolumn{1}{l}{\multirow{1}{*}{$2.79\pm 0.70^{\rm d}$}}\\
%%%%%%%%%%%%%%%%%%%%%%%%%%%%%%%%%%%%%%%%%%%%%%%%%%
%horizontal line:16
%subline:(16,1)
%\cline{1-1}
%subline:(16,2)
%\cline{2-2}
%subline:(16,3)
%\cline{3-3}
%subline:(16,4)
%\cline{4-4}
%subline:(16,5)
%\cline{5-5}
%subline:(16,6)
%\cline{6-6}
%subline:(16,7)
%\cline{7-7}
%subline:(16,8)
%\cline{8-8}
%%%%%%%%%%%%%%%%%%%%%%%%%%%%%%%%%%%%%%%%%%%%%%%%%%
%%%%%%%%%%%%%%%%%%%%%%%%%%%%%%%%%%%%%%%%%%%%%%%%%%
%  row:17
%%%%%%%%%%%%%%%%%%%%%%%%%%%%%%%%%%%%%%%%%%%%%%%%%%
%%%%%%%%%%%%%%%%%%%%%%%%%%%%%%%%%%%%%%%%%%%%%%%%%%
%subvertical line:(17,0)
     %%%%%%%%%%%%%%%%%%%%%%%%%%%%%%%%%%%%%%%%%%%%%
     %  column:1
     %%%%%%%%%%%%%%%%%%%%%%%%%%%%%%%%%%%%%%%%%%%%%
\multicolumn{1}{r}{\multirow{1}{*}{}}&
     %%%%%%%%%%%%%%%%%%%%%%%%%%%%%%%%%%%%%%%%%%%%%
     %  column:2
     %%%%%%%%%%%%%%%%%%%%%%%%%%%%%%%%%%%%%%%%%%%%%
\multicolumn{1}{l|}{\multirow{1}{*}{    }}&
     %%%%%%%%%%%%%%%%%%%%%%%%%%%%%%%%%%%%%%%%%%%%%
     %  column:3
     %%%%%%%%%%%%%%%%%%%%%%%%%%%%%%%%%%%%%%%%%%%%%
\multicolumn{1}{r}{\multirow{1}{*}{}}&
     %%%%%%%%%%%%%%%%%%%%%%%%%%%%%%%%%%%%%%%%%%%%%
     %  column:4
     %%%%%%%%%%%%%%%%%%%%%%%%%%%%%%%%%%%%%%%%%%%%%
\multicolumn{1}{l}{\multirow{1}{*}{
          }}&
     %%%%%%%%%%%%%%%%%%%%%%%%%%%%%%%%%%%%%%%%%%%%%
     %  column:5
     %%%%%%%%%%%%%%%%%%%%%%%%%%%%%%%%%%%%%%%%%%%%%
\multicolumn{1}{l|}{\multirow{1}{*}{}}&
     %%%%%%%%%%%%%%%%%%%%%%%%%%%%%%%%%%%%%%%%%%%%%
     %  column:6
     %%%%%%%%%%%%%%%%%%%%%%%%%%%%%%%%%%%%%%%%%%%%%
\multicolumn{1}{r}{\multirow{1}{*}{}}&
     %%%%%%%%%%%%%%%%%%%%%%%%%%%%%%%%%%%%%%%%%%%%%
     %  column:7
     %%%%%%%%%%%%%%%%%%%%%%%%%%%%%%%%%%%%%%%%%%%%%
\multicolumn{1}{l}{\multirow{1}{*}{$5.84 \pm 0.13 ^{\rm e} $}}&
     %%%%%%%%%%%%%%%%%%%%%%%%%%%%%%%%%%%%%%%%%%%%%
     %  column:8
     %%%%%%%%%%%%%%%%%%%%%%%%%%%%%%%%%%%%%%%%%%%%%
\multicolumn{1}{l}{\multirow{1}{*}{$2.58 \pm 0.06 ^{\rm e}$}}\\
%%%%%%%%%%%%%%%%%%%%%%%%%%%%%%%%%%%%%%%%%%%%%%%%%%
%  row:18
%%%%%%%%%%%%%%%%%%%%%%%%%%%%%%%%%%%%%%%%%%%%%%%%%%
%%%%%%%%%%%%%%%%%%%%%%%%%%%%%%%%%%%%%%%%%%%%%%%%%%
%subvertical line:(18,0)
     %%%%%%%%%%%%%%%%%%%%%%%%%%%%%%%%%%%%%%%%%%%%%
     %  column:1
     %%%%%%%%%%%%%%%%%%%%%%%%%%%%%%%%%%%%%%%%%%%%%
\multicolumn{1}{r}{\multirow{1}{*}{}}&
     %%%%%%%%%%%%%%%%%%%%%%%%%%%%%%%%%%%%%%%%%%%%%
     %  column:2
     %%%%%%%%%%%%%%%%%%%%%%%%%%%%%%%%%%%%%%%%%%%%%
\multicolumn{1}{l|}{\multirow{1}{*}{$1.00^{\rm a}$}}&
     %%%%%%%%%%%%%%%%%%%%%%%%%%%%%%%%%%%%%%%%%%%%%
     %  column:3
     %%%%%%%%%%%%%%%%%%%%%%%%%%%%%%%%%%%%%%%%%%%%%
\multicolumn{1}{r}{\multirow{1}{*}{}}&
     %%%%%%%%%%%%%%%%%%%%%%%%%%%%%%%%%%%%%%%%%%%%%
     %  column:4
     %%%%%%%%%%%%%%%%%%%%%%%%%%%%%%%%%%%%%%%%%%%%%
\multicolumn{1}{l}{\multirow{1}{*}{$1.61^{\rm c}$}}&
     %%%%%%%%%%%%%%%%%%%%%%%%%%%%%%%%%%%%%%%%%%%%%
     %  column:5
     %%%%%%%%%%%%%%%%%%%%%%%%%%%%%%%%%%%%%%%%%%%%%
\multicolumn{1}{l|}{\multirow{1}{*}{$1.37^{\rm c}$}}&
     %%%%%%%%%%%%%%%%%%%%%%%%%%%%%%%%%%%%%%%%%%%%%
     %  column:6
     %%%%%%%%%%%%%%%%%%%%%%%%%%%%%%%%%%%%%%%%%%%%%
\multicolumn{1}{r}{\multirow{1}{*}{}}&
     %%%%%%%%%%%%%%%%%%%%%%%%%%%%%%%%%%%%%%%%%%%%%
     %  column:7
     %%%%%%%%%%%%%%%%%%%%%%%%%%%%%%%%%%%%%%%%%%%%%
\multicolumn{1}{l}{\multirow{1}{*}{$5.83^{\rm c} $}}&
     %%%%%%%%%%%%%%%%%%%%%%%%%%%%%%%%%%%%%%%%%%%%%
     %  column:8
     %%%%%%%%%%%%%%%%%%%%%%%%%%%%%%%%%%%%%%%%%%%%%
\multicolumn{1}{l}{\multirow{1}{*}{$2.58^{\rm c}$}}\\
%%%%%%%%%%%%%%%%%%%%%%%%%%%%%%%%%%%%%%%%%%%%%%%%%%

%horizontal line:17
%subline:(17,1)
\cline{1-1}
%subline:(17,2)
\cline{2-2}
%subline:(17,3)
\cline{3-3}
%subline:(17,4)
\cline{4-4}
%subline:(17,5)
\cline{5-5}
%subline:(17,6)
\cline{6-6}
%subline:(17,7)
\cline{7-7}
%subline:(17,8)
\cline{8-8}
\end{tabular}
\end{center}
\end{table*}
%\end{document}

%%%%%% table %%%%%%%%%%%%%%%%%%%%%%
All values are within 
a 50\% deviation from the
 true value of $10^{-50}$ 
 with the more accurate estimates
 for the simulations having a large number of workers.
 Also, the true value is within one  standard deviation of the reported averages for 70\% of the data points, as is expected from the standard Gaussian confidence intervals. 
Figure~\ref{fig:scaling}a shows the scaling of the MD time (solid lines) and number of MC moves (dashed lines) of the MSVS simulations (orange) compared to linear scaling (black) and the expected scaling 
for standard replica exchange (REPEX) in which ensembles are updated in cohort (purple).

Whereas 
the number of "MD steps" and MC moves quickly levels off
to a nearly flat plateau in the standard approach due to workers being idle
as they need to wait for the slowest worker, 
the replica exchange approach developed in this article
shows a perfect linear scaling with respect to the MD time.
The number of MC moves in the new method shows
an even better than linear scaling
due to the fact that the ensembles with shorter 
"path lengths"
get simulated relatively more often with more workers, resulting in more MC moves per second. 
This in itself does not necessarily mean that the simulations converge much faster because the additional computational effort may 
not be targeted to the sampling where it is needed.
If we neglect the fact that path ensemble simulations are 
correlated via the replica exchange moves, we can write 
that the relative error in overall crossing probability
$\epsilon$ follows from the relative errors in each path ensemble $\epsilon_i$ via: $\epsilon^2=\sum_i \epsilon_i^2$. 
It is henceforth clear that additional
computational power should not aim to lower the error in a few path ensembles that were already low compared to other path ensembles.
We therefore measure the effectiveness 
of the additional workers by calculating computational 
efficiencies. The efficiency 
of a specific computational method
is here defined as the inverse
computer time, CPU- or wall-time, to obtain an overall 
relative error equal to 1: $\epsilon=1$. 

In figure~\ref{fig:scaling}d the efficiencies based on wall-time (solid) and CPU-time (dashed) are plotted for the MSVS process.
These plots depends on the ability
of computing reliable statistical errors 
in the overall crossing probability that is 
an extremely small number, $10^{-50}$.
The somewhat fluctuating behavior of these curves 
should hence be viewed as  statistical noise 
as the confidence interval
of these efficiencies
depends on the 
statistical error of this error.
Despite that, clear trends can be observed in which the CPU-time efficiency is more or less flat, while the wall-time efficiency shows an upward trend.
If we neglect the effect of replica exchange moves on the efficiency, we can relate these numerical results with theoretical ones~\cite{TISeff, Raffa}
for any possible division of a fixed total
CPU-time over the different ensembles.
A common sense approach would be to aim for the same error $\epsilon_i$ in each ensemble 
(which implies doing the same
number of MC moves per ensemble) or to divide the total CPU-time evenly over the ensembles. 
These two strategies correspond
to the case $K=1$ or standard RETIS and $K=N$ or standard TIS, respectively.
Ref.~\cite{TISeff} showed that these two strategies provide the same efficiency and
in the SI we derive that 
this leads to a wall-time efficiency 
as function of the number of
workers ($K$) equal to $K/56250$ which is the 
continuous
purple
line in figure~\ref{fig:scaling}d. The optimum division, however, would give 
a slightly better
wall-time efficiency equal to $K/50000$ which is the continuous black line in this figure.
Also shown in figure~\ref{fig:scaling}d are
the expected theoretical efficiencies based
on the numerical distribution of MC moves in each ensemble. This hybrid numerical/theoretical result is 
shown by the small purple dots.
This shows that $\infty$RETIS, at least 
for a system in which the path length grows linearly with the ensemble's rank, 
naturally provides a division of the computational resources that is even better than TIS ($K=N$) or RETIS $(K=1)$.
Yet, due to statistical inaccuracies
this is only evident for the $K=15$ case.
The best wall-time efficiency is obtained for the case $K=N$, which is essentially
equivalent of running independent TIS simulations (i.e. without doing any replica exchange moves). 
We do not expect this to apply to more complex systems where the replica exchange move is a proven weapon for efficient sampling.

\subsection*{Two-channel simulations}
In the middle column of table~\ref{tab:table1} we report the %found average 
calculated
crossing probabilities and permeabilities for 5 simulations for every number of workers. All simulations 
are somewhat higher, though still in good agreement with
the previous simulation from
Ref.~\citep{permeability}.  
We also evaluated the approximate result based
on Kramers' theory (see SI) which seem to confirm the results obtained in this paper.

\begin{figure*}[ht]
\centering
\subfigimg[width=.32\textwidth]{a)}{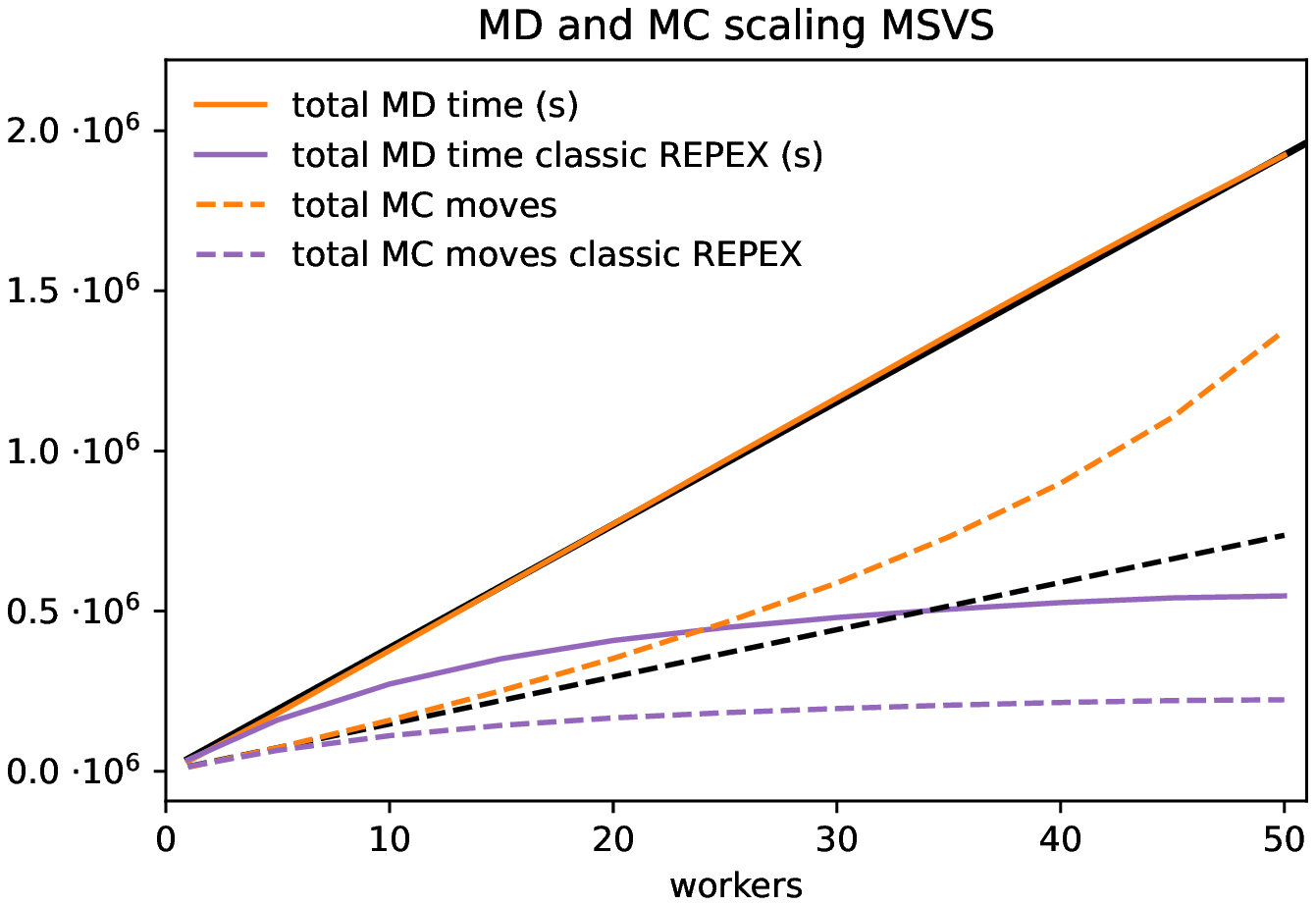}
\subfigimg[width=.32\textwidth]{b)}{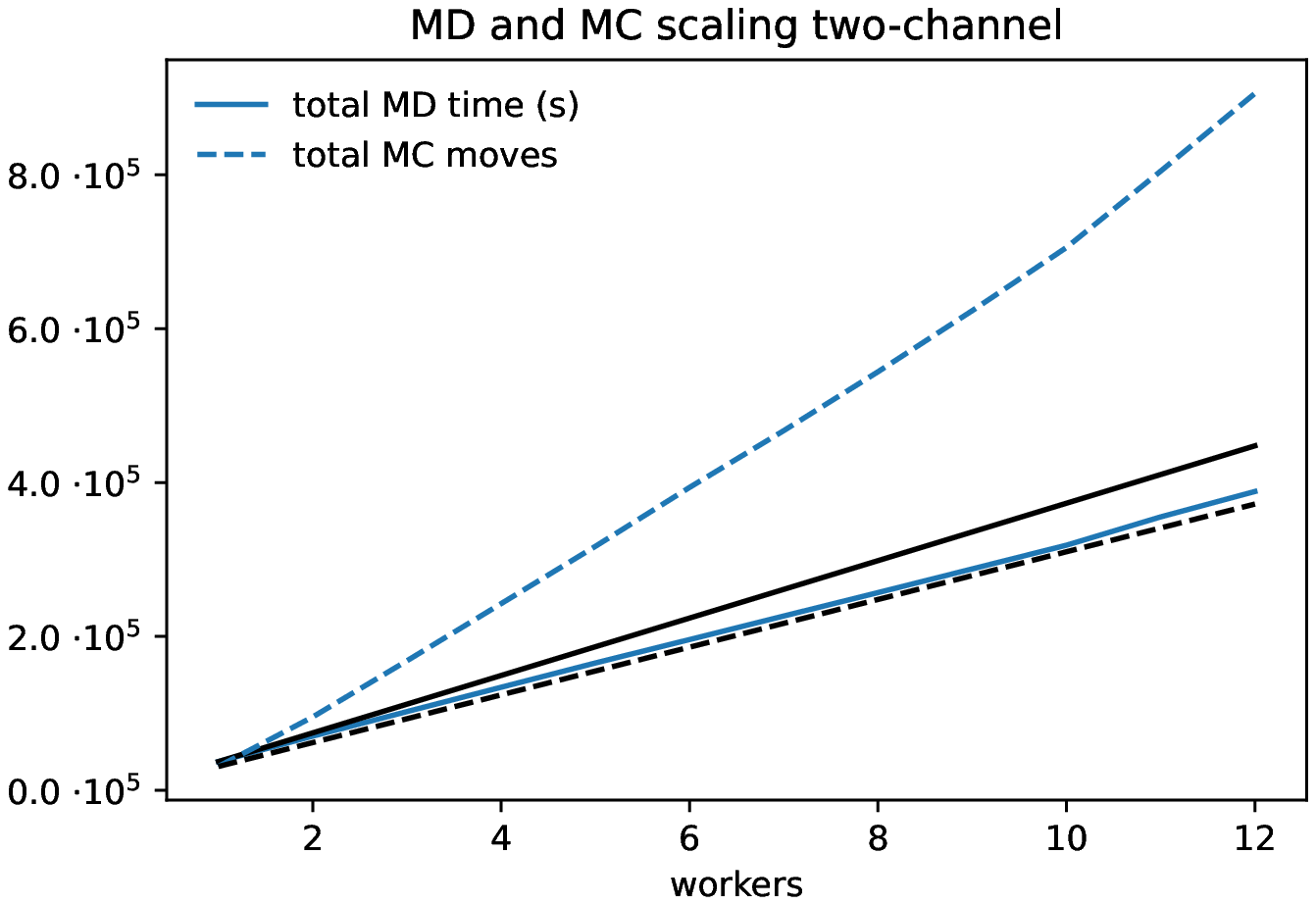}
\subfigimg[width=.32\textwidth]{c)}{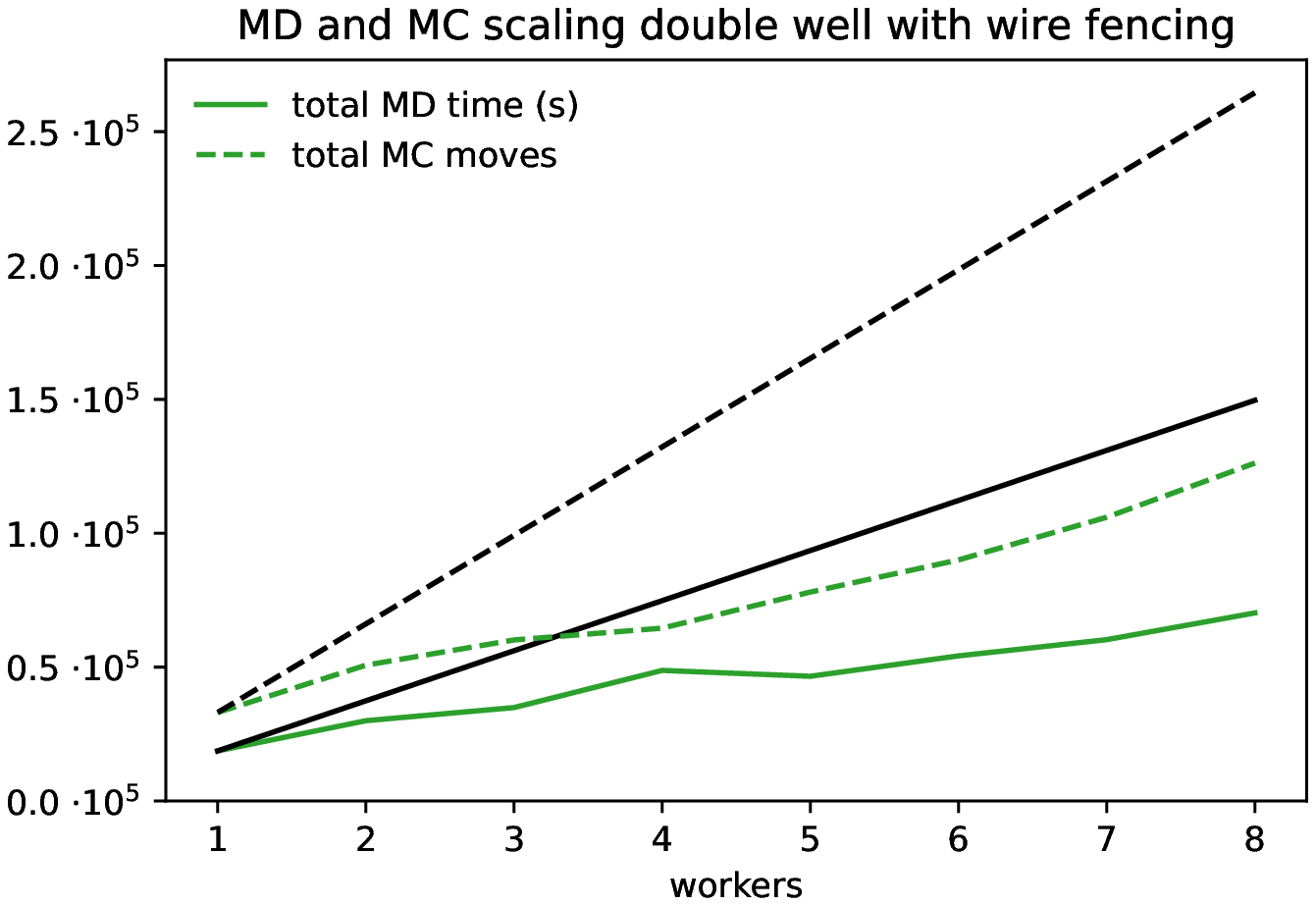}\\
\subfigimg[width=.32\textwidth]{d)}{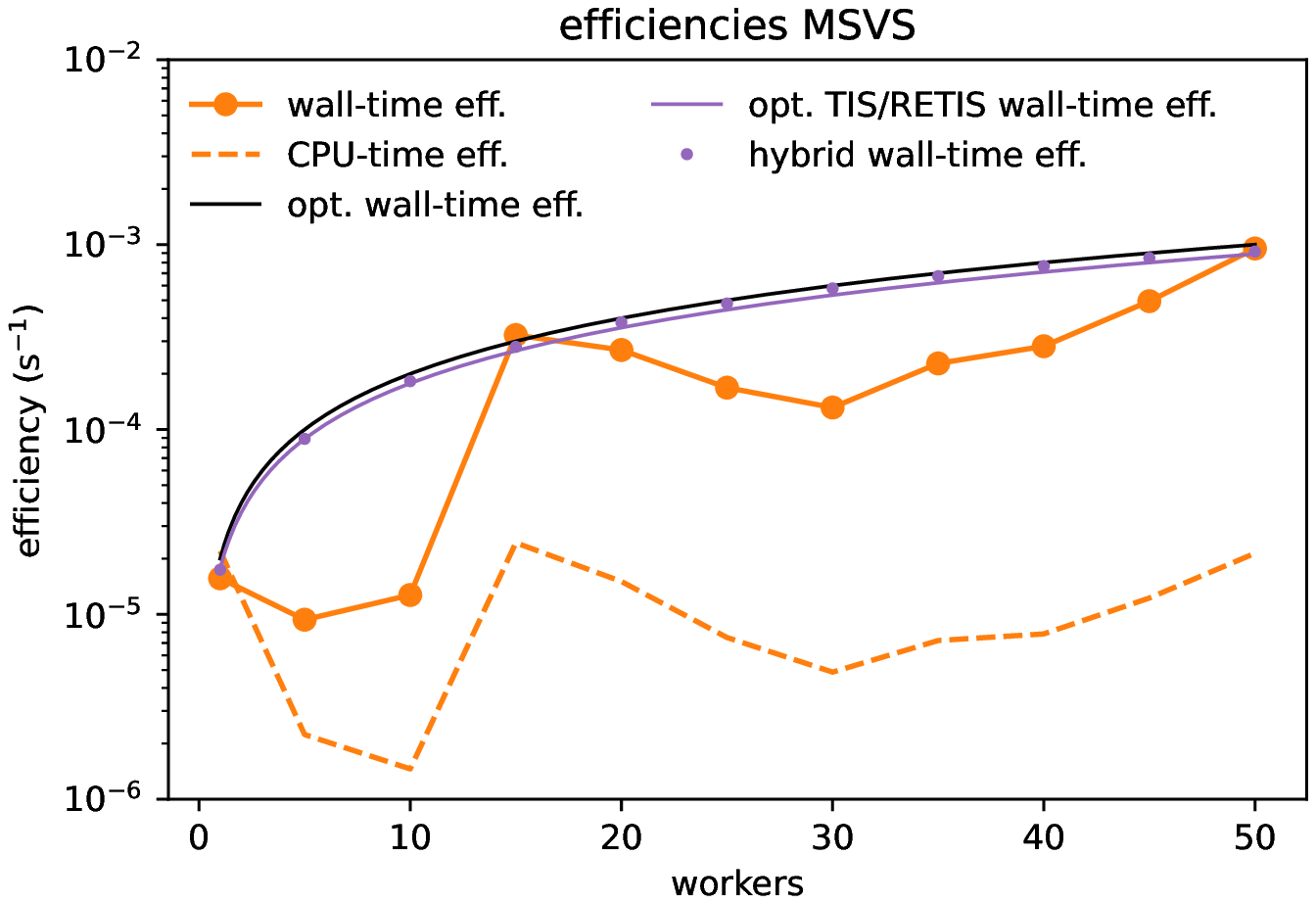}
\subfigimg[width=.32\textwidth]{e)}{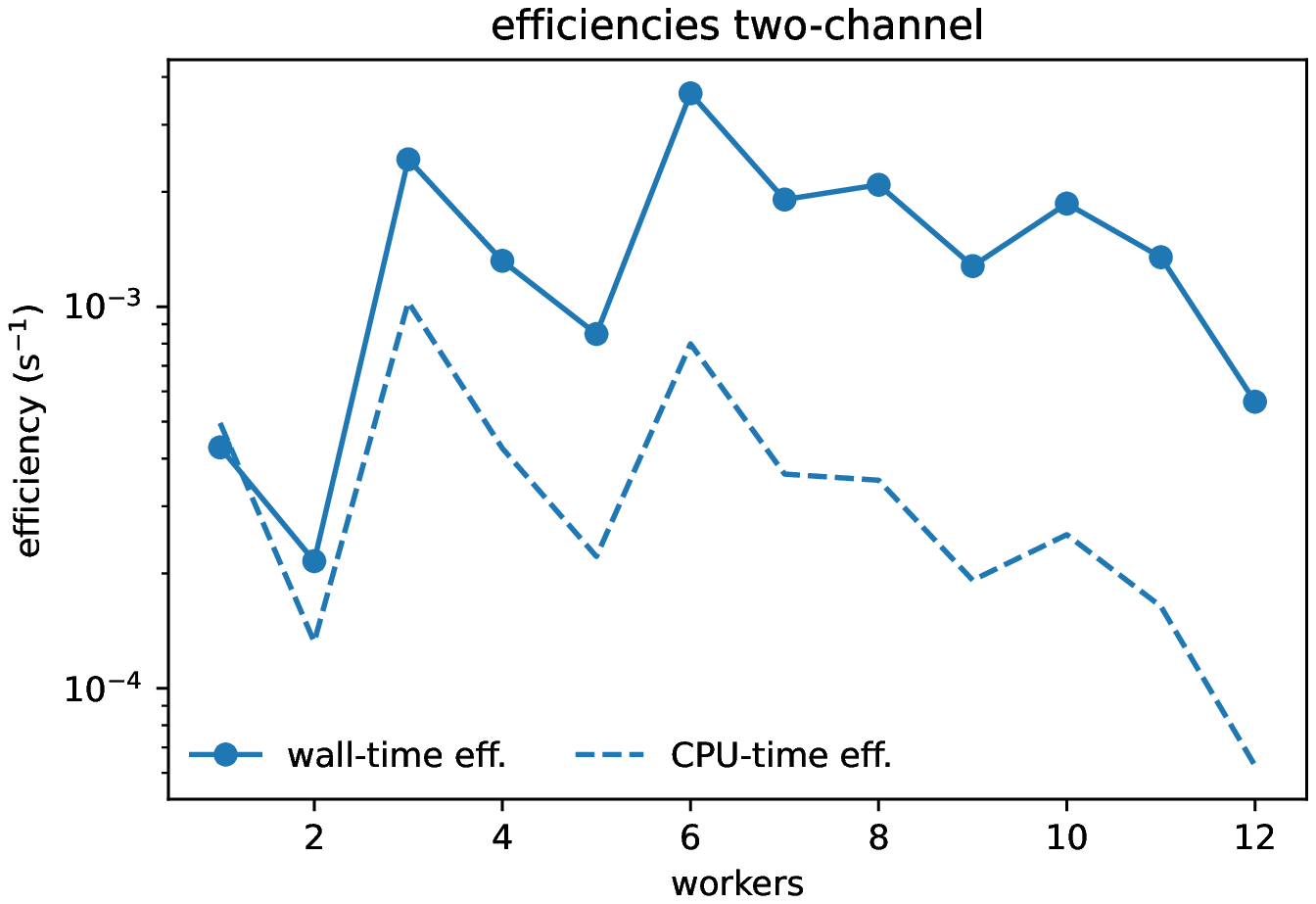}
\subfigimg[width=.32\textwidth]{f)}{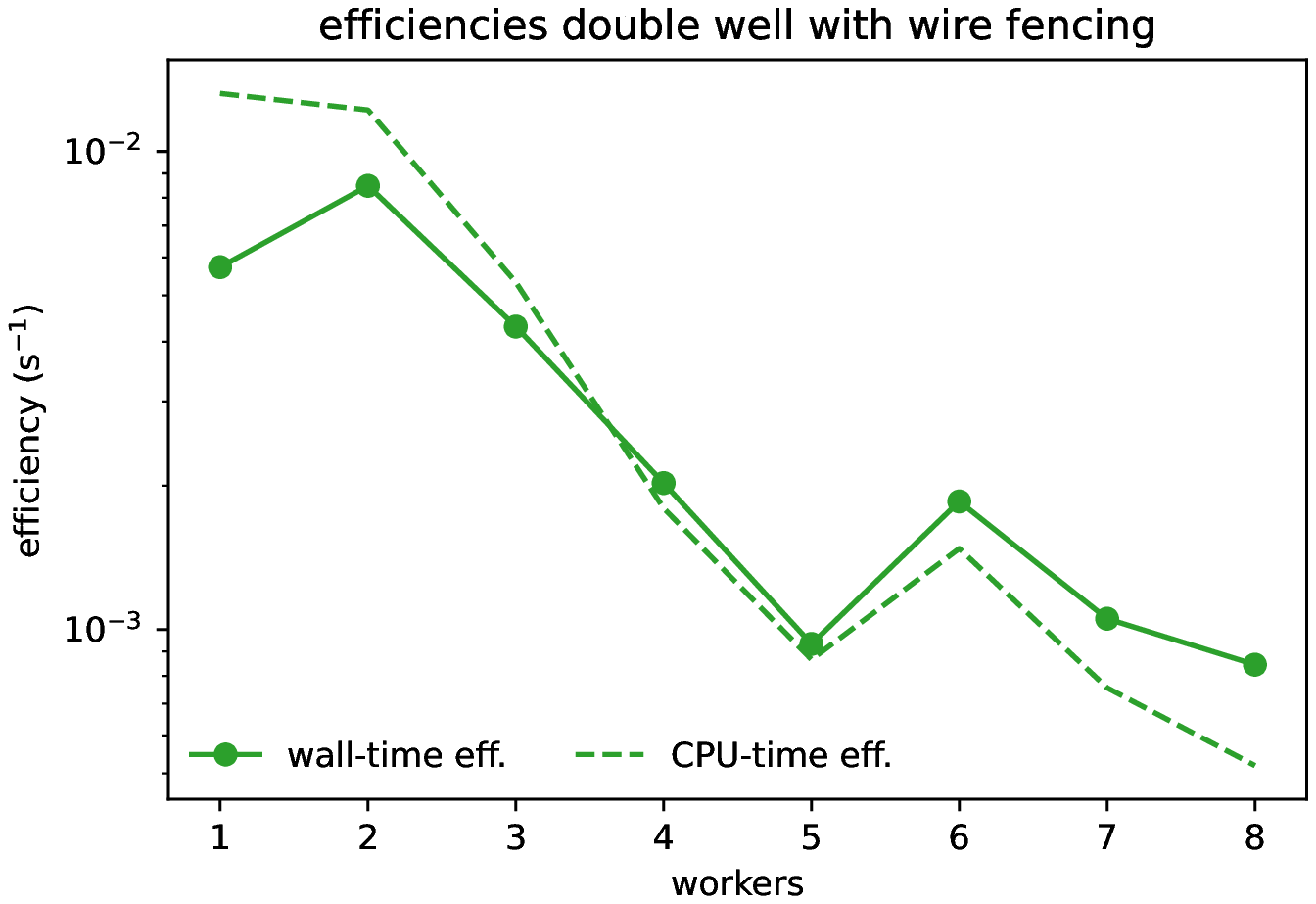}
\caption{The average scaling of total MD time (cumulative time spend by all the workers) (solid) and MC moves (dashed) (a-c) and wall-time (solid) and the CPU-time (dashed) efficiencies (d-f) for each number of workers. This is shown for the 
memoryless single variable 
stochastic (MSVS) process 
(a, d, orange), the two-channel system (b, e, blue) and the
double well with wire fencing (c, f, green) simulations. Each of the data points is based on 5 independent simulations. For the scaling plots, the black lines are guides for linear scaling from the 1 worker data-point. The purple lines in the scaling plot for the 
MSVS
simulations (a) show what the scaling would be if we had to wait for the slowest ensemble to finish for each MC move. The black line, purple line, and points in the efficiency plot of the MSVS process (d) show the optimal, optimal TIS/RETIS, and hybrid wall-time efficiency, respectively, as computed in the SI.
}
\label{fig:scaling}
\end{figure*}

Figure~\ref{fig:scaling}b shows the scaling of the MD time (solid lines) and number of MC moves (dashed lines) of the two-channel simulations (blue) compared to linear scaling (black). We see a slightly worse than linear scaling of the MD time, which might just be due to a small positive fluctuation of the 1 worker data-point. We also see a similar more than linear scaling in the number of MC moves as with the MSVS simulations, for the same reason. 
In figure~\ref{fig:scaling}e the efficiencies based on wall-time (solid) and CPU-time (dashed) are plotted for the two-channel system. The CPU-time efficiency is more or less flat until 8 workers after which it starts to drop off. The wall-time efficiency shows an upward trend until 10 workers after which it starts to drop off as well. 
We assign this drop to the reduction of
replica exchange moves which is an essential 
aspect for sampling this system efficiently~\cite{permeability}.
This is tangible from 
figure S1
in the SI where we plot fraction of trajectories, 
passing through $\lambda_{M-1}$, that are in the lower barrier channel.  
While from the average fraction it still looks like the simulations sampled both channels 
for any number of workers, 
4 out of the 5 simulations
in the $K=N=12$ case solely visited one of the two channels. This is in agreement with previous TIS results~\cite{permeability}.
The $K=11$ case already provides a dramatic improvement, but is still expected to be sub optimal due to the relatively low frequency
of replica exchange moves compared to $K<11$.
As reported in ref.~\cite{permeability}, this ratio requires many MC moves to converge to the theoretical value of 0.71 without the added MC moves introduced in that paper. We did not simulate with these added moves and thus see the same slow convergence for all of our simulations.
From this 2D system it would indicate that having half the number of ensembles as workers is a safe bet for optimum efficiency. 

%%%%%%%%%%%%%%%%%%%%%%%%%%%%%%%%%%%%%%%%%%%%%%%
%%%%%%%%%%%%%%%%%%%%%%%%%%%%%%%%%%%%%%%%%%%%%%%%%

\subsection*{Double well 1D barrier using wire fencing}
In the right column of table~\ref{tab:table1} we report the  
calculated
crossing probabilities and rates 
for the underdamped Langevin particle in the 1D double well potential. 
All simulations 
are in reasonable agreement with each other 
and the results of Refs.~\citep{wf} and
and~\citep{TitusRev}, as well as the 
approximate value based on Kramers' theory.
However, while these results confirm
the soundness of the method, the scaling and
efficiency are less convincing.
Figure~\ref{fig:scaling}b shows 
a significantly worse than linear scaling. On further inspection we found the average time per MC move was significantly smaller than our infinite-swapping goal of 1~s when the simulation was run with more than 2 workers. This results in a bottleneck on how many MC moves can be started per second, which is the reason for the observed bad scaling. It still is slightly positive instead of flat as the infinite swapping procedure becomes quicker with more workers due to the smaller $W$-matrix. 
The same bottleneck can be seen in figure~\ref{fig:scaling}f were both efficiencies plummet with more than 2 workers. 
The reported scaling deficiency 
is of little significance for actual molecular systems where the creation of a full path takes minutes to hours rather
than subseconds.

\section*{Conclusions}
%In conclusion 
We developed a 
new generic replica exchange method that is able to  effectively deal with MC moves with varying CPU costs, for instance due to the algorithmic complexity of the MC moves.
An essential aspect 
of the method is that 
the number of workers, who 
execute the ensemble's specific MC moves in parallel, is less than the 
number of ensembles. 
Once a worker is finished with its move, replica exchange moves are carried out solely between those ensembles that are not occupied by a worker.
This implies that the ensembles are updated at irregular intervals and a different number of MC moves will  be executed
for each ensemble.
As a result, the conceptual viewpoint 
in which the set of replica's are viewed 
as a single superstate is no longer valid and
the existence of some kind of detailed-balance relation is no longer trivial.
To prove the exactness of our approach, 
we introduced some new conceptual 
views on the replica exchange methodology
that is different from the common superstate
principle.
Instead, we show that 
the distributions
in the new approach
are conserved for
 each ensemble individually via a 
twisted detailed balance relation in which the
other ensembles  constitute an environment that is potentially actively involved in the MC move 
of the ensemble considered.
In addition,
the method can be combined with an
infinite swapping 
approach without the factorial scaling  
based on a mathematical reformulation 
using permanents. 

We applied 
the novel replica exchange 
technique
on a 
path sampling
algorithm, RETIS, 
which is a prototype of algorithm
where the costs for a Monte Carlo move can vary enormously. 
The resulting new path sampling algorithm, coined $\infty$RETIS, was thereafter 
tested on three model systems.
The results of these simulations show that
the number of MD steps increase linearly with the number of workers invoked as long as the ensemble's MC 
move has a lower computational cost than the replica exchange move carried out by the scheduler. The number 
of executed 
MC moves shows an even better than linear scaling. 
Moreover, 
the efficiency increases linearly 
with the number of workers
for a low-dimensional 
system
in which the replica exchange has little effect, while it has an optimum 
in more complex systems
as the number of successful 
replica exchange moves decreases when the number 
of workers is close to the number of ensembles.

In summary, the replica exchange method discussed in this paper has
a clear potential to accelerate 
present path sampling simulations,
but
can also be combined with many other 
complex algorithms including those that are yet to be invented.
With continuing trend to
to run 
progressively more massively-parallel computing
jobs, our algorithm is likely
to gain  importance 
and will open up new avenues
in the field of molecular simulations and beyond.

\subsection*{Supporting Information Appendix (SI)}
This article contains Supplementary Information
 %\href{https://www.pnas.org/}{SI-infRETIS}. 

\section*{Materials and Methods}
\subsection*{Simulation methods}  
The implementation
of $\infty$RETIS was structured as follows.
We start $1 \leq K \leq N$ 
worker- and 1 scheduler-process. 
Each of the worker-processes is going to process  
ensemble specific MC moves while the scheduler-process will do all the replica exchange moves and submits new jobs to the workers. All ensembles/trajectories that are currently being updated by a worker-process are not considered for MC moves by the scheduler, essentially being 'locked'. This means that no data is written for those ensembles and they are not valid targets for swapping moves. After a worker is done, it submits the result to the scheduler, the scheduler then unlocks the returned ensemble/trajectory and 
executes the replica exchange moves
on all ensembles/trajectories that 
that are not locked.
It then submits a new job to the freed worker for performing 
a new MC move in a
randomly chosen free ensemble
(or two ensembles in case of a point exchange move)
and locks the involved ensembles/trajectories.

In the $\infty$RETIS method there are two kind of ensemble moves 
that involve MD steps.
The first one is the shooting move (either standard shooting~\cite{shoot} or
the more recent sub-trajectory moves~\cite{riccardi2017fast, wf}) 
in which
 a new path is being generated from an old path within a single ensemble.
The second one is the point exchange move between $[0^-]$ and $[0^+]$. 
If a worker is assigned to this task, it means 
that both $[0^-]$ and $[0^+]$ are occupied by this worker. 
The scheduler ensures that there is never more than 1 worker considered free
at a given time.
When the free worker is assigned to perform a new  MC move,
each of  the ensembles have an equal probability to be selected.
 If $[0^+]$ or $[0^-]$ is selected and the other is also free, there is a 50\% chance to perform a 
 $[0^-] \leftrightarrow [0^+]$ 
 point exchange move 
 instead of a shooting move in the selected ensemble.

\subsection*{Memoryless single variable stochastic  (MSVS) process}
No actual MD is run for the 
MSVS
simulations. Instead, we directly sample two random numbers, $r_1$
and $r_2$ from an uniform distribution $\in [0,1)$
to set the
path's progress and the path length.
A path in ensemble $[k^+]$
is assumed to cross 
interface
$\lambda_{k+l}$ 
if $r_1 < (0.1)^l$. After this, we 
wait a random time, $t = 0.2 \, r_2\, k+0.1$ in seconds.
This was done to simulate both the increasing average simulation time and variance for outer ensembles. This setup means that we have no history dependence and allows us to compute the theoretical values show in figure~\ref{fig:scaling}. 5 independent $\infty$RETIS simulations were run with $1,5,10,15,\ldots,45,50$ workers.

\subsection*{Double channel simulations}
In order to investigate the effect of our algorithm on the ergodicity of the sampling, a 2D 
two-channel simulation was run
as described in reference~\citep{permeability}. The new RETIS moves introduced in that paper (mirror-move and target-swap move) were not used. Instead, MD was only run to do shooting moves or the $[0^-] \leftrightarrow [0^+]$ point exchanges. As the MD for this system completed too fast, every worker was set to wait $9$ times the time it took to run the MD before returning the result. 5 independent $\infty$RETIS simulations were run with $1, 2, \ldots, 11, 12$ workers.

\subsection*{1D double well with wire fencing}
In order to investigate the accuracy with a $W$ matrix that contains more numbers than $0$s or $1$s we simulated a 1D double-well system~\cite{TitusRev} together with the high-acceptance version of a novel path-sampling algorithm, wire fencing. The algorithm is described in reference~\citep{wf}, but for us the relevant part is that the high-acceptance weight is the number of frames that a path has outside the interface for each ensemble times an extra factor 2 if the path ends at the last interface. As for the 
two-channel system, a worker was set to wait 9 times the time it took to complete the MD move before returning the result. 
5 independent $\infty$RETIS simulations were run with $1, 2, \ldots, 7, 8$ workers with interfaces placed at $ [-0.99, -0.8, -0.7, -0.6, -0.5, -0.4, -0.3, 1.0]$.

\section*{ACKNOWLEDGMENTS}
        We acknowledge funding from the Research Council of Norway through FRINATEK Project No. 275506.

\subsubsection*{The authors declare no conflict of interest}

\bibliography{InfinityRETIS}

\begin{thebibliography}{31}%
\makeatletter
\providecommand \@ifxundefined [1]{%
 \@ifx{#1\undefined}
}%
\providecommand \@ifnum [1]{%
 \ifnum #1\expandafter \@firstoftwo
 \else \expandafter \@secondoftwo
 \fi
}%
\providecommand \@ifx [1]{%
 \ifx #1\expandafter \@firstoftwo
 \else \expandafter \@secondoftwo
 \fi
}%
\providecommand \natexlab [1]{#1}%
\providecommand \enquote  [1]{``#1''}%
\providecommand \bibnamefont  [1]{#1}%
\providecommand \bibfnamefont [1]{#1}%
\providecommand \citenamefont [1]{#1}%
\providecommand \href@noop [0]{\@secondoftwo}%
\providecommand \href [0]{\begingroup \@sanitize@url \@href}%
\providecommand \@href[1]{\@@startlink{#1}\@@href}%
\providecommand \@@href[1]{\endgroup#1\@@endlink}%
\providecommand \@sanitize@url [0]{\catcode `\\12\catcode `\$12\catcode
  `\&12\catcode `\#12\catcode `\^12\catcode `\_12\catcode `\%12\relax}%
\providecommand \@@startlink[1]{}%
\providecommand \@@endlink[0]{}%
\providecommand \url  [0]{\begingroup\@sanitize@url \@url }%
\providecommand \@url [1]{\endgroup\@href {#1}{\urlprefix }}%
\providecommand \urlprefix  [0]{URL }%
\providecommand \Eprint [0]{\href }%
\providecommand \doibase [0]{https://doi.org/}%
\providecommand \selectlanguage [0]{\@gobble}%
\providecommand \bibinfo  [0]{\@secondoftwo}%
\providecommand \bibfield  [0]{\@secondoftwo}%
\providecommand \translation [1]{[#1]}%
\providecommand \BibitemOpen [0]{}%
\providecommand \bibitemStop [0]{}%
\providecommand \bibitemNoStop [0]{.\EOS\space}%
\providecommand \EOS [0]{\spacefactor3000\relax}%
\providecommand \BibitemShut  [1]{\csname bibitem#1\endcsname}%
\let\auto@bib@innerbib\@empty
%</preamble>
\bibitem [{\citenamefont {Metropolis}\ \emph {et~al.}(1953)\citenamefont
  {Metropolis}, \citenamefont {Rosenbluth}, \citenamefont {Rosenbluth},
  \citenamefont {Teller},\ and\ \citenamefont {Teller}}]{Metroplois}%
  \BibitemOpen
  \bibfield  {author} {\bibinfo {author} {\bibfnamefont {N.}~\bibnamefont
  {Metropolis}}, \bibinfo {author} {\bibfnamefont {A.}~\bibnamefont
  {Rosenbluth}}, \bibinfo {author} {\bibfnamefont {M.}~\bibnamefont
  {Rosenbluth}}, \bibinfo {author} {\bibfnamefont {A.}~\bibnamefont {Teller}},\
  and\ \bibinfo {author} {\bibfnamefont {E.}~\bibnamefont {Teller}},\
  }\bibfield  {title} {\bibinfo {title} {Equation of state calculations by fast
  computing machines},\ }\href@noop {} {\bibfield  {journal} {\bibinfo
  {journal} {J. Chem. Phys.}\ }\textbf {\bibinfo {volume} {21}},\ \bibinfo
  {pages} {1087} (\bibinfo {year} {1953})}\BibitemShut {NoStop}%
\bibitem [{\citenamefont {Hastings}(1970)}]{Hastings}%
  \BibitemOpen
  \bibfield  {author} {\bibinfo {author} {\bibfnamefont {W.}~\bibnamefont
  {Hastings}},\ }\bibfield  {title} {\bibinfo {title} {{M}onte-{C}arlo sampling
  methods using {M}arkov chains and their applications},\ }\href
  {https://doi.org/10.2307/2334940} {\bibfield  {journal} {\bibinfo  {journal}
  {Biometrika}\ }\textbf {\bibinfo {volume} {57}},\ \bibinfo {pages} {97}
  (\bibinfo {year} {1970})}\BibitemShut {NoStop}%
\bibitem [{\citenamefont {Swendsen}\ and\ \citenamefont {Wang}(1986)}]{RE2}%
  \BibitemOpen
  \bibfield  {author} {\bibinfo {author} {\bibfnamefont {R.~H.}\ \bibnamefont
  {Swendsen}}\ and\ \bibinfo {author} {\bibfnamefont {J.~S.}\ \bibnamefont
  {Wang}},\ }\bibfield  {title} {\bibinfo {title} {Replica monte-carlo
  simulation of spin-glasses},\ }\href@noop {} {\bibfield  {journal} {\bibinfo
  {journal} {Phys. Rev. Lett.}\ }\textbf {\bibinfo {volume} {57}},\ \bibinfo
  {pages} {2607} (\bibinfo {year} {1986})}\BibitemShut {NoStop}%
\bibitem [{\citenamefont {Marinari}\ and\ \citenamefont
  {Parisi}(1992)}]{marinari92}%
  \BibitemOpen
  \bibfield  {author} {\bibinfo {author} {\bibfnamefont {E.}~\bibnamefont
  {Marinari}}\ and\ \bibinfo {author} {\bibfnamefont {G.}~\bibnamefont
  {Parisi}},\ }\bibfield  {title} {\bibinfo {title} {Simulated tempering - a
  new monte-carlo scheme},\ }\href@noop {} {\bibfield  {journal} {\bibinfo
  {journal} {Europhysics Lett.}\ }\textbf {\bibinfo {volume} {19}},\ \bibinfo
  {pages} {451} (\bibinfo {year} {1992})}\BibitemShut {NoStop}%
\bibitem [{\citenamefont {Sugita}\ and\ \citenamefont
  {Okamoto}(1999)}]{Sugita1999REMD}%
  \BibitemOpen
  \bibfield  {author} {\bibinfo {author} {\bibfnamefont {Y.}~\bibnamefont
  {Sugita}}\ and\ \bibinfo {author} {\bibfnamefont {Y.}~\bibnamefont
  {Okamoto}},\ }\bibfield  {title} {\bibinfo {title} {Replica-exchange
  molecular dynamics method for protein folding},\ }\href@noop {} {\bibfield
  {journal} {\bibinfo  {journal} {Chem. Phys. Lett.}\ }\textbf {\bibinfo
  {volume} {314}},\ \bibinfo {pages} {141 } (\bibinfo {year}
  {1999})}\BibitemShut {NoStop}%
\bibitem [{\citenamefont {Siepmann}\ and\ \citenamefont
  {Frenkel}(1992)}]{Siepmann92}%
  \BibitemOpen
  \bibfield  {author} {\bibinfo {author} {\bibfnamefont {J.~I.}\ \bibnamefont
  {Siepmann}}\ and\ \bibinfo {author} {\bibfnamefont {D.}~\bibnamefont
  {Frenkel}},\ }\bibfield  {title} {\bibinfo {title} {Configurational bias
  {M}onte-{C}arlo - a new sampling scheme for flexible chains},\ }\href@noop {}
  {\bibfield  {journal} {\bibinfo  {journal} {Molecular Physics}\ }\textbf
  {\bibinfo {volume} {75}},\ \bibinfo {pages} {59} (\bibinfo {year}
  {1992})}\BibitemShut {NoStop}%
\bibitem [{\citenamefont {Vlugt}\ \emph {et~al.}(1999)\citenamefont {Vlugt},
  \citenamefont {Krishna},\ and\ \citenamefont {Smit}}]{Vlugt99}%
  \BibitemOpen
  \bibfield  {author} {\bibinfo {author} {\bibfnamefont {T.}~\bibnamefont
  {Vlugt}}, \bibinfo {author} {\bibfnamefont {R.}~\bibnamefont {Krishna}},\
  and\ \bibinfo {author} {\bibfnamefont {B.}~\bibnamefont {Smit}},\ }\bibfield
  {title} {\bibinfo {title} {Molecular simulations of adsorption isotherms for
  linear and branched alkanes and their mixtures in silicalite},\ }\href@noop
  {} {\bibfield  {journal} {\bibinfo  {journal} {J. Phys. Chem. B}\ }\textbf
  {\bibinfo {volume} {103}},\ \bibinfo {pages} {1102} (\bibinfo {year}
  {1999})}\BibitemShut {NoStop}%
\bibitem [{\citenamefont {Frenkel}\ and\ \citenamefont
  {Smit}(2002)}]{FrenkelBook}%
  \BibitemOpen
  \bibfield  {author} {\bibinfo {author} {\bibfnamefont {D.}~\bibnamefont
  {Frenkel}}\ and\ \bibinfo {author} {\bibfnamefont {B.}~\bibnamefont {Smit}},\
  }\href@noop {} {\emph {\bibinfo {title} {Understanding molecular simulations
  from algorithms to applications}}}\ (\bibinfo  {publisher} {Academic press},\
  \bibinfo {address} {San Diego, California, U.S.A.},\ \bibinfo {year}
  {2002})\BibitemShut {NoStop}%
\bibitem [{\citenamefont {Dellago}\ \emph
  {et~al.}(1998{\natexlab{a}})\citenamefont {Dellago}, \citenamefont {Bolhuis},
  \citenamefont {Csajka},\ and\ \citenamefont {Chandler}}]{TPS98}%
  \BibitemOpen
  \bibfield  {author} {\bibinfo {author} {\bibfnamefont {C.}~\bibnamefont
  {Dellago}}, \bibinfo {author} {\bibfnamefont {P.~G.}\ \bibnamefont
  {Bolhuis}}, \bibinfo {author} {\bibfnamefont {F.~S.}\ \bibnamefont
  {Csajka}},\ and\ \bibinfo {author} {\bibfnamefont {D.}~\bibnamefont
  {Chandler}},\ }\bibfield  {title} {\bibinfo {title} {Transition path sampling
  and the calculation of rate constants},\ }\href@noop {} {\bibfield  {journal}
  {\bibinfo  {journal} {J. Chem. Phys.}\ }\textbf {\bibinfo {volume} {108}},\
  \bibinfo {pages} {1964} (\bibinfo {year} {1998}{\natexlab{a}})}\BibitemShut
  {NoStop}%
\bibitem [{\citenamefont {Swendsen}\ and\ \citenamefont
  {Wang}(1987)}]{SwendsenWang}%
  \BibitemOpen
  \bibfield  {author} {\bibinfo {author} {\bibfnamefont {R.~H.}\ \bibnamefont
  {Swendsen}}\ and\ \bibinfo {author} {\bibfnamefont {J.-S.}\ \bibnamefont
  {Wang}},\ }\bibfield  {title} {\bibinfo {title} {Nonuniversal critical
  dynamics in monte carlo simulations},\ }\href
  {https://doi.org/10.1103/PhysRevLett.58.86} {\bibfield  {journal} {\bibinfo
  {journal} {Phys. Rev. Lett.}\ }\textbf {\bibinfo {volume} {58}},\ \bibinfo
  {pages} {86} (\bibinfo {year} {1987})}\BibitemShut {NoStop}%
\bibitem [{\citenamefont {Peters}\ and\ \citenamefont
  {de~With}(2012)}]{Peters2012}%
  \BibitemOpen
  \bibfield  {author} {\bibinfo {author} {\bibfnamefont {E.~A. J.~F.}\
  \bibnamefont {Peters}}\ and\ \bibinfo {author} {\bibfnamefont
  {G.}~\bibnamefont {de~With}},\ }\bibfield  {title} {\bibinfo {title}
  {Rejection-free monte carlo sampling for general potentials},\ }\href
  {https://doi.org/10.1103/PhysRevE.85.026703} {\bibfield  {journal} {\bibinfo
  {journal} {Phys. Rev. E}\ }\textbf {\bibinfo {volume} {85}},\ \bibinfo
  {pages} {026703} (\bibinfo {year} {2012})}\BibitemShut {NoStop}%
\bibitem [{\citenamefont {Michel}\ \emph {et~al.}(2014)\citenamefont {Michel},
  \citenamefont {Kapfer},\ and\ \citenamefont {Krauth}}]{Michel2014}%
  \BibitemOpen
  \bibfield  {author} {\bibinfo {author} {\bibfnamefont {M.}~\bibnamefont
  {Michel}}, \bibinfo {author} {\bibfnamefont {S.~C.}\ \bibnamefont {Kapfer}},\
  and\ \bibinfo {author} {\bibfnamefont {W.}~\bibnamefont {Krauth}},\
  }\bibfield  {title} {\bibinfo {title} {Generalized event-chain {M}onte
  {C}arlo: Constructing rejection-free global-balance algorithms from
  infinitesimal steps},\ }\href {https://doi.org/10.1063/1.4863991} {\bibfield
  {journal} {\bibinfo  {journal} {J. Chem. Phys.}\ }\textbf {\bibinfo {volume}
  {140}},\ \bibinfo {pages} {054116} (\bibinfo {year} {2014})}\BibitemShut
  {NoStop}%
\bibitem [{\citenamefont {Plattner}\ \emph {et~al.}(2011)\citenamefont
  {Plattner}, \citenamefont {Doll}, \citenamefont {Dupuis}, \citenamefont
  {Wang}, \citenamefont {Liu},\ and\ \citenamefont
  {Gubernatis}}]{plattner2011}%
  \BibitemOpen
  \bibfield  {author} {\bibinfo {author} {\bibfnamefont {N.}~\bibnamefont
  {Plattner}}, \bibinfo {author} {\bibfnamefont {J.~D.}\ \bibnamefont {Doll}},
  \bibinfo {author} {\bibfnamefont {P.}~\bibnamefont {Dupuis}}, \bibinfo
  {author} {\bibfnamefont {H.}~\bibnamefont {Wang}}, \bibinfo {author}
  {\bibfnamefont {Y.}~\bibnamefont {Liu}},\ and\ \bibinfo {author}
  {\bibfnamefont {J.~E.}\ \bibnamefont {Gubernatis}},\ }\bibfield  {title}
  {\bibinfo {title} {An infinite swapping approach to the rare-event sampling
  problem},\ }\href {https://doi.org/10.1063/1.3643325} {\bibfield  {journal}
  {\bibinfo  {journal} {The Journal of Chemical Physics}\ }\textbf {\bibinfo
  {volume} {135}},\ \bibinfo {pages} {134111} (\bibinfo {year} {2011})},\
  \Eprint {https://arxiv.org/abs/https://doi.org/10.1063/1.3643325}
  {https://doi.org/10.1063/1.3643325} \BibitemShut {NoStop}%
\bibitem [{\citenamefont {Plattner}\ \emph {et~al.}(2013)\citenamefont
  {Plattner}, \citenamefont {Doll},\ and\ \citenamefont
  {Meuwly}}]{infswap2013}%
  \BibitemOpen
  \bibfield  {author} {\bibinfo {author} {\bibfnamefont {N.}~\bibnamefont
  {Plattner}}, \bibinfo {author} {\bibfnamefont {J.~D.}\ \bibnamefont {Doll}},\
  and\ \bibinfo {author} {\bibfnamefont {M.}~\bibnamefont {Meuwly}},\
  }\bibfield  {title} {\bibinfo {title} {Overcoming the rare event sampling
  problem in biological systems with infinite swapping},\ }\href@noop {}
  {\bibfield  {journal} {\bibinfo  {journal} {J. Chem. Theory Comput.}\
  }\textbf {\bibinfo {volume} {9}},\ \bibinfo {pages} {4215} (\bibinfo {year}
  {2013})}\BibitemShut {NoStop}%
\bibitem [{\citenamefont {Lu}\ and\ \citenamefont
  {Vanden-Eijnden}(2019)}]{Lu_2019}%
  \BibitemOpen
  \bibfield  {author} {\bibinfo {author} {\bibfnamefont {J.}~\bibnamefont
  {Lu}}\ and\ \bibinfo {author} {\bibfnamefont {E.}~\bibnamefont
  {Vanden-Eijnden}},\ }\bibfield  {title} {\bibinfo {title} {Methodological and
  computational aspects of parallel tempering methods in the infinite swapping
  limit},\ }\href {https://doi.org/10.1007/s10955-018-2210-y} {\bibfield
  {journal} {\bibinfo  {journal} {J Stat Phys}\ }\textbf {\bibinfo {volume}
  {174}},\ \bibinfo {pages} {715} (\bibinfo {year} {2019})}\BibitemShut
  {NoStop}%
\bibitem [{\citenamefont {Balasubramanian}(1980)}]{balasubramanian_1984}%
  \BibitemOpen
  \bibfield  {author} {\bibinfo {author} {\bibfnamefont {K.}~\bibnamefont
  {Balasubramanian}},\ }\emph {\bibinfo {title} {Combinatorics and diagonals of
  matrices}},\ \href@noop {} {Ph.D. thesis},\ \bibinfo  {school} {Loyola
  College}, \bibinfo {address} {Madras, India} (\bibinfo {year}
  {1980})\BibitemShut {NoStop}%
\bibitem [{\citenamefont {Bax}(1998)}]{bax_1998}%
  \BibitemOpen
  \bibfield  {author} {\bibinfo {author} {\bibfnamefont {E.}~\bibnamefont
  {Bax}},\ }\emph {\bibinfo {title} {Finite-difference Algorithms for Counting
  Problems}},\ \href@noop {} {Ph.D. thesis},\ \bibinfo  {school} {California
  Institute of Technology}, \bibinfo {address} {Pasadena, United States of
  America} (\bibinfo {year} {1998})\BibitemShut {NoStop}%
\bibitem [{\citenamefont {Glynn}(2010)}]{Glynn_2010}%
  \BibitemOpen
  \bibfield  {author} {\bibinfo {author} {\bibfnamefont {D.~G.}\ \bibnamefont
  {Glynn}},\ }\bibfield  {title} {\bibinfo {title} {The permanent of a square
  matrix},\ }\href {https://doi.org/10.1016/j.ejc.2010.01.010} {\bibfield
  {journal} {\bibinfo  {journal} {European Journal of Combinatorics}\ }\textbf
  {\bibinfo {volume} {31}},\ \bibinfo {pages} {1887} (\bibinfo {year}
  {2010})}\BibitemShut {NoStop}%
\bibitem [{\citenamefont {{van Erp}}(2007)}]{RETIS}%
  \BibitemOpen
  \bibfield  {author} {\bibinfo {author} {\bibfnamefont {T.}~\bibnamefont {{van
  Erp}}},\ }\bibfield  {title} {\bibinfo {title} {Reaction rate calculation by
  parallel path swapping},\ }\href@noop {} {\bibfield  {journal} {\bibinfo
  {journal} {Phys. Rev. Lett.}\ }\textbf {\bibinfo {volume} {98}},\ \bibinfo
  {pages} {268301} (\bibinfo {year} {2007})}\BibitemShut {NoStop}%
\bibitem [{\citenamefont {Cabriolu}\ \emph {et~al.}(2017)\citenamefont
  {Cabriolu}, \citenamefont {Refsnes}, \citenamefont {Bolhuis},\ and\
  \citenamefont {van Erp}}]{Raffa}%
  \BibitemOpen
  \bibfield  {author} {\bibinfo {author} {\bibfnamefont {R.}~\bibnamefont
  {Cabriolu}}, \bibinfo {author} {\bibfnamefont {K.~M.~S.}\ \bibnamefont
  {Refsnes}}, \bibinfo {author} {\bibfnamefont {P.~G.}\ \bibnamefont
  {Bolhuis}},\ and\ \bibinfo {author} {\bibfnamefont {T.~S.}\ \bibnamefont {van
  Erp}},\ }\bibfield  {title} {\bibinfo {title} {{Foundations and latest
  advances in replica exchange transition interface sampling}},\ }\href
  {https://doi.org/{10.1063/1.4989844}} {\bibfield  {journal} {\bibinfo
  {journal} {J. Chem. Phys.}\ }\textbf {\bibinfo {volume} {{147}}},\ \bibinfo
  {pages} {152722} (\bibinfo {year} {{2017}})}\BibitemShut {NoStop}%
\bibitem [{\citenamefont {Dellago}\ \emph
  {et~al.}(1998{\natexlab{b}})\citenamefont {Dellago}, \citenamefont
  {Bolhuis},\ and\ \citenamefont {Chandler}}]{shoot}%
  \BibitemOpen
  \bibfield  {author} {\bibinfo {author} {\bibfnamefont {C.}~\bibnamefont
  {Dellago}}, \bibinfo {author} {\bibfnamefont {P.~G.}\ \bibnamefont
  {Bolhuis}},\ and\ \bibinfo {author} {\bibfnamefont {D.}~\bibnamefont
  {Chandler}},\ }\bibfield  {title} {\bibinfo {title} {Efficient transition
  path sampling: Application to lennard-jones cluster rearrangements},\ }\href
  {https://doi.org/10.1063/1.476378} {\bibfield  {journal} {\bibinfo  {journal}
  {The Journal of Chemical Physics}\ }\textbf {\bibinfo {volume} {108}},\
  \bibinfo {pages} {9236} (\bibinfo {year} {1998}{\natexlab{b}})},\ \Eprint
  {https://arxiv.org/abs/https://doi.org/10.1063/1.476378}
  {https://doi.org/10.1063/1.476378} \BibitemShut {NoStop}%
\bibitem [{\citenamefont {van Erp}\ \emph {et~al.}(2003)\citenamefont {van
  Erp}, \citenamefont {Moroni},\ and\ \citenamefont {Bolhuis}}]{TIS}%
  \BibitemOpen
  \bibfield  {author} {\bibinfo {author} {\bibfnamefont {T.~S.}\ \bibnamefont
  {van Erp}}, \bibinfo {author} {\bibfnamefont {D.}~\bibnamefont {Moroni}},\
  and\ \bibinfo {author} {\bibfnamefont {P.~G.}\ \bibnamefont {Bolhuis}},\
  }\bibfield  {title} {\bibinfo {title} {A novel path sampling method for the
  sampling of rate constants},\ }\href@noop {} {\bibfield  {journal} {\bibinfo
  {journal} {J. Chem. Phys.}\ }\textbf {\bibinfo {volume} {118}},\ \bibinfo
  {pages} {7762} (\bibinfo {year} {2003})}\BibitemShut {NoStop}%
\bibitem [{\citenamefont {Swenson}\ \emph {et~al.}(2019)\citenamefont
  {Swenson}, \citenamefont {Prinz}, \citenamefont {Noe}, \citenamefont
  {Chodera},\ and\ \citenamefont {Bolhuis}}]{OPS2}%
  \BibitemOpen
  \bibfield  {author} {\bibinfo {author} {\bibfnamefont {D.~W.~H.}\
  \bibnamefont {Swenson}}, \bibinfo {author} {\bibfnamefont {J.-H.}\
  \bibnamefont {Prinz}}, \bibinfo {author} {\bibfnamefont {F.}~\bibnamefont
  {Noe}}, \bibinfo {author} {\bibfnamefont {J.~D.}\ \bibnamefont {Chodera}},\
  and\ \bibinfo {author} {\bibfnamefont {P.~G.}\ \bibnamefont {Bolhuis}},\
  }\bibfield  {title} {\bibinfo {title} {Openpathsampling: A python framework
  for path sampling simulations. 2. building and customizing path ensembles and
  sample schemes},\ }\href@noop {} {\bibfield  {journal} {\bibinfo  {journal}
  {J. Chem. Theory Comput.}\ }\textbf {\bibinfo {volume} {15}},\ \bibinfo
  {pages} {837} (\bibinfo {year} {2019})}\BibitemShut {NoStop}%
\bibitem [{\citenamefont {Riccardi}\ \emph {et~al.}(2020)\citenamefont
  {Riccardi}, \citenamefont {Lervik}, \citenamefont {Roet}, \citenamefont
  {Aaroen},\ and\ \citenamefont {van Erp}}]{PyRETIS2}%
  \BibitemOpen
  \bibfield  {author} {\bibinfo {author} {\bibfnamefont {E.}~\bibnamefont
  {Riccardi}}, \bibinfo {author} {\bibfnamefont {A.}~\bibnamefont {Lervik}},
  \bibinfo {author} {\bibfnamefont {S.}~\bibnamefont {Roet}}, \bibinfo {author}
  {\bibfnamefont {O.}~\bibnamefont {Aaroen}},\ and\ \bibinfo {author}
  {\bibfnamefont {T.~S.}\ \bibnamefont {van Erp}},\ }\bibfield  {title}
  {\bibinfo {title} {Pyretis 2: An improbability drive for rare events},\
  }\href@noop {} {\bibfield  {journal} {\bibinfo  {journal} {J. Comput. Chem.}\
  }\textbf {\bibinfo {volume} {41}},\ \bibinfo {pages} {370} (\bibinfo {year}
  {2020})}\BibitemShut {NoStop}%
\bibitem [{\citenamefont {Riccardi}\ \emph {et~al.}(2017)\citenamefont
  {Riccardi}, \citenamefont {Dahlen},\ and\ \citenamefont {van
  Erp}}]{riccardi2017fast}%
  \BibitemOpen
  \bibfield  {author} {\bibinfo {author} {\bibfnamefont {E.}~\bibnamefont
  {Riccardi}}, \bibinfo {author} {\bibfnamefont {O.}~\bibnamefont {Dahlen}},\
  and\ \bibinfo {author} {\bibfnamefont {T.~S.}\ \bibnamefont {van Erp}},\
  }\bibfield  {title} {\bibinfo {title} {Fast decorrelating monte carlo moves
  for efficient path sampling},\ }\href@noop {} {\bibfield  {journal} {\bibinfo
   {journal} {J. Phys. Chem. Lett.}\ }\textbf {\bibinfo {volume} {8}},\
  \bibinfo {pages} {4456} (\bibinfo {year} {2017})}\BibitemShut {NoStop}%
\bibitem [{\citenamefont {Zhang}\ \emph {et~al.}(2021)\citenamefont {Zhang},
  \citenamefont {Riccardi},\ and\ \citenamefont {van Erp}}]{wf}%
  \BibitemOpen
  \bibfield  {author} {\bibinfo {author} {\bibfnamefont {D.~T.}\ \bibnamefont
  {Zhang}}, \bibinfo {author} {\bibfnamefont {E.}~\bibnamefont {Riccardi}},\
  and\ \bibinfo {author} {\bibfnamefont {T.~S.}\ \bibnamefont {van Erp}},\
  }\bibfield  {title} {\bibinfo {title} {Path sampling with sub-trajectory
  moves},\ }\href@noop {} {\bibfield  {journal} {\bibinfo  {journal} {In
  preparation}\ }\textbf {\bibinfo {volume} {xx}},\ \bibinfo {pages} {xx}
  (\bibinfo {year} {2021})}\BibitemShut {NoStop}%
\bibitem [{\citenamefont {Au}\ and\ \citenamefont {Beck}(2001)}]{subset}%
  \BibitemOpen
  \bibfield  {author} {\bibinfo {author} {\bibfnamefont {S.-K.}\ \bibnamefont
  {Au}}\ and\ \bibinfo {author} {\bibfnamefont {J.~L.}\ \bibnamefont {Beck}},\
  }\bibfield  {title} {\bibinfo {title} {Estimation of small failure
  probabilities in high dimensions by subset simulation},\ }\href
  {https://doi.org/https://doi.org/10.1016/S0266-8920(01)00019-4} {\bibfield
  {journal} {\bibinfo  {journal} {Probabilistic Eng. Mech.}\ }\textbf {\bibinfo
  {volume} {16}},\ \bibinfo {pages} {263} (\bibinfo {year} {2001})}\BibitemShut
  {NoStop}%
\bibitem [{\citenamefont {Torrie}\ and\ \citenamefont {Valleau}(1977)}]{US}%
  \BibitemOpen
  \bibfield  {author} {\bibinfo {author} {\bibfnamefont {G.}~\bibnamefont
  {Torrie}}\ and\ \bibinfo {author} {\bibfnamefont {J.}~\bibnamefont
  {Valleau}},\ }\bibfield  {title} {\bibinfo {title} {Nonphysical sampling
  distributions in {M}onte {C}arlo free-energy estimation: Umbrella samping},\
  }\href@noop {} {\bibfield  {journal} {\bibinfo  {journal} {J. Comp. Phys.}\
  }\textbf {\bibinfo {volume} {23}},\ \bibinfo {pages} {187} (\bibinfo {year}
  {1977})}\BibitemShut {NoStop}%
\bibitem [{\citenamefont {Ghysels}\ \emph {et~al.}(2021)\citenamefont
  {Ghysels}, \citenamefont {Roet}, \citenamefont {Davoudi},\ and\ \citenamefont
  {van Erp}}]{permeability}%
  \BibitemOpen
  \bibfield  {author} {\bibinfo {author} {\bibfnamefont {A.}~\bibnamefont
  {Ghysels}}, \bibinfo {author} {\bibfnamefont {S.}~\bibnamefont {Roet}},
  \bibinfo {author} {\bibfnamefont {S.}~\bibnamefont {Davoudi}},\ and\ \bibinfo
  {author} {\bibfnamefont {T.~S.}\ \bibnamefont {van Erp}},\ }\bibfield
  {title} {\bibinfo {title} {Exact non-markovian permeability from rare event
  simulations},\ }\href {https://doi.org/10.1103/PhysRevResearch.3.033068}
  {\bibfield  {journal} {\bibinfo  {journal} {Phys. Rev. Research}\ }\textbf
  {\bibinfo {volume} {3}},\ \bibinfo {pages} {033068} (\bibinfo {year}
  {2021})}\BibitemShut {NoStop}%
\bibitem [{\citenamefont {{van Erp}}(2012)}]{TitusRev}%
  \BibitemOpen
  \bibfield  {author} {\bibinfo {author} {\bibfnamefont {T.}~\bibnamefont {{van
  Erp}}},\ }\bibfield  {title} {\bibinfo {title} {Dynamical rare event
  simulation techniques for equilibrium and nonequilibrium systems},\
  }\href@noop {} {\bibfield  {journal} {\bibinfo  {journal} {Adv. Chem. Phys.}\
  }\textbf {\bibinfo {volume} {151}},\ \bibinfo {pages} {27} (\bibinfo {year}
  {2012})}\BibitemShut {NoStop}%
\bibitem [{\citenamefont {van Erp}(2006)}]{TISeff}%
  \BibitemOpen
  \bibfield  {author} {\bibinfo {author} {\bibfnamefont {T.~S.}\ \bibnamefont
  {van Erp}},\ }\bibfield  {title} {\bibinfo {title} {Efficiency analysis of
  reaction rate calculation methods using analytical models {I:} {The}
  two-dimensional sharp barrier},\ }\href@noop {} {\bibfield  {journal}
  {\bibinfo  {journal} {J. Chem. Phys.}\ }\textbf {\bibinfo {volume} {125}},\
  \bibinfo {pages} {174106} (\bibinfo {year} {2006})}\BibitemShut {NoStop}%
\end{thebibliography}%


\begin{thebibliography}{5}%
\makeatletter
\providecommand \@ifxundefined [1]{%
 \@ifx{#1\undefined}
}%
\providecommand \@ifnum [1]{%
 \ifnum #1\expandafter \@firstoftwo
 \else \expandafter \@secondoftwo
 \fi
}%
\providecommand \@ifx [1]{%
 \ifx #1\expandafter \@firstoftwo
 \else \expandafter \@secondoftwo
 \fi
}%
\providecommand \natexlab [1]{#1}%
\providecommand \enquote  [1]{``#1''}%
\providecommand \bibnamefont  [1]{#1}%
\providecommand \bibfnamefont [1]{#1}%
\providecommand \citenamefont [1]{#1}%
\providecommand \href@noop [0]{\@secondoftwo}%
\providecommand \href [0]{\begingroup \@sanitize@url \@href}%
\providecommand \@href[1]{\@@startlink{#1}\@@href}%
\providecommand \@@href[1]{\endgroup#1\@@endlink}%
\providecommand \@sanitize@url [0]{\catcode `\\12\catcode `\$12\catcode
  `\&12\catcode `\#12\catcode `\^12\catcode `\_12\catcode `\%12\relax}%
\providecommand \@@startlink[1]{}%
\providecommand \@@endlink[0]{}%
\providecommand \url  [0]{\begingroup\@sanitize@url \@url }%
\providecommand \@url [1]{\endgroup\@href {#1}{\urlprefix }}%
\providecommand \urlprefix  [0]{URL }%
\providecommand \Eprint [0]{\href }%
\providecommand \doibase [0]{https://doi.org/}%
\providecommand \selectlanguage [0]{\@gobble}%
\providecommand \bibinfo  [0]{\@secondoftwo}%
\providecommand \bibfield  [0]{\@secondoftwo}%
\providecommand \translation [1]{[#1]}%
\providecommand \BibitemOpen [0]{}%
\providecommand \bibitemStop [0]{}%
\providecommand \bibitemNoStop [0]{.\EOS\space}%
\providecommand \EOS [0]{\spacefactor3000\relax}%
\providecommand \BibitemShut  [1]{\csname bibitem#1\endcsname}%
\let\auto@bib@innerbib\@empty
%</preamble>
\bibitem [{\citenamefont {Frenkel}\ and\ \citenamefont
  {Smit}(2002)}]{FrenkelBook}%
  \BibitemOpen
  \bibfield  {author} {\bibinfo {author} {\bibfnamefont {D.}~\bibnamefont
  {Frenkel}}\ and\ \bibinfo {author} {\bibfnamefont {B.}~\bibnamefont {Smit}},\
  }\href@noop {} {\emph {\bibinfo {title} {Understanding molecular simulations
  from algorithms to applications}}}\ (\bibinfo  {publisher} {Academic press},\
  \bibinfo {address} {San Diego, California, U.S.A.},\ \bibinfo {year}
  {2002})\BibitemShut {NoStop}%
\bibitem [{\citenamefont {Ghysels}\ \emph {et~al.}(2021)\citenamefont
  {Ghysels}, \citenamefont {Roet}, \citenamefont {Davoudi},\ and\ \citenamefont
  {van Erp}}]{permeability}%
  \BibitemOpen
  \bibfield  {author} {\bibinfo {author} {\bibfnamefont {A.}~\bibnamefont
  {Ghysels}}, \bibinfo {author} {\bibfnamefont {S.}~\bibnamefont {Roet}},
  \bibinfo {author} {\bibfnamefont {S.}~\bibnamefont {Davoudi}},\ and\ \bibinfo
  {author} {\bibfnamefont {T.~S.}\ \bibnamefont {van Erp}},\ }\bibfield
  {title} {\bibinfo {title} {Exact non-markovian permeability from rare event
  simulations},\ }\href {https://doi.org/10.1103/PhysRevResearch.3.033068}
  {\bibfield  {journal} {\bibinfo  {journal} {Phys. Rev. Research}\ }\textbf
  {\bibinfo {volume} {3}},\ \bibinfo {pages} {033068} (\bibinfo {year}
  {2021})}\BibitemShut {NoStop}%
\bibitem [{\citenamefont {{van Erp}}(2012)}]{TitusRev}%
  \BibitemOpen
  \bibfield  {author} {\bibinfo {author} {\bibfnamefont {T.}~\bibnamefont {{van
  Erp}}},\ }\bibfield  {title} {\bibinfo {title} {Dynamical rare event
  simulation techniques for equilibrium and nonequilibrium systems},\
  }\href@noop {} {\bibfield  {journal} {\bibinfo  {journal} {Adv. Chem. Phys.}\
  }\textbf {\bibinfo {volume} {151}},\ \bibinfo {pages} {27} (\bibinfo {year}
  {2012})}\BibitemShut {NoStop}%
\bibitem [{\citenamefont {Zhang}\ \emph {et~al.}(2021)\citenamefont {Zhang},
  \citenamefont {Riccardi},\ and\ \citenamefont {van Erp}}]{wf}%
  \BibitemOpen
  \bibfield  {author} {\bibinfo {author} {\bibfnamefont {D.~T.}\ \bibnamefont
  {Zhang}}, \bibinfo {author} {\bibfnamefont {E.}~\bibnamefont {Riccardi}},\
  and\ \bibinfo {author} {\bibfnamefont {T.~S.}\ \bibnamefont {van Erp}},\
  }\bibfield  {title} {\bibinfo {title} {Path sampling with sub-trajectory
  moves},\ }\href@noop {} {\bibfield  {journal} {\bibinfo  {journal} {In
  preparation}\ }\textbf {\bibinfo {volume} {xx}},\ \bibinfo {pages} {xx}
  (\bibinfo {year} {2021})}\BibitemShut {NoStop}%
\bibitem [{\citenamefont {van Erp}(2006)}]{TISeff}%
  \BibitemOpen
  \bibfield  {author} {\bibinfo {author} {\bibfnamefont {T.~S.}\ \bibnamefont
  {van Erp}},\ }\bibfield  {title} {\bibinfo {title} {Efficiency analysis of
  reaction rate calculation methods using analytical models {I:} {The}
  two-dimensional sharp barrier},\ }\href@noop {} {\bibfield  {journal}
  {\bibinfo  {journal} {J. Chem. Phys.}\ }\textbf {\bibinfo {volume} {125}},\
  \bibinfo {pages} {174106} (\bibinfo {year} {2006})}\BibitemShut {NoStop}%
\end{thebibliography}%

\end{document}

\title{Supporting Information for: \\ Exchanging replicas with unequal cost, infinitely and permanently}
\author{Sander Roet}
\author{Daniel T. Zhang}%
\author{Titus S. van Erp}
 \email{titus.van.erp@ntnu.no}
\affiliation{%
Department of Chemistry, Norwegian University of Science and Technology (NTNU),\\ N-7491 Trondheim, Norway
}

\maketitle

\section{Supporting Information Text}
This Supplementary Information contains the following data,  derivations, and numerical examples.
In Sec.~\ref{sec:detbal}, we provide a
proof that the replica exchange method with cost unbalanced replicas conserves the equilibrium distribution at the individual ensemble level.
Instead of the superstate principle, the derivation is based on the
individual ensemble's perspective where
the other ensembles serve as an environment,
which finally leads to a twisted detailed-balance relation.
In Sec.\ref{sec:oneszeros}, we show a ${\mathcal O}(n^2)$ algorithm 
for computing the $P$-matrix from a $W$-matrix
for the case that the $W$-matrix consists of
rows having a series with ones, followed by zeros. This is the type of matrix that
is relevant for RETIS simulations based on the standard shooting move.
Sec.~\ref{sec:Kramer} presents the derivations of the
theoretical results on the crossing probabilities, rate constant, and permeability via Kramer's theory that are shown
in table 1 of the main article.
In Sec.~\ref{sec:efficiencies} the computational efficiencies, including the derivations for the most optimal efficiencies, are discussed.
Finally, in Sec.~\ref{sec:add} we provide some additional
simulation results on the relative transition 
probabilities through the lower and higher barrier channel.

\section{Detailed-balance relations}
\label{sec:detbal}
In this section, we will derive
 detailed-balance relations 
for parallel replica's
that 
are not based on the common superstate viewpoint.
These alternative relations 
can be used to validate the
replica exchange algorithm for replica's with unequal 
CPU cost.
Our derivation is based on the finite swapping approach,
though the infinite swapping version follows automatically from this
when the probability to perform a swap goes
to unity ($P_{\rm RE}\rightarrow 1$) as explained in the main text.
To simplify matters, we assume
that we have one type of replica exchange 
move that is low in CPU cost and one type of ensemble move that operates within one ensemble and has a high CPU cost. 
The relations that we derive are, however, by no means limited to that. In fact, in the RETIS algorithm there is 
also a point exchange move between the $[0^-]$ and 
$[0^+]$ ensemble. In 
previous publications this move, annotated as $[0^-]\leftrightarrow [0^+]$, was 
categorized as a special type of swapping/replica 
exchange move. 
In this article we reserve the name swap or replica exchange
to an operation that involves the swapping of full paths, which does not require any MD steps.
In contrast, the $[0^-]\leftrightarrow [0^+]$ point exchange 
implies the exchange of time slices
at the end and start of the paths that are then extended at the other side of the $\lambda_0$ interface. In our implementation, this $[0^-]\leftrightarrow [0^+]$ move is carried out by a single worker that locks both the 
$[0^-]$ and $[0^+]$ ensembles during this move. As 
the $[0^-]$ paths can never be swapped with any of the other paths,  we can view the point exchange move as 
an ensemble move in ensemble $[0^+]$.

As explained in the main article, 
the replica exchange algorithm that we propose is 
based on a set of workers and a set of ensembles. The number of workers $K$ is less than the number of ensembles $N$.
Most of the time the worker is performing a CPU intensive single-ensemble move. The ensemble  in which the worker operates is considered occupied/locked. Once a worker has completed a CPU intensive move, the move will be accepted or rejected, after which either a replica exchange move will be carried out with any of the unoccupied ensembles
or the worker will be assigned to do a new single-ensemble move at a randomly picked free ensemble. 

In order to indicate the difference between occupied and unoccupied ensembles,
we introduce a new state vector that indicates 
both the available ensembles as in the main text and the occupied ensembles with a bar, e. g:
$S=(s_1, s_2, \overline{s_{3}}, s_4, \overline{s_{5}})$ to show
that there are 5 ensembles of which ensemble 3 and 5 are occupied by a worker.
For both occupied and unoccupied ensembles, the 
$s_i$-terms reflect the most recent state that was sampled in the $i$th ensemble. 
Now our sole aim is to ensure that if we just count the instances 
that an ensemble $i$ is updated with a new sample (which could be a copy of the previous sample in case of a rejected move), these should be distributed according the correct probability density $\rho_i$. 

It is important to note that 
the time between two updates can vary and depends on the state that
was most recently 
sampled. However,  the 
 waiting time between an update of a specific ensemble and the point 
in time
that this ensemble gets occupied by a worker  
will depend on the states of all other ensembles, but  \emph{not}  on the state
in the ensemble
considered. Since the ensembles are independent,
this waiting time will be the same on average irrespective to this sampled state.
This has as a consequence that if we  take "photographs" of the state vector, at intervals or randomly,  
evenly distributed over time, we should again obtain the correct distributions $\rho_i$, for all $i$, of the states in ensemble $i$ as long as we ignore the instances that this ensemble is occupied.    
In other words, we can write for the previous example state vector
\begin{align}
\label{eq:rho5}
\rho(S)=\rho(s_1, s_2, \overline{s_{3}}, s_4, \overline{s_{5}})=
\rho_1(s_1) \rho_2(s_2)  \rho_3^u(\overline{s_{3}})  \rho_4(s_4)
\rho_5^u(\overline{s_{5}}) 
\end{align}
where $\rho_i(\cdot)$ is the statistically correct distribution of ensemble $i$, and 
$\rho_j^u(\cdot)$ an unknown distribution for occupied ensemble $j$ that has no clear
physical interpretation. For instance, it can happen that a state $s$ is 
relatively 
unlikely to exist in ensemble $i$, low $\rho_i(s)$, but that any MC move starting from that state
takes a very long time, resulting in a high  
$\rho_i^u(s)$.

Now, let's consider the Markov chain from the perspective of ensemble 1
where we monitor its state at the point that a new MC is initiated from an old state $s_1$. From the viewpoint of ensemble 1, the other ensembles
are viewed as an "environment" (${\mathcal E}=(s_2, \overline{s_{3}}, s_4, \overline{s_{5}})$ in the aforementioned example), that might or might not influence the MC move. The probability of state $s_1$ in ensemble 1 can be written as an integral of the conditional probability given an environment: 
\begin{align}
\rho_1(s_1)=\int \rho_1(s_1|{\mathcal E}) \rho({\mathcal E}) {\mathrm d}{\mathcal E}.
\end{align}
As the ensembles are independent we can write
\begin{align}
\label{eq:s1independ}
\rho_1(s_1|{\mathcal E})=\rho_1(s_1), 
\end{align}
but we temporary keep the condition to  clarify the logical structure of the upcoming derivation.

As stated, we assume that we employ two types of moves: 1) a CPU intensive move that modifies $s_1$ without using the environment ${\mathcal E}$ and  2) a swapping move. In addition, the environment might influence the relative selection probabilities for choosing either 1) or 2). 
Typically, this selection probability will 
depend on 
 $N_a({\mathcal E})$, the number of unoccupied 
ensembles in ${\mathcal E}$. Further, we need to keep in mind that during the execution
of the MC move in ensemble 1, the environment  changes. 
How much the environment changes will depend on how long it takes to fully 
execute the move involving ensemble 1.

To derive detailed-balance
relations for the replica exchange
method for cost unbalanced ensembles,  we start with the more general balance concept;
if we have an infinite number of states distributed according to the equilibrium distribution, all of which make a MC move at the same time, then we have to get the equilibrium distribution again.
This means that
the flux out off $s_1$ should be equal to the flux into $s_1$ which  can be written as
\begin{align}
&\int \rho_1(s_1|{\mathcal E})  \rho({\mathcal E}) \pi(s_1, {\mathcal E} \rightarrow s_1', {\mathcal E}') {\mathrm d}{\mathcal E}
{\mathrm d}{\mathcal E}' {\mathrm d}s_1'
=
%\nonumber \\
%&
\int \rho_1(s_1''|{\mathcal E}'') \rho({\mathcal E}'')
\pi(s_1'', {\mathcal E}'' \rightarrow s_1, {\mathcal E}''') {\mathrm d}{\mathcal E}''
{\mathrm d}{\mathcal E}''' {\mathrm d}s_1'' \label{eq:balance}
\end{align}
The transition probability $\pi(\cdot)$ can be split into the transitions via the different types moves
(that we will indicate with the Greek letter $\alpha$)
which will be 
selected with a probability $P_\alpha^{\rm sel}({\mathcal E})$ that
can depend on the environment ${\mathcal E}$:
\begin{align}
 &\pi(s_1, {\mathcal E} \rightarrow s_1', {\mathcal E}')=
 \sum_\alpha P_\alpha^{\rm sel} ({\mathcal E})
 \pi_\alpha(s_1, {\mathcal E} \rightarrow s_1', {\mathcal E}') \label{eq:palpha}
\end{align}
This shows another complicating factor 
as in standard detailed-balance we need to consider the probability 
that the exact reverse move will be executed once the new state has been established.
However,
as the environment could have changed,
the  reverse move might involve different selection probabilities.

By substituting Eq.~\ref{eq:palpha}
into Eq.~\ref{eq:balance}, we 
get an extra summation over $\alpha$ in addition 
to the integrals:
\begin{align}
 &\sum_\alpha
 &\int \rho_1(s_1|{\mathcal E})  \rho({\mathcal E}) 
  P_\alpha^{\rm sel} ({\mathcal E})
 \pi_\alpha(s_1, {\mathcal E} \rightarrow s_1', {\mathcal E}') {\mathrm d}{\mathcal E}
{\mathrm d}{\mathcal E}' {\mathrm d}s_1'
=
%\nonumber \\
%&
 &\sum_\alpha 
\int \rho_1(s_1''|{\mathcal E}'') \rho({\mathcal E}'')
P_\alpha^{\rm sel} ({\mathcal E''})
\pi_\alpha(s_1'', {\mathcal E}'' \rightarrow s_1, {\mathcal E}''') {\mathrm d}{\mathcal E}''
{\mathrm d}{\mathcal E}''' {\mathrm d}s_1'' \label{eq:sumalpha}
\end{align}
But
at this point, we apply the first level of "detailedness"
by requiring the equation to hold for 
\emph{each} $\alpha$:
\begin{align}
\label{eq:eachalpha} 
&\int \rho_1(s_1|{\mathcal E})  \rho({\mathcal E})
P_\alpha^{\rm sel}({\mathcal E})
\pi_\alpha(s_1, {\mathcal E} \rightarrow s_1', {\mathcal E}') {\mathrm d}{\mathcal E}
{\mathrm d}{\mathcal E}' {\mathrm d}s_1'
=  
%\\
%& 
\int \rho_1(s_1''|{\mathcal E}'') \rho({\mathcal E}'')
P_\alpha^{\rm sel}({\mathcal E}'')
\pi_\alpha(s_1'', {\mathcal E}'' \rightarrow s_1, {\mathcal E}''') {\mathrm d}{\mathcal E}''
{\mathrm d}{\mathcal E}''' {\mathrm d}s_1'' 
\end{align}
So now we can evaluate the different moves separately.
We further simplify this expression 
by integration out the variables 
${\mathcal E}'$ and ${\mathcal E}'''$
using the following 
relation:
\begin{align}
\int    \pi_\alpha(s, {\mathcal E} \rightarrow s', {\mathcal E}')
\mathrm{d} {\mathcal E}'=
\pi_\alpha(s, {\mathcal E} \rightarrow s', 
%^a \hskip -3.5pt {\mathcal E}
\textrm{$^a \hskip -2pt {\mathcal E}$}
)
\label{eq:anyE}
\end{align}
where 
$^a \hskip -2pt {\mathcal E}$
refers to \emph{any}
possible environment.
Substitution of Eq.~\ref{eq:anyE}
in Eq.~\ref{eq:eachalpha} gives:
\begin{align}
\label{eq:eachalpha2} 
&\int \rho_1(s_1|{\mathcal E})  \rho({\mathcal E})
P_\alpha^{\rm sel}({\mathcal E})
\pi_\alpha(s_1, {\mathcal E} \rightarrow s_1', 
\textrm{$^a \hskip -2pt {\mathcal E}$}
) {\mathrm d}{\mathcal E}
 {\mathrm d}s_1'
=  
%\\
%& 
\int \rho_1(s_1''|{\mathcal E}'') \rho({\mathcal E}'')
P_\alpha^{\rm sel}({\mathcal E}'')
\pi_\alpha(s_1'', {\mathcal E}'' \rightarrow s_1, 
\textrm{$^a \hskip -2pt {\mathcal E}$}
) {\mathrm d}{\mathcal E}''
 {\mathrm d}s_1'' 
\end{align}

First, we consider $\alpha=1$ referring the CPU intensive move
that only operates in  ensemble 1.
For this move we  substitute 
$\alpha=1$
in Eq.~\ref{eq:eachalpha2} and
replace ${\mathcal E}''$ and $s_1''$ with 
respectively
${\mathcal E}$ and $s_1'$, which is allowed
since these are dummy integration variables
\begin{align}
&\int \rho_1(s_1|{\mathcal E})  \rho({\mathcal E})
P_1^{\rm sel}({\mathcal E})
\pi_1(s_1, {\mathcal E} \rightarrow s_1', 
\textrm{$^a \hskip -2pt {\mathcal E}$}
) {\mathrm d}{\mathcal E}
 {\mathrm d}s_1'
=  %\label{eq:bal_ens1} 
%\\
%& 
\int \rho_1(s_1'|{\mathcal E}) \rho({\mathcal E})
P_1^{\rm sel}({\mathcal E})
\pi_1(s_1', {\mathcal E} \rightarrow s_1, 
\textrm{$^a \hskip -2pt {\mathcal E}$}
) {\mathrm d}{\mathcal E}
 {\mathrm d}s_1' \nonumber
\end{align}
Then, we
fix another level of detailedness by requiring 
that the integrands at the left and right side of equality sign to be 
identical for any ${\mathcal E}$ and $s_1'$.
As a result, 
$\rho({\mathcal E})
P_\alpha^{\rm sel}({\mathcal E})$ will cancel out such that we can write
\begin{align}
\label{eq:rhopi}
&\rho(s_1|{\mathcal E})  
%\rho({\mathcal E})
%P_\alpha({\mathcal E})
\pi_1(s_1, {\mathcal E} \rightarrow s_1', 
\textrm{$^a \hskip -2pt {\mathcal E}$}
)=
%\\&
\rho(s_1'|{\mathcal E}) 
%\rho({\mathcal E})
%P_\alpha({\mathcal E})
\pi_1(s_1', {\mathcal E} \rightarrow s_1, 
\textrm{$^a \hskip -2pt {\mathcal E}$}
) 
\end{align}
Since in move 1) the ensembles progress independently from each other, we have
\begin{align}
\label{eq:pi1}
%$
\pi_1(s_1, {\mathcal E} \rightarrow s_1', 
\textrm{$^a \hskip -2pt {\mathcal E}$}
)=
\pi_1(s_1 \rightarrow s_1')\pi_1({\mathcal E} \rightarrow 
\textrm{$^a \hskip -2pt {\mathcal E}$}
)
%$. 
\end{align}
The subscript "1" in $\pi_1({\mathcal E} \rightarrow$
$^a \hskip -2pt {\mathcal E}
)$
might seem contradictory to 
the previous statement on independent progression,
but it just indicates that the points in time at which
 the environment is evaluated
relates the duration of the MC move in ensemble 1: ${\mathcal E}$ 
is the environment
at the start of the MC move in ensemble 1, and  
$^a \hskip -2pt {\mathcal E}$ 
is that when the move is completed.
As the time for a
$s_1 \rightarrow s_1'$
move is likely not the same
as the time for a
$s_1' \rightarrow s_1$ move,
the final environments
are likely not the same.
However,
$^a \hskip -2pt {\mathcal E}$ 
refers to \emph{any} environment.
Hence, by substituting
Eq.~\ref{eq:pi1} into Eq.~\ref{eq:rhopi}, 
$\pi_1({\mathcal E} \rightarrow$
$^a \hskip -2pt {\mathcal E}
)$ does not only cancel as it appears at both sides of the equals sign, it is also equal to one. We therefore have not just one, but two very good reasons to eliminate this term such that:
\begin{align}
&\rho_1(s_1|{\mathcal E})  
\pi_1(s_1 \rightarrow s_1')=
\rho_1(s_1'|{\mathcal E}) 
\pi_1(s_1' \rightarrow s_1) 
\end{align}
or, via Eq.~\ref{eq:s1independ}:
\begin{align}
&\rho_1(s_1)  
\pi_1(s_1 \rightarrow s_1')=
\rho_1(s_1') 
\pi_1(s_1' \rightarrow s_1) 
\end{align}
This equation essentially the same as the standard detailed balance equation such that we can adapt our acceptance
according to
\begin{align}
P_{\rm acc}(
s_1\rightarrow s_1'
)={\rm min}\left[ 1, \frac{\rho_1(s_1') P_{\rm gen}(s_1'\rightarrow s_1)}{
\rho_1(s_1) P_{\rm gen}(s_1\rightarrow s_1')
}\right] \label{eq:Pacc1}
\end{align}
which is exactly the same as in standard Metropolis-Hastings.
Still, the underlying philosophy is different from a super-state perspective as the number of transitions from old to new,
$S^{(o)} \rightarrow S^{(n)}$,
is not the same as from new to old, $S^{(n)} \rightarrow S^{(o)}$.
Instead, by writing $S=(s_1,{\mathcal E})$ we have that 
the number of 
$(s_1^{(o)}, {\mathcal E}^{(o)}) \rightarrow 
(s_1^{(n)},
\textrm{$^a \hskip -2pt {\mathcal E}$}^{(n)}
)$ transitions should be equal to 
the number of 
$(s_1^{(n)}, {\mathcal E}^{(o)}) \rightarrow 
(s_1^{(o)}, 
\textrm{$^a \hskip -2pt {\mathcal E}$}^{(n)}
)$ transitions.
In addition,  as at the end of the move we only update 
ensemble 1, and not those that are here considered as environment,
the number of sampled states in the ensembles do not increase in cohort. Sampling all states simultaneously like in a true 
superstate move would imply that distributions get mixed 
with the unknown and unphysical $\rho_i^u$ distributions.

For the swapping move we just consider the example of an attempted 
$1\leftrightarrow2$ swap as all other swaps 
$i \leftrightarrow j$ are completely analogous.
We start again at Eq.~\ref{eq:eachalpha}
with $\alpha=1\leftrightarrow2$, and further 
we split the environment
${\mathcal E}=\{s_2, {\mathcal E}_{\cancel{2}}\}$
into the part that participates in the swap move, $s_2$, and the rest, ${\mathcal E}_{\cancel{2}}$:
\begin{align}
&\int \rho_1(s_1|s_2, {\mathcal E}_{\cancel{2}})  \rho_2(s_2)
\rho({\mathcal E}_{\cancel{2}})
P_{1\leftrightarrow2}^{\rm sel}({\mathcal E}_{\cancel{2}}) \times 
%\nonumber \\
%&
\pi_{1\leftrightarrow2}(s_1, s_2, {\mathcal E}_{\cancel{2}} \rightarrow s_1', s_2', {\mathcal E}_{\cancel{2}}') {\mathrm d}s_2 {\mathrm d} {\mathcal E}_{\cancel{2}}
{\mathrm d}s_2' {\mathrm d} {\mathcal E}_{\cancel{2}}'
 {\mathrm d}s_1'
=  
\nonumber \\
&
\int \rho_1(s_1''|s_2'', {\mathcal E}_{\cancel{2}}'')  \rho_2(s_2'')
\rho({\mathcal E}_{\cancel{2}}'')
P_{1\leftrightarrow2}^{\rm sel}({\mathcal E}_{\cancel{2}}'') 
\times 
%\nonumber \\
%&
\pi_{1\leftrightarrow2}(s_1'', s_2'', {\mathcal E}_{\cancel{2}}'' \rightarrow s_1, s_2''', {\mathcal E}_{\cancel{2}}''') {\mathrm d}s_2'' {\mathrm d} {\mathcal E}_{\cancel{2}}''
{\mathrm d}s_2''' {\mathrm d} {\mathcal E}_{\cancel{2}}'''
 {\mathrm d}s_1''
 \label{eq:re}
\end{align}
Here, we assume that
the selection probability 
$P_{1\leftrightarrow2}^{\rm sel}$
depends on ${\mathcal E}_{\cancel{2}}$. The chance to do a replica exchange move 
equals $P_{\rm RE}$, but once it is decided to perform a replica exchange move, all 
possible swaps $i \leftrightarrow j$ compete to be selected with an equal probability.
Hence, the probability for 
the $1 \leftrightarrow 2$ swap to be selected depends on the number of available
ensembles, which is the total number of ensembles minus the number of occupied ones.
This latter information is contained in ${\mathcal E}_{\cancel{2}}$

The swapping transition probability 
$\pi_{1\leftrightarrow2}$ 
relates to a move that has only one possible outcome, namely the one in which 
the states in ensemble 1 and 2 are exchanged. Therefore, 
$\pi_{1\leftrightarrow2}(s_1, s_2, {\mathcal E}_{\cancel{2}} \rightarrow s_1', s_2', {\mathcal E}_{\cancel{2}}')$ 
is vanishing if $s_1' \neq s_2$ and $s_2' \neq s_1$.
Likewise, 
$\pi_{1\leftrightarrow2}(s_1'', s_2'', {\mathcal E}_{\cancel{2}}'' \rightarrow s_1, s_2''', {\mathcal E}_{\cancel{2}}''')$ vanishes if
$s_2'' \neq s_1$ and $s_1'' \neq s_2'''$.
We can, therefore, write 
\begin{align}
\pi_{1\leftrightarrow2}(s_1, s_2, {\mathcal E}_{\cancel{2}} \rightarrow s_1', s_2', {\mathcal E}_{\cancel{2}}')&=
\hat{\pi}_{1\leftrightarrow2}(s_1, s_2, {\mathcal E}_{\cancel{2}} \rightarrow s_2, s_1, {\mathcal E}_{\cancel{2}}')
\delta(s_2-s_1')  \delta(s_1 - s_2')
\nonumber \\
\pi_{1\leftrightarrow2}(s_1'', s_2'', {\mathcal E}_{\cancel{2}}'' \rightarrow s_1, s_2''', {\mathcal E}_{\cancel{2}}''')
&=
\hat{\pi}_{1\leftrightarrow 2}
(s_2''', s_1, {\mathcal E}_{\cancel{2}}'' 
\rightarrow s_1, s_2''', {\mathcal E}_{\cancel{2}}''')
\delta(s_2'''-s_1'')  \delta(s_1 - s_2'')
\label{eq:delta}
\end{align}
where the 
transition probability with the hat,
$\hat{\pi}_{1\leftrightarrow 2}$,
differs from transition probability without the hat,
${\pi}_{1\leftrightarrow 2}$,
by the fact that the latter considers any potential (even if impossible) result of
the swapping operation, while the former actually 
relates to
the probability of successfully executing the move in practice in which $s_1$ and $s_2$ change places.  
Substitution of Eqs.~\ref{eq:delta} in Eq.~\ref{eq:re} allows us to eliminate 
the integrals over $s_1'$, $s_2'$, $s_1''$, and $s_2''$
via the delta-function integration property.
\begin{align}
&\int \rho_1(s_1|s_2, {\mathcal E}_{\cancel{2}})  \rho_2(s_2)
\rho({\mathcal E}_{\cancel{2}})
P_{1\leftrightarrow2}^{\rm sel}({\mathcal E}_{\cancel{2}}) 
%\times 
%\nonumber \\
%&
\hat{\pi}_{1\leftrightarrow2}(s_1, s_2, {\mathcal E}_{\cancel{2}} \rightarrow s_2, s_1, {\mathcal E}_{\cancel{2}}') {\mathrm d}s_2 {\mathrm d} {\mathcal E}_{\cancel{2}}
 {\mathrm d} {\mathcal E}_{\cancel{2}}'
=  \nonumber \\
&\int \rho_1(s_2'''|s_1, {\mathcal E}_{\cancel{2}}'')  \rho_2(s_1)
\rho({\mathcal E}_{\cancel{2}}'')
P_{1\leftrightarrow2}^{\rm sel}({\mathcal E}_{\cancel{2}}'') 
%\times 
%\nonumber \\
%&
\hat{\pi}_{1\leftrightarrow2}(s_2''', s_1, {\mathcal E}_{\cancel{2}}'' \rightarrow s_1, s_2''', {\mathcal E}_{\cancel{2}}''')  {\mathrm d} {\mathcal E}_{\cancel{2}}''
{\mathrm d}s_2''' {\mathrm d} {\mathcal E}_{\cancel{2}}'''
 \label{eq:re2}
\end{align}
We then eliminate the integrals
over 
${\mathcal E}_{\cancel{2}}'$
and ${\mathcal E}_{\cancel{2}}'''$ 
 using a similar expression as Eq.~\ref{eq:anyE}.
\begin{align}
&\int \rho_1(s_1|s_2, {\mathcal E}_{\cancel{2}})  \rho_2(s_2)
\rho({\mathcal E}_{\cancel{2}})
P_{1\leftrightarrow2}^{\rm sel}({\mathcal E}_{\cancel{2}}) 
%\times 
%\nonumber \\
%&
\hat{\pi}_{1\leftrightarrow2}(s_1, s_2, {\mathcal E}_{\cancel{2}} \rightarrow s_2, s_1, 
\textrm{$^a \hskip -2pt {\mathcal E}_{\cancel{2}}$}
) {\mathrm d}s_2 {\mathrm d} {\mathcal E}_{\cancel{2}}
=  \nonumber \\
&\int \rho_1(s_2'''|s_1, {\mathcal E}_{\cancel{2}}'')  \rho_2(s_1)
\rho({\mathcal E}_{\cancel{2}}'')
P_{1\leftrightarrow2}^{\rm sel}({\mathcal E}_{\cancel{2}}'') 
%\times 
%\nonumber \\
%&
\hat{\pi}_{1\leftrightarrow2}(s_2''', s_1, {\mathcal E}_{\cancel{2}}'' \rightarrow s_1, s_2''', 
\textrm{$^a \hskip -2pt {\mathcal E}_{\cancel{2}}$}
)  {\mathrm d} {\mathcal E}_{\cancel{2}}''
{\mathrm d}s_2''' 
 \label{eq:re3}
\end{align}

In the next step, we change some of the dummy 
integration variable
names: 
$s_2'''$ to $s_2$
and 
${\mathcal E}_{\cancel{2}}''$
to 
${\mathcal E}_{\cancel{2}}$.
\begin{align}
&\int \rho_1(s_1|s_2, {\mathcal E}_{\cancel{2}})  \rho_2(s_2)
\rho({\mathcal E}_{\cancel{2}})
P_{1\leftrightarrow2}^{\rm sel}({\mathcal E}_{\cancel{2}}) 
%\times 
%\nonumber \\
%&
\hat{\pi}_{1\leftrightarrow2}(s_1, s_2, {\mathcal E}_{\cancel{2}} \rightarrow s_2, s_1, 
\textrm{$^a \hskip -2pt {\mathcal E}_{\cancel{2}}$}
) {\mathrm d}s_2 {\mathrm d} {\mathcal E}_{\cancel{2}}
=  \nonumber \\
&\int \rho_1(s_2|s_1, {\mathcal E}_{\cancel{2}})  \rho_2(s_1)
\rho({\mathcal E}_{\cancel{2}})
P_{1\leftrightarrow2}^{\rm sel}({\mathcal E}_{\cancel{2}}) 
%\times 
%\nonumber \\
%&
\hat{\pi}_{1\leftrightarrow2}(s_2, s_1, {\mathcal E}_{\cancel{2}} \rightarrow s_1, s_2, 
\textrm{$^a \hskip -2pt {\mathcal E}_{\cancel{2}}$}
)  {\mathrm d} {\mathcal E}_{\cancel{2}}
{\mathrm d}s_2
 \label{eq:re3}
\end{align}
and use a detailed-balance principle by stating that the
equality does not only hold when integrated, but is true 
for any pair $s_2, {\mathcal E}_{\cancel{2}}$.  
\begin{align}
 &\rho_1(s_1|s_2, {\mathcal E}_{\cancel{2}})  \rho_2(s_2)
\hat{\pi}_{1\leftrightarrow2}(s_1, s_2, {\mathcal E}_{\cancel{2}} \rightarrow s_2, s_1, 
\textrm{$^a \hskip -2pt {\mathcal E}_{\cancel{2}}$}
) 
=
%\nonumber \\
%
%&
\rho_1(s_2|s_1, {\mathcal E}_{\cancel{2}})  \rho_2(s_1)
\hat{\pi}_{1\leftrightarrow2}(s_2, s_1, {\mathcal E}_{\cancel{2}} \rightarrow s_1, s_2, 
\textrm{$^a \hskip -2pt {\mathcal E}_{\cancel{2}}$}
)  
 \label{eq:re4}
\end{align}
We further simplify $\rho_1(s_1|s_2, {\mathcal E}_{\cancel{2}})$ by
$\rho_1(s_1)$ using Eq.~\ref{eq:s1independ}, and split
$\hat{\pi}_{1\leftrightarrow2}(s_1, s_2, {\mathcal E}_{\cancel{2}} \rightarrow s_2, s_1, 
\textrm{$^a \hskip -2pt {\mathcal E}_{\cancel{2}}$}
)$
 into 
 $\hat{\pi}_{1\leftrightarrow2}(s_1, s_2 \rightarrow s_2, s_1)
 \times 
 \pi_{1\leftrightarrow2}({\mathcal E}_{\cancel{2}} \rightarrow 
 \textrm{$^a \hskip -2pt {\mathcal E}_{\cancel{2}}$}
 )$ where the latter term cancels like before:
 \begin{align}
 &\rho_1(s_1)  \rho_2(s_2)
\hat{\pi}_{1\leftrightarrow2}(s_1, s_2  \rightarrow s_2, s_1 ) 
= 
%\nonumber \\
%
%&
\rho_1(s_2)  \rho_2(s_1)
\hat{\pi}_{1\leftrightarrow2}(s_2, s_1 \rightarrow s_1, s_2 )
 \label{eq:rex}
\end{align}
Since $\hat{\pi}_{1\leftrightarrow2}(s_2, s_1 \rightarrow s_1, s_2 )$ is the transition probability
from $(s_1, s_2)$ 
to $(s_2, s_1)$ in the first two ensembles
given that the $1\leftrightarrow2$ swap move was selected,
and given that there are no other possible outcomes of 
this swap ($P_{\rm gen}=1$),  the transition probability equals the
acceptance probability:
 \begin{align}
 &\rho_1(s_1)  \rho_2(s_2)
P_{\rm acc}(s_1, s_2  \rightarrow s_2, s_1 ) 
= 
%\nonumber \\
%
%&
\rho_1(s_2)  \rho_2(s_1)
P_{\rm acc}(s_2, s_1 \rightarrow s_1, s_2 )
 \label{eq:rex2}
\end{align}
To satisfy this relation, Eq.~(4) of the main article suffices. 
\begin{align}
P_{\rm acc}
%(S^{(o)}\rightarrow S^{(n)})
=
{\rm min}\left[ 1, \frac{\rho_1(s_2^{}) \rho_2(s_1^{})}{
\rho_1(s_1^{}) \rho_2(s_2^{})
}\right] \label{eq:PaccRE}
\end{align}

So also here, 
the standard replica exchange acceptance rule applies. The main difference is that ensembles are not updated in 
cohort. After the $1 \leftrightarrow 2$ swap move 
we only update ensembles 1 and 2. Alternatively, 
after the $1 \leftrightarrow 2$ swap all 
other free ensembles will be updated as well with "null moves". 
In the example of Eq.~\ref{eq:rho5} this would mean that besides, ensemble 1 and 2, also ensemble
4 would be updated. 
As the state in this ensemble 
is not changing in a $1 \leftrightarrow 2$ swap, 
this would imply recounting the existing $s_4$ state. 
Hence, this
could be viewed as 
a superstate move, but then  
without the occupied states.
Resampling $s_4$ is allowed as the chance for resampling is independent of the content of ensemble 4.
However, the sampling of the  ensembles 3 and 5 should, while occupied, at all cost be avoided 
since the time that 
ensembles 3 and 5 remain occupied can correlate with the values of $s_3$ and $s_5$, respectively.

Like in Eq.~\ref{eq:Pacc1}, 
the acceptance rule of Eq.~\ref{eq:PaccRE} is based
on a twisted 
detailed balance 
relation: 
we require that, given an equilibrium distribution, 
the number of 
$(s_1^{(o)}, s_2^{(o)}, {\mathcal E}_{\cancel{2}}^{(o)}) \rightarrow 
(s_1^{(n)}, s_2^{(n)}, 
\textrm{$^a \hskip -2pt {\mathcal E}_{\cancel{2}}$}^{(n)}
)$ transitions should be equal to 
the number of 
$(s_1^{(n)}, s_2^{(n)},{\mathcal E}_{\cancel{2}}^{(o)}) \rightarrow 
(s_1^{(o)}, s_2^{(o)}, 
\textrm{$^a \hskip -2pt {\mathcal E}_{\cancel{2}}$}^{(n)}
)$ transitions, where 
$s_1^{(o)}=s_2^{(n)}=s_1$ and 
$s_2^{(o)}=s_1^{(n)}=s_2$.
So in this section, we proved that standard acceptance-rejection rules can be applied in a parallel 
scheme in which replica exchange moves occur only between unoccupied ensembles, such that 
ensembles are not updated in cohort. 

\section{Matrices with consecutive ones and zeros}
\label{sec:oneszeros}
If the high-acceptance approach is not applied, $w_i(X)=1$ in Eq.~(6)
%Hard-coded "6"
of the main article
and the $W$-matrix has 
rows consisting of a sequence of ones, followed a sequence of zeros.
The $P$-matrix can then be determined from Eq.~(7)
of the main article which has an ${\mathcal O}(n^2)$ scaling.
In this section we provide the proof of this equation.

Let $n_i$ be the number of ones in row $i$.
The first step to order the rows with increasing order of $n_i$.
For instance in the following $5\times 5$ matrix\\
$$
W =
\bordermatrix{ & e_1 & e_2 & e_3 & e_4 & e_5\cr
       s_1       & 1   & 1   & 0   & 0   &  0\cr
       s_2       & 1   & 1   & 1   & 1   &  0\cr
       s_3       & 1   & 1   & 1   & 0   &  0  \cr
       s_4       & 1   & 1   & 1   & 1   &  0 \cr
       s_5       & 1   & 1   & 1   & 1   &  1}
$$
we see that $s_2$, originating from an MC move in ensemble $e_2$,
is also valid
for $e_3$ and $e_4$. State $s_3$ that was created in $e_3$ only
reaches the minimal condition for that ensemble. In path sampling, 
where $s_2$ and $s_3$ are paths and 
 $e_2$, $e_3$ and $e_4$ refer to path ensembles
$[k^+]$, $[l^+]$ and $[m^+]$ with $m>l >k$,
it would mean that path $s_3$ crosses $\lambda_l$, but not $\lambda_m$, while path $s_2$
crosses at least $m-k$ 
more
additional interfaces than strictly needed 
for being a
valid trajectory in 
$e_2=[k^+]$.
As a result, 
the third row has fewer ones than
the second row. After reordering, the $W$-matrix looks as follows:
$$
W =
\bordermatrix{ & e_1 & e_2 & e_3 & e_4 & e_5 \cr
       s_1'=s_1      & 1   & 1   & 0   & 0   &  0\cr
       s_2'=s_3      & 1   & 1   & 1   & 0   &  0\cr
       s_3=s_2       & 1   & 1   & 1   & 1   &  0  \cr
       s_4'=s_4      & 1   & 1   & 1   & 1   &  0 \cr
       s_5'=s_5       & 1   & 1   & 1   & 1   &  1}
       =W[n_1,n_2, n_3, n_4, n_5]=W[2,3, 4, 4, 5]
$$
where we introduced the bracket notation
$W[\cdot]$
indicating the
number of ones in each row in which $1 \leq n_1 \leq n_2 \leq n_3 \ldots 
\leq n_{n}=n$. Likewise, we always have $n_i\geq i$.

Based on the recursive relation,
${\rm perm}(W)=\sum_{j} W_{1j}{\rm perm}(W{\{1j\}})$, and the fact that
the matrix after removing row 1 and column $j$, $W{\{1j\}}$, is identical
for any $j\leq n_1$, we can write
\begin{align}
    {\rm perm}(W[n_1, n_2, n_3, \ldots, n_n])=n_1 \times
    {\rm perm}(W[n_2-1, n_3-1, \ldots, n_n-1])
\end{align}
The permanent of the remaining matrix 
$W[n_2-1, n_3-1, \ldots, n_n-1]$
can again be written as
$(n_2-1) \times {\rm perm}(W[n_3-2, \ldots, n_n-2])$ and so on.
The permanent is, hence, equal to
\begin{align}
    {\rm perm}(W[n_1, n_2, \ldots, n_n])=
    \prod_{i=1}^n (n_i+1-i) 
\end{align}

The $P$-matrix follows from Eq.~(5) of the main article: $P_{ij}=
W_{ij} {\rm perm}(W\{ij\}/{\rm perm}(W)$. This means that 
$P_{ij}=0$ whenever $W_{ij}=0$. If $W_{ij}=1$, and $n_{i-1} <j$ or $i=1$, we have that 
for a matrix
$W[n_1, n_2,  \ldots, n_{i-1}, n_i,  n_{i+1},\ldots, n_n]$ the following matrix remains after removal of 
row $i$ and column $j$:
\begin{align}
W\{ij\}=W[n_1, n_2,  \ldots, n_{i-1}, n_{i+1}-1, \ldots, n_n-1]
\end{align}
and the permanent
\begin{align}
{\rm perm}(W\{ij\})&=
\left(\prod_{i'=1}^{i-1} (n_{i'}+1-i')\right)
\left(\prod_{i'=i+1}^{n}
(n_{i'}-1+1-(i'-1)) \right) \nonumber \\ 
&
=
\left(\prod_{i'=1}^{i-1} (n_{i'}+1-i')\right)
\left(\prod_{i'=i+1}^{n}
(n_{i'}+1-i') \right)
=\frac{{\rm perm}(W)}{(n_i+1-i)} %=P_{ij} {\rm perm}(W)
\end{align}
and, therefore, for this case we have
\begin{align}
P_{ij}=\frac{1 \times {\rm perm}(W\{ij\})}{{\rm perm}(W)}=
\frac{1}{(n_i+1-i)}.
\end{align}

If for some $k<i$, $n_k \geq j$, while 
$n_{k-1}<j$ or ${k}=1$,
we have that for a matrix $W[n_1, n_2, \ldots, 
n_{k-1}, n_k, \ldots, n_i, n_{i+1}, \ldots, n_n]$
the following matrix remains 
after removal of 
row $i$ and column $j$:
\begin{align}
W\{ij\}=W[n_1, n_2,  \ldots, 
n_{k-1}, n_k-1, n_{k+1}-1, \ldots, n_{i-1}-1, 
n_{i+1}-1, \ldots, n_n-1]
\end{align}
Therefore, the permanent of $W\{ij\}$ can
be written as
\begin{align}
{\rm perm}(W\{ij\})&=\left(\prod_{i'=1}^{k-1} (n_{i'}+1-i')\right)
\left(\prod_{i'=k}^{i-1} (n_{i'}-1+1-i')\right)
\left(\prod_{i'=i+1}^{n}
(n_{i'}+1-1-(i'-1)) \right) 
\nonumber \\
&=\left(\prod_{i'=1}^{k-1} (n_{i'}+1-i')\right)
\left(\prod_{i'=k}^{i-1} (n_{i'}-i')\right)
\left(\prod_{i'=i+1}^{n}
(n_{i'}+1-i') \right) 
\nonumber \\
&=\frac{{\rm perm}(W)}{(n_i+1-i)}
\prod_{i'=k}^{i-1} \frac{(n_{i'}-i')}{n_{i'}+1-i'}
%=P_{ij} {\rm perm}(W)
\end{align}
This gives for $P_{ij}$:
\begin{align}
P_{ij}=
\frac{1}{(n_i+1-i)}
\prod_{i'=k}^{i-1} \frac{(n_{i'}-i')}{n_{i'}+1-i'}
\end{align}
We can compare this result 
with that of one row below (row $i+1$):
\begin{align}
P_{(i+1)j} =\frac{1}{(n_{i+1}+1-(i+1))}
\prod_{i'=k}^{i} \frac{(n_{i'}-i')}{n_{i'}+1-i'}
=
\frac{P_{ij}(n_i+1-i)}{(n_{i+1}-i)}
\frac{(n_{i}-i)}{n_{i}+1-i}
=
P_{ij}
\frac{
n_i-i
}{
(n_{i+1}-i)
}
\end{align}
Therefore, we have following recursive relations
\begin{align}
P_{ij} =
\begin{cases}
  0,  & \text{ if } W_{ij}=0 \\
  \frac{1}{n_i+1-i},  &  \text{ if }  W_{ij}=1 \text{ and } [W_{(i-1)j}=0
  \text{ or } i=1] \\
  \left( \frac{n_{i-1}+1-i}{n_i+1-i}\right) P_{(i-1)j}, &  \text{ otherwise } 
  \end{cases}
\end{align}
For the example given above, this relation gives the following 
$P$-matrix:\\
$$
P =
\bordermatrix{ & e_1 & e_2 & e_3 & e_4 & e_5 \cr
       s_1'=s_1 & \frac{1}{2}  & \frac{1}{2}  & 0   & 0   &  0\cr
       s_2'=s_3 & \frac{1}{4}  & \frac{1}{4}  & \frac{1}{2}& 0   &  0\cr
       s_3=s_2  & \frac{1}{8}  & \frac{1}{8}  & \frac{1}{4}& \frac{1}{2}   &  0  \cr
       s_4'=s_4 & \frac{1}{8}  & \frac{1}{8}  & \frac{1}{4}& \frac{1}{2}   &  0 \cr
       s_5'=s_5 & 0   & 0   & 0   & 0   &  1}
$$\\
This ${\mathcal O}(n^2)$ algorithm can be done
within a second for $n \leq 3500$, bigger than any foreseeable RETIS
simulation, 
without even leveraging the block-diagonalization.
One could swap again the second and third row to get them ordered according to the original $s_i$-states, though there is in principle no need for this. This is because 
it is  irrelevant to connect the existing states to the ensembles 
in which they were originally created. 

\section{Kramer's theory}
\label{sec:Kramer}
For Langevin dynamics,
Kramer's relation provides
a way to improve 
upon 
transition state theory 
via an 
approximate
expression for the transmission coefficient:
\begin{align}
\kappa=(1/\omega_b)\left(
-\gamma/2+\sqrt{\gamma^2/4+ w_b^2}
\right)
\label{eq:kappa}
\end{align}
Here, $\gamma$ is the friction coefficient 
of the Langevin dynamics
and
$\omega_b=\sqrt{k/m}$
with $m$ the particle's mass and 
$k$ the  curvature along the reaction coordinate at the transition state.
The rate constant is then
the product of the transmission coefficient times the transition state 
theory  expression for the rate:
\begin{align}
k=\kappa k^{\rm TST}
\label{eq:kappakTST}
\end{align}
For a one-dimensional motion along 
a coordinate
$z$, the  
transition state theory expression can be expressed 
as~\cite{FrenkelBook}:
\begin{align}
k^{\rm TST}=
\sqrt{\frac{k_B T}{2 \pi m}}
\frac{e^{-\beta V(0)}}{
\int_{-\infty}^0 e^{-\beta V(z)} \mathrm{d}z
}
\label{eq:kTST}
\end{align}
where $V(\cdot)$ is the underlying potential, $T$ the temperature, $k_B$ the
Boltzmann constant, and $\beta=1/k_B T$.
The transition state is here assumed to be located at $z=0$
and the system is in state $A$,
the reactant state, if $z<0$.

The Kramer's approximation
for the rate constant $k$ follows from
Eqs.~\ref{eq:kappa}-\ref{eq:kTST}.  However, other properties like crossing probabilities
and the permeability through a membrane 
can be derived from the transmission coefficient as well.

The crossing probability $P_A(\lambda_B|\lambda_A)$ from 
interface $\lambda_A$ to interface
$\lambda_B$ follows from the
main TIS/RETIS rate equation:
\begin{align}
    k=f_A P_A(\lambda_B|\lambda_A)
\end{align}
where $f_A$ is the conditional
flux through $\lambda_A$ given the system is in state $A$. Here, $\lambda_A$ and $\lambda_B$ 
correspond to the first, $\lambda_0$,
and last interface, $\lambda_M$, respectively. 
The flux $f_A$ through 
$\lambda_A$
is similar to $k^{\rm TST}$, the flux through
the transition state without recrossing correction, as it counts 
all positive crossings and is based on the same normalization (integration over state $A$):
\begin{align}
    f_A=
\sqrt{\frac{k_B T}{2 \pi m}}
\frac{e^{-\beta V(\lambda_A)}}{
\int_{-\infty}^0 e^{-\beta V(z)
}
\mathrm{d}z
}
\label{eq:fA}
\end{align}
From Eqs.~\ref{eq:kappa}-\ref{eq:fA} we end up with an equation for the 
crossing probability:
\begin{align}
    P_A(\lambda_B|\lambda_A)
    =\frac{\kappa e^{-\beta V(0)}}{
    e^{-\beta V(\lambda_A)}}
\label{eq:Pcross}
\end{align}
Hence, based on the underlying potential and Kramer's expression, Eq.~\ref{eq:kappa}, one can obtain an approximate value for the crossing probability.
Likewise, for a membrane system
we can derive a Kramer's expression for the permeability $P$ starting from Eq.~18 in Ref.\cite{permeability}:
\begin{align}
    P=\frac{k}{(\rho_{\rm ref})_A}=
    \frac{f_A P_A(\lambda_B|\lambda_A)}{(\rho_{\rm ref})_A}
    \label{eq:perm}
\end{align}
where $\rho_{\rm ref}$ refers to the probability density for a permeant at a location away from the membrane, $z_{\rm ref}$, where $V(\cdot)$ is considered to be flat, and the subscript $(\cdot)_A$ indicates that
it is normalized over the reactant state region $A$:
\begin{align}
   (\rho_{\rm ref})_A=\frac{e^{-\beta V(
   z_{\rm ref}
   )}}{
   \int_{-\infty}^0 e^{-\beta V(z)} 
   \mathrm{d}z
   }
   \label{eq:rhoref}
\end{align}
Note that the integral in 
the denominator of
Eqs.~\ref{eq:fA} and~\ref{eq:rhoref} is usually diverging since the underlying potential $V(\cdot)$ is generally flat away from the barrier in a membrane system. Fortunately, this integral term cancels in Eq.~\ref{eq:perm}:
\begin{align}
    P=
    \sqrt{\frac{k_B T}{2 \pi m}}
    \left(
\frac{
e^{-\beta V(\lambda_A)}
}
    {
    e^{-\beta V(z_{\rm ref})}
    }
    \right)
    P_A(\lambda_B|\lambda_A)=
    \sqrt{\frac{k_B T}{2 \pi m}}
    \left(
\frac{\kappa
e^{-\beta V(0)}
}
    {
    e^{-\beta V(z_{\rm ref})}
    }
    \right)
    \label{eq:perm2}
\end{align}
where in the second equality we substituted $P_A(\lambda_B|\lambda_A)$
using Eq.~\ref{eq:Pcross}.
Hence, based on Eq.~\ref{eq:kappa} and Eq.~\ref{eq:perm2}, we can obtain a value for the permeability based on Kramer's theory.

The aforementioned equations can be generalized for multidimensional systems by replacing the $V(z)$ terms with the Landau free energy $F(z)$. That is, for one additional degree of freedom $y$:
\begin{align}
    F(z)=-k_B T \ln\left( 
\int e^{-\beta V(y,z)} {\mathrm d}y
\right)
\label{eq:Landau}
\end{align}
In addition, if multiple reaction channels
yield competing parallel saddle points in the potential energy surface, these need to summed up as we will do in the next section.

\subsection{Kramer's relation for crossing probability of a two-channel system}
The potential energy surface described in Ref.~\cite{permeability} is the following
\begin{align}
V(y,z)&=e^{-c z^2}
\left( 
V_1+A+A\sin\left(\frac{2\pi y}{L_y}\right)+B+B\cos\left(\frac{4\pi y}{Ly} \right)
\right)
\textrm{ with }
\nonumber \\
A&=(V_2-V_1)/2, \, B=V_{\rm max}/2-V_1/4-V_2/4,
%\nonumber \\
\, V_1=10, \, V_2=11, \, V_{\rm max}=20, \, c=1, \, L_y=6
%\nonumber \\
%&\gamma =5, \quad T=m=k_B=\beta=1
\end{align}
Note that the potential is periodic
along the $y$-direction such that 
$V(y,z)=V(y+L_y,z)$ and that it is zero in the limit $|z|\rightarrow \infty$.
Further, the following mass, Langevin friction coefficient and thermodynamic parameters were set in dimensionless reduced units:
$\gamma =5, \quad T=m=k_B=\beta=1$. 
The first and last interfaces were set at:
$\lambda_A=-1.5$ and $\lambda_B=1.2$.
In this case, we have two saddle points at
$(-L_y/4,0)$ and at $(+L_y/4,0)$ where the former
is slightly lower in potential energy by 
$1 k_B T$ ($V_1$ and $V_2$, respectively).
The curvatures 
can be obtained by
applying a second order Taylor expansion around $z=0$:
\begin{align}
 &V(-L_y/4,z)\approx V_1-cV_1 z^2   \Rightarrow k_1= 2 c V_1 \nonumber \\
  &V(+L_y/4,z)\approx V_2-cV_2 z^2   \Rightarrow k_2= 2 c V_2  \nonumber
\end{align}
which gives $w_{b,1}= \sqrt{20}$ and 
$w_{b,2}=\sqrt{22}$. As a result 
$\kappa_1= 0.5866, \quad
\kappa_2= 0.6002$ via Eq.~\ref{eq:kappa}. 
From this we can compute the crossing probability
based on essentially Eq.~\ref{eq:Pcross}, but using
the Landau free energy, $F(\cdot)$, by Eq.~\ref{eq:Landau}, instead of 
the potential energy,
$V(\cdot)$, and 
using both transmission coefficients for the parts along the orthogonal coordinate, $y$, where they are relevant: 
\begin{align}
P_A(\lambda_B|\lambda_A) \approx
\frac{\kappa_1 \int_{-3}^0 e^{-\beta V(y,0)} {\rm d}y
+\kappa_2 \int_0^3 e^{-\beta V(y,0)} {\rm d}y
}{\int_{-3}^3 e^{-\beta V(y,\lambda_A)} {\mathrm d}y} 
=
1.61 \cdot 10^{-5}
\end{align}
where the integrals over $y$ are taken over one period. 
Note that
the system in Ref.~\cite{permeability} actually contains 3 particles that move in this 
2D potential energy surface such that the dimension of the system is actually 6. 
However, since we follow one single target permeant and the other particles are assumed to have no influence on the 
target (the interparticle interaction was set to 0~\cite{permeability}), the effective
dimension for our analysis is 2 
with coordinates
$y$ and $z$.

The permeability then follows from Eq.~\ref{eq:perm2} with $V(\cdot)$ replaced by $F(\cdot)$, where we used 
the expression based on the crossing probability
to have the effect of the two different transmission coefficients directly included:
\begin{align}
P=
  \sqrt{\frac{k_B T}{2 \pi m}}
    \left(
\frac{\int_{-3}^{3}
e^{-\beta V(y,\lambda_A)}
\mathrm{d}y
}
    {
    \int_{-3}^{3}
    e^{-\beta V(y,z_{\rm ref})}
    \mathrm{d}y
    }
    \right)
    P_A(\lambda_B|\lambda_A)=
    \frac{1}{6} 
    \sqrt{\frac{k_B T}{2 \pi m}}
    \left( \int_{-3}^{3}
e^{-\beta V(y,\lambda_A)}
\mathrm{d}y \right)
    P_A(\lambda_B|\lambda_A) =1.37\cdot 10^{-6}
\end{align}
where we assumed that $z_{\rm ref}$ is taken far away from the membrane at $z=0$ such that
$z_{\rm ref} \ll 0$
and
$V(y,z_{\rm ref})\approx 0$.

\subsection{Kramer's relation for crossing probability of double well potential}
The double well potential is given by~\cite{TitusRev}
\begin{align}
    V(z)=k_1 z^4-k_2 z^2 \textrm{ with } k_1=1, \quad k_2=2
\end{align}
which has a transition state at $z=0$ and minima at
$z=-1$ and $z=1$.
Further is given that $T=0.07$ and  $k_B=m=1$
such that the transition state theory expression for the rate, Eq.~\ref{eq:kTST}, equals~\cite{TitusRev}:  
%\begin{align}
    $k^{\rm TST}=2.776 \cdot 10^{-7}$.
%\end{align}

The curvature at the transition state equals $2 k_2=4$ such that $w_b=2$. Together with the friction coefficient of $\gamma=0.3$, Kramer's relation, Eq.~\ref{eq:kappa}, provides a transmission coefficient: 
%\begin{align}
    $\kappa=0.9278$.
%\end{align}
%
Henceforth, by Eq.~\ref{eq:kappakTST} the
rate constant based on Kramer's theory equals:
$k=2.58\cdot 10^{-7}$.

The crossing probability follows from Eq.~\ref{eq:Pcross} where
in this case $\lambda_A=-0.99$~\cite{wf}.
From the previously determined
value for $\kappa$, we get:
$P_A(\lambda_B|\lambda_A)=
5.83\cdot 10^{-7}$

\section{Computational efficiencies}
\label{sec:efficiencies}
In this paper, the computational efficiency 
is defined as
\begin{align}
    \textrm{efficiency}=\frac{1}{\tau^{\rm eff}}
    \label{eq:eff}
\end{align}
where $\tau^{\rm eff}$ is the efficiency time~\cite{TISeff}, which is equal to the computational cost that is needed to get a statistical relative error equal to $1$ for the 
property that is computed. Here, 
$\tau^{\rm eff}$ could be expressed as the number of MD steps in path sampling 
simulations of large systems or path sampling simulations based on Ab Initio MD where the
number of force calculations 
completely
determines  
the total CPU cost. Expressing the efficiency time is this way has the advantage that it is  hardware independent.
In this article, however, we express the efficiency time in actual CPU- or wall-time seconds in order to include also the computational cost for calculating the permanents in the replica exchange move.

When a simulation is completed after a certain time $\tau$ and the relative error $\epsilon$ has been obtained via, e.g. independent runs, block averaging or bootstrapping, the 
efficiency time is estimated by
\begin{align}
    \tau^{\rm eff} = \epsilon^2 \tau
    \label{eq:taueff1}
\end{align}
Note that for serial simulations this property is in principle independent
of the simulation length $\tau$. If we increase the simulation by a certain factor, the error should reduce by the square root of this factor such that $\tau^{\rm eff}$ remains unchanged.
However, we should realize that there
is a rather large statistical uncertainty in the estimated values for  $\tau^{\rm eff}$  due the fact that the statistical error in the error is generally large.

In the following,
unless stated otherwise, we will refer
to the CPU-time and CPU-based efficiency time
when referring to $\tau$ and $\tau^{\rm eff}$.
However, let us shortly discuss the 
wall-time efficiency that follows from the 
same equation, Eq.~\ref{eq:taueff1}, but
with $\tau$ being the wall-time instead of CPU-time. 
In all our simulations, we fixed the wall-time to 5$\times$12 hours with 5 independent runs. 
So the wall-time is constant and independent 
to the number of workers that is used.
However, with $K$ workers instead of 1, the 
CPU-time increases by a factor $K$.
 This means that if the 
 error would follow the same trend as  in a serial run, 
the use of $K$ instead of 1 worker 
would 
 result in a $\sqrt{K}$ reduction
of the error. Yet, with  
 $\tau$ in Eq.~\ref{eq:taueff1} 
 being the wall-time instead of CPU-time, 
 the reduction in the error is not canceled
 by an increase in $\tau$ and 
 the efficiency, Eq.~\ref{eq:eff}, 
 would increase linearly with $K$.
 This would mean that we can write:
 \begin{align}
    \textrm{efficiency(wall-time)}=K \times
    \textrm{efficiency(CPU-time)}
    \label{eq:eff2}
\end{align}
if the parallel
run uses the total CPU-time as effectively
as a serial
simulation that
runs 
$K\times 5\times 12$ hours long.
However,
our parallel algorithm will introduce changes 
in the relative CPU-time that is used for 
MC moves in the different ensembles. This effect 
was investigated for the memoryless single variable stochastic (MSVS) process. 
In the next subsection, we give the meaning and derivation of the continuous curves shown in Fig.~1
of the main article.

\subsection{Theoretical efficiencies for the MSVS process}

The efficiency time can also be calculated for
for specific parts of the calculation.
In specific,  TIS/RETIS consists of different path ensemble simulations that compute a local crossing probability. In the path ensemble $[k^+]$
which consists of paths that at least cross $\lambda_k$, this 
local crossing probability equals the fraction of paths that cross $\lambda_{k+1}$ as well.
Based on the expected error in the local crossing probability, the CPU-based efficiency time of ensemble $[k^+]$ can be expressed as~\cite{TISeff}:
\begin{align}
    \tau^{\rm eff}_k=
    \frac{1-p_k}{p_k} {\mathcal N}_k 
\xi_k
L_k
\end{align}
where $p_k$ is the local crossing probability of
ensemble $[k^+]$, $L_k$ is the average 
path length (expressed in MD steps or CPU seconds), and
$\xi_k$ is the ratio of the average cost of a MC move
to $L_k$. In other words, $\xi_k L_k$ is the average 
computational cost for doing a MC move (creation of a trial path that might then be accepted or rejected). Finally, ${\mathcal N}_k$
is a measure of the effective correlations between MC moves also called the "statistical inefficiency".
Paths can be correlated due to rejections, which implies that the old path is recounted, or because of
similarities between accepted paths. In practice,
${\mathcal N}_k$ tends to be significantly larger than 1 while $\xi_k$ is often smaller than 1 as many rejections occur without that a trial path
needs to be fully completed.
In addition, some MC moves like the replica exchange move or the time-reversal move do not require
any MD steps.
 
In the following,  we will
neglect the effect that the replica exchange moves 
have on the errors and on the CPU-time. 
Under this assumption, 
the successive MC moves  are completely independent.
In addition, the ensemble moves are memoryless
(hence ${\mathcal N}=1$). 
The overall error can thus be computed from the  errors in the individual ensembles using standard error propagation rules 
for independent estimates. Except for the replica exchange part, the MSVS simulation 
is rejection-free such that we also have $\xi=1$. 
In addition, the 
random
artificial 
MD time for a path in ensemble in ensemble
$[k^+]$ was on average $0.1 \, k+0.1$ seconds. 
To simplify our analysis, we 
neglect the final 0.1 addition, and 
state that $L_k =a k$ with $a=0.1$.
Finally, we fixed the local crossing probability to $p_k=p=1/10$ for all ensembles 
$[k^+]$ such that
\begin{align}
    \tau^{\rm eff}_k=a\frac{1-p}{p} 
    k
    \label{eq:taueffk}
\end{align}

The relative error in 
estimate of the local crossing 
probability 
of ensemble $[k^+]$
follows from Eq.~\ref{eq:taueff1}:
\begin{align}
    \epsilon_k=
    \sqrt{
    \frac{\tau^{\rm eff}_k}{
    \tau_k
    }}
\end{align}
with $\tau_k$ the CPU-time that is spend to ensemble $[k^+]$.
Given a certain 
division of  the total 
simulation time $\tau$ into
the times 
 $(\tau_0, \tau_1, \ldots, 
\tau_{N-1})$, we can compute
the total efficiency time
by Eq.~\ref{eq:taueff1} with
\begin{align}
    \epsilon^2=\sum_{k=0}^{N-1} \epsilon_k^2 =
    \sum_{k=0}^{N-1}
    \frac{\tau^{\rm eff}_k}{
    \tau_k
    }\quad
    \textrm{ and }\quad \tau=\sum_{k=0}^{N-1} \tau_k
    \label{eq:epsandtau}
\end{align}
The first expression 
is the standard 
error propagation rule 
for the error in a final estimate that is 
obtained from
a product of independent
estimates. 

Now let us first consider standard 
TIS or the $N=K$ case.
In this simulation we would have an equal number of workers
as ensembles. Each worker is solely designated 
to a single ensemble
such that an equal amount of CPU-time is spend per ensemble
when the simulation is stopped.
So we can simply put $\tau_k=1$
such that $\tau=N$ and
\begin{align}
    \epsilon^2=\sum_{k=0}^{N-1} \tau^{\rm eff}_k=
    a\frac{1-p}{p} \sum_{k=0}^{N-1} k=
    a\frac{1-p}{p}
    \frac{1}{2}
    (N-1)N \approx \frac{a}{2}  \frac{1-p}{p} N^2
\end{align}
where in the last equality we assumed $N \gg 1$.
The efficiency time for  TIS is hence
\begin{align}
    \tau^{\rm eff} 
    \approx \frac{1}{2} a \frac{1-p}{p} N^3,
    \quad \textrm{ for TIS or } K=N
\end{align}
For serial RETIS, each ensemble is updated by a MC move before a next cycle of moves is started. As
a result, in each ensemble the same number of
MC moves are carried out such that $\tau_k \propto L_k
\propto k$. By taking $\tau_k=k$, we get that $\tau=(N-1)N/2\approx N^2/2$ and
\begin{align}
    \epsilon^2=\sum_{k=0}^{N-1} 
    \frac{
    \tau^{\rm eff}_k
    }{\tau_k}=a N \frac{1-p}{p} 
\end{align}
and the CPU-based efficiency time is exactly the same
\begin{align}
    \tau^{\rm eff} 
    \approx \frac{1}{2} a \frac{1-p}{p} N^3,
    \quad \textrm{ for RETIS or } K=1
\end{align}
This is in agreement with Ref.~\cite{TISeff}
which stated that an equal division of CPU-time or
aiming for the same error in each ensemble gives the same efficiency. Since the local crossing probability
 is the same for each ensemble, $p_k=p$,
 aiming for the same error in each ensemble is equivalent to having the same number of MC moves per 
 ensemble (if the statistical inefficiencies, ${\mathcal N}_k$, are the same).
 The optimal division of CPU-time over the 
 different ensembles is, however, $\tau_k \propto
 \sqrt{\tau^{\rm eff}_k}$~\cite{TISeff}.
 By taking $\tau_k =\sqrt{k}$, the total CPU-time becomes
 \begin{align}
    \tau=\sum_{k=0}^{N-1} \sqrt{k} 
    \approx \int_0^N \sqrt{x} {\mathrm d}x =\frac{2}{3} N^{3/2}
\end{align}
and the total error
 \begin{align}
 \epsilon^2=
  \sum_{k=0}^{N-1} 
    \frac{
    \tau^{\rm eff}_k
    }{\tau_k}=a  \frac{1-p}{p} \sum_{k=0}^{N-1} \sqrt{k} \approx a \frac{1-p}{p}  \frac{2}{3}  N^{3/2}
\end{align}
which by Eq.~\ref{eq:taueff1} results
in a slightly lower efficiency time than for TIS/RETIS:
\begin{align}
    \tau^{\rm eff} 
    \approx \frac{4}{9} a \frac{1-p}{p} N^3,
    \quad \textrm{ for an optimal division } 
\end{align}
Based on $a=p=0.1$ and $N=50$, 
the efficiency times are 
$\tau^{\rm eff}=56250$ for TIS/RETIS 
and $\tau^{\rm eff}=50000$ for the optimal division.
Naturally, the 
corresponding
CPU-time efficiencies by Eq.~\ref{eq:eff}
are 1/56250 and 1/50000. Furthermore, based on Eq.~\ref{eq:eff2},
the 
optimal wall-time efficiency and the
optimal TIS/RETIS 
wall-time efficiency are given by
$K/50000$ and $K/56250$,
respectively. These
are the 
continuous black and purple
curves in Fig.1d of the main
article.

It is interesting to observe that the optimal
TIS/RETIS CPU-time efficiency is only 
12.5\% lower than the optimal CPU-time
efficiency. 
This seems to suggest that it is difficult to 
improve the CPU-time efficiency of TIS and RETIS 
unless the division of CPU-time is exactly targeted
to do so. On the other hand,
one can easily get a much worse CPU-time efficiency
when errors in some ensembles are reduced to unnecessary small values while 
the other ensemble
errors are ignored.
Based on the fact that 
$\tau_k\propto \sqrt{\tau^{\rm eff}_k}$ gives the optimum,
the optimum division of MC moves 
is obtained when in ensembles $[k^+]$ 
the number of MC moves is
proportional to $\sqrt{\tau^{\rm eff}_k}/L_k$. For the MSVS system this means that
the number of executed MC moves in each ensemble should optimally be taken as $\propto 1/\sqrt{k}$ for $k=1, 2, \ldots, M-1$ (to account for 
$k=0$ we should have kept 
the neglected 0.1 addition in the path length
to avoid divergence). This means that it is actually good to execute more MC moves at the lower rank  ensembles (low $k$) than at the higher rank (high $k$). 
However, this should not be exaggerated since 
too many MC moves in the low ranked ensembles
will 
just result
in inefficient use of CPU-time as discussed above.
Based on the numerical sampling ratios,
we determined the CPU-time spend in each ensemble, $\tau_k$,
by multiplying these ratios by $L_k=ak$.
We then estimated the error based on Eqs.~\ref{eq:epsandtau} and \ref{eq:taueffk}. The resulting  efficiency,
based on the
actual
sampling ratios of $\infty$RETIS,
turned out to give a slightly better CPU-time efficiency than that of TIS/RETIS.
The resulting wall-time efficiencies of this hybrid theoretical/numerical 
result is shown by the purple dots if Fig.1d as well.
This shows that 
$\infty$RETIS
can actually improve both the CPU- and wall-time efficiency compared to TIS/RETIS.
The latter
is expected based on the brute force
principle that more CPU power is 
used per second. The former is more subtle and related
to the fact that $\infty$RETIS 
leads to a more efficient 
distribution of the CPU-time 
among the different
ensembles compared to TIS or RETIS.

\section{Additional simulation results}
\label{sec:add}
\subsection{Ratios of channels crossings}
\phantom{a} % here to force figure placement inside the subsection
\begin{figure}[hbt!]
\centering
\includegraphics[width=.8\textwidth]{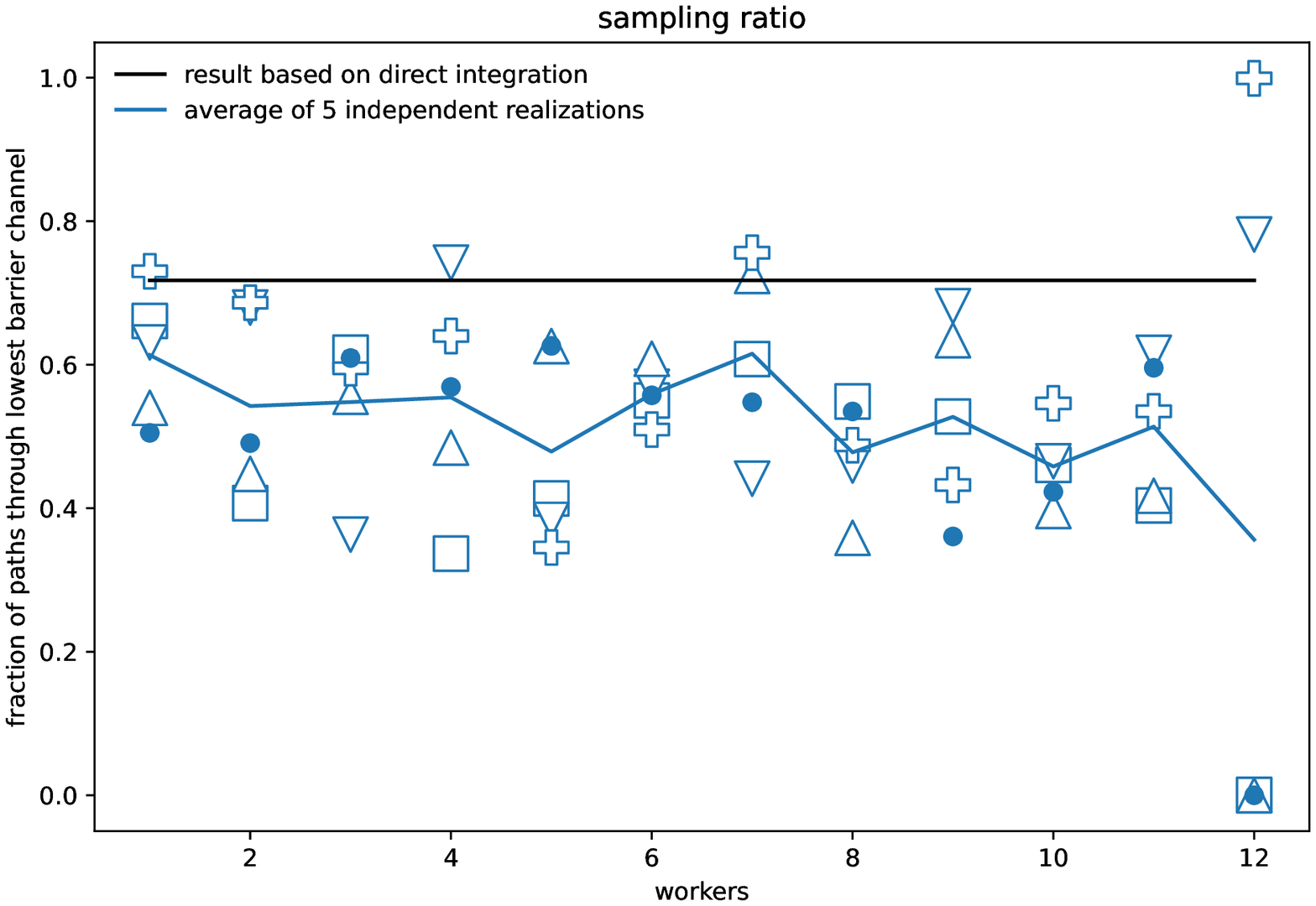}
\caption{The ratio of first crossings points for the last ensemble in the more favorable channel. The blue icons shows the sampled ratio for each simulation, the blue line is the average of 5 simulations for each amount of workers and the black line is the expected value from direct integration 
of $\exp(-\beta V(y,z))$
over $y$ with $z$ fixed at $z=\lambda_{10}=-0.2$. The $3$ icons overlapping at $0.0$ for $12$ workers is the result of the known ergodicity issues of the TIS algorithm due to the lack of replica exchange moves.}
\label{fig:ratios}
\end{figure}
\FloatBarrier % Force inside subsection
%\clearpage

\bibliography{infinityRETIS.bib}